%% file: main.tex
\documentclass[sigconf,table,xcolor,dvipsnames]{acmart}
\AtBeginDocument{%
  \providecommand\BibTeX{{%
    \normalfont B\kern-0.5em{\scshape i\kern-0.25em b}\kern-0.8em\TeX}}}

\setcopyright{acmcopyright}
\copyrightyear{2018}
\acmYear{2018}
\acmDOI{10.1145/1122445.1122456}

\acmConference[Woodstock '18]{Woodstock '18: ACM Symposium on Neural
  Gaze Detection}{June 03--05, 2018}{Woodstock, NY}
\acmBooktitle{Woodstock '18: ACM Symposium on Neural Gaze Detection,
  June 03--05, 2018, Woodstock, NY}
\acmPrice{15.00}
\acmISBN{978-1-4503-XXXX-X/18/06}

\renewcommand\footnotetextcopyrightpermission[1]{} 

\input{macros}

\newcommand{\reva}[1]{{\leavevmode\color{black}{#1}}} 
\newcommand{\revb}[1]{{\leavevmode\color{black}{#1}}}
\newcommand{\revc}[1]{{\leavevmode\color{black}{#1}}}
\newcommand{\common}[1]{{\leavevmode\color{black}{#1}}}

\begin{document}
\fancyhead{} 

\title{\sys: Hypothetical Reasoning With What-If and How-To Queries Using a Probabilistic Causal Approach}

\settopmatter{authorsperrow=4}
\author{Sainyam Galhotra}
\authornote{Both authors contributed equally to this research.}
\affiliation{%
\institution{University of Chicago}
\country{}
}
\email{sainyam@uchicago.edu}
\author{Amir Gilad}
\authornotemark[1]
\affiliation{%
  \institution{Duke University}
  \country{}
}
\email{agilad@cs.duke.edu}

\author{Sudeepa Roy}
\affiliation{%
  \institution{Duke University}
  \country{}
}
\email{sudeepa@cs.duke.edu }

\author{Babak Salimi}
\affiliation{%
  \institution{University of California, San Diego}
  \country{}
}
\email{bsalimi@ucsd.edu}

\begin{abstract} 
What-if (provisioning for an update to a database) and how-to (how to modify the database to achieve a goal) analyses provide insights to users who wish to examine hypothetical scenarios without making actual changes to a database and thereby help plan strategies in their fields. 
Typically, such analyses are done by testing the effect of an update in the existing database on a specific view created by a query of interest. 
In real-world scenarios, however, an update to a particular part of the database may affect tuples and attributes in a completely different part due to implicit semantic dependencies. 
To allow for hypothetical reasoning while accommodating such 
dependencies, we develop \sys, a framework that supports what-if and how-to queries accounting for 
probabilistic dependencies among attributes captured by a probabilistic causal model.
We 
extend the SQL syntax to include the necessary operators for expressing these hypothetical queries, define their semantics, devise efficient algorithms and optimizations to compute their results using concepts from causality and probabilistic databases, and  
evaluate the effectiveness of our approach experimentally.
\end{abstract}

\maketitle

\input{intro}
\input{model2}

\input{whatif}
\input{howto}
\input{experiments}
\input{related}

\input{conclusions}


\clearpage
\balance
\bibliographystyle{abbrv}
 \bibliography{bibtex}

\clearpage
 \input{proofs/decomposition-proof}
  \input{proofs/pre-post-split-lemma}

\input{proofs/singleblockproof}
  \input{proofs/multi-relation-extension}
\input{proofs/implementation}
\end{document}

%% file: macros.tex
\usepackage{multirow}
\usepackage{graphicx}
\usepackage{listings}
\usepackage{commath}
\usepackage{hyperref}
\usepackage{url}
\usepackage{subcaption}
\usepackage[skip=0.5pt]{caption}
\usepackage{tabularx}
\usepackage[linesnumbered,ruled,vlined]{algorithm2e}
\usepackage{tikz}
\usepackage{bbm}
\usepackage{paralist}
\usepackage{ulem}
\usepackage{xcolor}

\usetikzlibrary{trees,arrows.meta,chains,fit,shapes,calc}

\lstnewenvironment{VerbatimText}[1][]{
    
    \lstset{fancyvrb=true,basicstyle=\footnotesize,captionpos=b,xleftmargin=2em,#1}
}{}
\newcommand{\paratitle}[1]{\vspace{1mm}\noindent\textbf{{#1}.}}

\definecolor{mygreen}{rgb}{0,0.6,0}
\definecolor{mygray}{rgb}{0.5,0.5,0.5}
\definecolor{mymauve}{rgb}{0.58,0,0.82}

\newcommand{\key}{{\mathbf {key}}}

\newcommand{\Pa}{\mb{Pa}}

\DeclareMathOperator*{\argmax}{argmax}

\newcommand{\pr}{{\tt \mathrm{Pr}}}
\newcommand{\sys}{{\textsc{HypeR}}}

\newcommand{\mc}[1]{\mathcal{#1}}

\newcommand{\ignore}[1]{}

\newcommand*{\rom}[1]{\expandafter\@slowromancap\romannumeral #1@}

\newcommand{\babak}[1]{}
  \newcommand{\amir}[1]{}
\newcommand{\sainyam}[1]{}
\newcommand{\sg}[1]{}

\newcommand{\indep}{\mbox{$\perp\!\!\!\perp$}}

\usepackage{enumitem}

\def\Dom{\texttt{Dom}}
\def\attr{\texttt{Attr}}

\def\DDom{\mathbf{Dom}}

\newcommand{\pred}{{\mu}}
\newcommand{\expectation}{\mathbb{E}}
\newcommand{\exo}{\mc V} 
\newcommand{\cview}{{{\mathcal V}^{rel}}}
\newcommand{\cviewpwd}{{{\mathcal V_I}^{rel}}}

\newcommand{\augmentedtable}{{relevant view}}

\newcommand{\valwhatif}{{{{\tt val}}_{\tt whatif}}}

\newcommand{\candwhatif}{{{\mathcal Q}_{whatif}}}

\newcommand{\when}{{\textsc{When}}}
\newcommand{\select}{{\textsc{Select}}}
\newcommand{\pre}{{\tt {\textsc{Pre}}}}
\newcommand{\post}{{\textsc{Post}}}
\newcommand{\update}{{\textsc{Update}}}
\newcommand{\howto}{{\tt {\textsc{HowToUpdate}}}}

\newcommand{\suchthat}{{\tt {\textsc{Limit}}}}
\newcommand{\inht}{{\tt {\textsc{In}}}}
\newcommand{\maxi}{{\tt {\textsc{ToMaximize}}}}
\newcommand{\mini}{{\tt {\textsc{ToMinimize}}}}
\newcommand{\where}{{\textsc{Where}}}
\newcommand{\with}{{{\textsc{With}}}}
\newcommand{\outputw}{{\textsc{Output}}}
\newcommand{\forw}{{\textsc{ For}}}
\newcommand{\use}{{{\textsc{Use}}}}

\newcommand{\groupby}{{\tt \textsc{Group By}}}

\newcommand{\from}{{\tt \textsc{From}}}
\newcommand{\ct}{{\tt \textsc{Count}}}
\newcommand{\avg}{{\tt \textsc{Avg}}}
\newcommand{\sumsql}{{\tt \textsc{Sum}}}

\newcommand{\as}{{\tt \textsc{As}}}

\newcommand{\sqland}{{\tt \textsc{And}}}
\newcommand{\sqlor}{{\tt \textsc{Or}}}

\newcommand{\true}{{\tt true}}
\newcommand{\false}{{\tt false}}

\newcommand{\RNum}[1]{\uppercase\expandafter{\romannumeral #1\relax}}

\newcommand{\mb}[1]{{\mathbf{#1}}}

\newtheorem{theorem}{Theorem}
\newtheorem{proposition}{Proposition}
\newtheorem{example}{Example}

\newtheorem{definition}{Definition}

\newcommand{\proj}[1]{{\Pi}}
\newcommand{\sel}[1]{{\sigma}}

\newcommand{\cut}[1]{}
\newcommand{\eat}[1]{}

%% file: intro.tex
\section{Introduction}
Hypothetical reasoning is a crucial element in decision-making and risk assessment in business \cite{GolfarelliR09a,singh2013use,ZhangCSW07}, healthcare \cite{Qureshi14,RamakrishnanNDSCE04}, real estate \cite{donner2018digital}, etc. 
Such analysis is split by previous work into two categories: what-if analysis and how-to analysis. 
What-if analysis \cite{BalminPP00,LakshmananRS08,HerodotouB11} is usually meant for 
testing assumptions and projections on a particular outcome 
by allowing users to pose queries about hypothetical updates in the database and examining their effect on a query result. 
Users detail a specific hypothetical scenario whose effect they wish to examine on their view of choice and the system computes the view as if the update has been performed in the database. 
On the other hand, how-to analysis \cite{MeliouGS11,MeliouS12} has the reverse goal; users specify a target effect that they want to achieve and the system computes the appropriate hypothetical updates that have to be performed in the database to fulfill the goal.

\begin{example}\label{ex:intro1}
Consider a simplified version of the Amazon product database \cite{HeM16} shown in Figure \ref{fig:database} describing product details and product reviews. Each tuple has a unique tuple identifier next to it for clarity.
Now, consider an analyst who wants to examine the effect of laptop prices on their Amazon ratings. 
She may ask ``what would be the effect of increasing the price of Asus laptops by 10\% on their average ratings?''. 
This what-if query asks about the effect of the hypothetical update on the database (increasing the Price) on a specific view (average Rating). She may also be interested in ``what fraction of Asus laptops would have rating more than 4.0 if their price drops by \$100?'' or ``What would be the average sentiment in the reviews for cameras if their color was changed to red?". A different analyst may also be interested in maximizing the average rating of laptops reviews by changing their price. She may ask 
``how to maximize the average rating of laptops and cameras by updating the price of laptops so that it will not drop below 500 and increase above 800, and will be at most 100 away from it original value?'' or 
``How to increase average sentiment in the reviews for cameras by changing their color?"
Both queries are forms of hypothetical reasoning that can assist analysts and decision-makers in gaining insights about their products and their marketing strategies.
\end{example}


Multiple works in the database community have studied hypothetical reasoning.
A substantial part of these \cite{MeliouGS11,MeliouS12,DeutchIMT13,DeutchMT15,ArabG17,DeutchMR19} has focused on provenance updates and view manipulation as a main component for answering such queries. 
Therein, hypothetical updates are captured by changing values in the provenance and thus updating the view generated by the query of interest. However, in many real world situations, due to complex probabilistic causal dependencies between attributes of tuples that are relationally connected, updating an attribute of a tuple has collateral effects on other attributes of the same tuple, as well as attributes of other tuples. 
Such dependencies cannot be expressed and captured by provenance. We illustrate with an example.


\begin{example}
Reconsider Example \ref{ex:intro1}. The provenance of the average rating of Asus laptops will not change if the price of the laptops is augmented. Similarly, for the how-to query, the provenance of the average rating of laptops and cameras will not be affected by the change in price. Thus, previous work in databases fails to account for the collateral effect that increasing the price of a laptop may have on the user's ratings. Note that due to our lack of knowledge about the underlying process that leads to the user's ratings, we may only reason about the probabilistic effect of increasing the price on user's ratings. 
Figure \ref{fig:causal-graph} gives an intuitive description of potential dependencies between the attributes of the database in Figure \ref{fig:database}. 
For example, changing the Price of a laptop may affect its Rating (denoted as the edge from the blue Price node to the blue Rating node in Figure \ref{fig:causal-graph}). 
Furthermore, increasing the Price of Asus laptops may affect the Rating of Vaio laptops and vice versa (denoted as the edge from the red Price node to the blue Rating node in Figure \ref{fig:causal-graph}). 
In general, a directed edge stands for an effect of the outbound node on the inbound node, e.g., Price affects Rating. 
Accounting for such dependencies is crucial for sound hypothetical reasoning.  
\end{example}

\begin{figure}[t]
    \centering
	\begin{footnotesize}
    \begin{subfigure}{1\linewidth}
		\centering
        \begin{tabular}{c|c|c|c|c|c|c|c| } 
         \cline{2-7} 
         & \cellcolor[HTML]{C0C0C0} PID  & \cellcolor[HTML]{C0C0C0} Category & \cellcolor[HTML]{C0C0C0} Price &  \cellcolor[HTML]{C0C0C0} Brand & \cellcolor[HTML]{C0C0C0} Color & \cellcolor[HTML]{C0C0C0} Quality\\ 
         \hline
         $p_1$ & 1 &  Laptop & 999 & Vaio & Silver & 0.7\\ 
         $p_2$ & 2 &  Laptop & 529 & Asus & Black & 0.65 \\ 
         $p_3$ & 3 &  Laptop & 599 & HP & Silver & 0.5 \\ 
         $p_4$ & 4 &  DSLR Camera  & 549 & Canon & Black & 0.75\\ 
         $p_5$ & 5 & Sci Fi eBooks & 15.99  &Fantasy Press & Blue & 0.4 \\ 
         \hline
        \end{tabular}
        \caption{Product}\label{tbl:amazondatabase}
		\end{subfigure}
    \begin{subfigure}{1\linewidth}
        \centering
		\begin{tabular}{ c|c|c|c|c| } 
        \cline{2-5}
         & \cellcolor[HTML]{C0C0C0} PID & \cellcolor[HTML]{C0C0C0} ReviewID & \cellcolor[HTML]{C0C0C0} Sentiment & \cellcolor[HTML]{C0C0C0} Rating\\ 
         \hline
         $r_1$ & 1 & 1 & -0.95 & 2\\ 
         $r_2$ & 2 & 2 & 0.7 & 4\\ 
         $r_3$ & 2 &  3 & -0.2 & 1 \\
         $r_4$ & 3 &  3 & 0.23 & 3 \\
         $r_5$ & 3 &  5 & 0.95 & 5 \\
         $r_6$ & 4 &  5 & 0.7 & 4 \\
         \hline
        \end{tabular}
        \caption{Review}\label{tbl:amazonreview}
		\end{subfigure}
	\end{footnotesize}
	\caption{Amazon product database 
	}\label{fig:database}
\end{figure}



\begin{figure}[t]
    \centering
\begin{tikzpicture}[scale=0.6]
\begin{scope}[every node/.style={shape=rectangle, rounded corners,thick,draw,, top color=white, bottom color=blue!25}, 
other/.style={shape=rectangle, rounded corners,thick,draw,, top color=white, bottom color=red!25}]
    \node (Q) at (2.7,5.5) {\footnotesize Quality};
    \node (A) at (.3,4) {\footnotesize Category};
    \node (B) at (0,5.5) {\footnotesize Brand};
    \node (K) at (4,4) {\footnotesize Color};
    \node (C) at (6.2,4) {\footnotesize Price};
    \node (F) at (3,3) {\footnotesize Rating};
    \node (G) at (1,2) {\footnotesize Sentiment};
    \node[other] (P) at (6,2.5) {\footnotesize Price};
    \node[other] (Q2) at (5.2,5.5) {\footnotesize Quality};
\end{scope}

\begin{scope}[>={Stealth[black]},
              every edge/.style={draw=black, thick}]
    \path [->] (A) edge node {} (F);
    \path [->] (B) edge node {} (F);
    \path [->] (K) edge node {} (F);
    \path [->] (C) edge node {} (F);
    \path [->] (A) edge node {} (G);
    \path [<-] (F) edge node {} (G);
    \path [->] (B) edge node {} (K);
    \path [->] (B) edge node {} (C);
    \path [->] (B) edge node {} (A);
    
    \path [->] (Q) edge node {} (C);
    \path [->] (Q) edge node {} (F);
    \path [->] (B) edge node {} (Q);
    \path [->,bend right=70] (B) edge node {} (G);
    \path [->] (Q) edge node {} (G);
    
    \path [->,dashed] (P) edge node {} (F);
    \path [->,dashed,bend right] (Q2) edge node {} (F);
\end{scope}
\end{tikzpicture}
    \caption{A graph showing the dependencies between the attributes in the database in Figure \ref{fig:database}. Blue nodes are attributes of the same tuple and the red node is an attribute of a different tuple. A dashed edge denotes a dependency between attributes of different tuples 
    }\label{fig:causal-graph}
\end{figure}
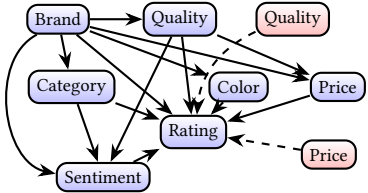

{\em In this paper, we propose a novel probabilistic framework for hypothetical reasoning in relational databases that accounts for collateral effects of hypothetical updates on the entire data.}
Our system, \sys\ ({\bf Hyp}oth{\bf e}tical {\bf R}easoning), allows users to ask complex relational what-if and how-to queries using a SQL-like declarative language. The underlying inference mechanism, then, internally accounts for the probabilistic causal effect of hypothetical updates and computes probabilistic answers to such hypothetical queries. Our framework brings together techniques from probabilistic databases \cite{,DalviS07,AntovaKO07a},  and recent advancements in inference from relational data \cite{SalimiPKGRS20,vanderweele2013social, zheleva2021causal}, to provide a principled approach for computing complex what-if and how-to queries from relational databases. Specifically, \sys\ relies on causal reasoning to capture background knowledge on probabilistic causal dependencies between attributes and interprets hypothetical updates as real world actions that potentially affect the other attributes.   

Our framework supports a rich class of what-if queries
that involve joins and aggregations to support complex real-world what-if scenarios in relational domains. \sys\ captures what-if queries through a novel model that can accommodate complex probabilistic dependencies, and computes their results efficiently by employing optimizations 
from probabilistic databases and causal inference. In addition, our framework supports complex how-to queries and frames them as an optimization problem on the search space of consistent what-if queries, and searches for a hypothetical update that
optimizes the desired query result.
\sys\ employs an efficient routine to solve this optimization problem, by expressing it as an Integer Program (IP) that can be efficiently handled using the existing IP solvers.

Our main contributions can be summarized as follows:
\begin{itemize}[leftmargin=0.2in]
    \item We propose a formal probabilistic model for hypothetical what-if and how-to queries in relational domains that combines notions from probabilistic databases and causality. Our model assigns a probability to each possible world \cite{DalviS07} that can be obtained after a hypothetical update according to the underlying probabilistic causal dependencies. We further define a probabilistic possible world semantics for complex what-if and how-to queries that support joins and aggregations.
    \item We develop a declarative language that extends the standard SQL syntax with new operators that capture hypothetical reasoning in relational domains and allow users to succinctly formulate complex probabilistic what-if and how-to queries. 
    \item Evaluating hypothetical queries in a naive manner can be inefficient due to the need to iterate over all possible worlds, or explore the space of all possible hypothetical updates. To address these, we develop a suite of  optimizations that allows \sys\ to efficiently evaluate hypothetical queries:
    \begin{itemize}
        \item We use the model of block-independent databases \cite{ReS07}, i.e., the database can be partitioned into blocks of tuples where the tuples in different blocks are independent, meaning there are no causal dependencies between the tuples across different blocks (without background knowledge, we assume tuple independence). 
        We then show that what-if queries can be evaluated independently within each block and 
        the results can be combined to get the result over the entire database.
        \item We further show that under some assumptions complex what-if queries in relational domains can be evaluated using the existing techniques in causal inference and machine leaning.
        \item We frame how-to queries as an optimization problem and develop an efficient mechanism to solve this optimization problem, by expressing it as an Integer Program (IP) that can be efficiently handled using the existing IP solvers. 
    \end{itemize}
    \item We perform an extensive experimental evaluation of \sys\ on both real and synthetic data. On real datasets, we show that the query output by \sys\ matches the conclusions from prior studies in fair and explainable AI~\cite{GalhotraPS21}. On synthetic datasets, we show that  \sys's query output is accurate as compared to other baselines. Running time analysis shows that both what-if and how-to components of \sys\ are highly efficient. 
\end{itemize}

%% file: model2.tex
\section{Probabilistic Updates in \sys}\label{sec:model}






In this section we describe our notations and then define 
the  probabilistic hypothetical update model in \sys\ (Section~\ref{sec:model-hyp-update})
that serve as the basis for probabilistic what-if and how-to queries in the following sections. Then in Section~\ref{sec:model-PCM}, we review necessary concepts from probabilistic causal models \cite{pearl2009causality} that capture the propagation of the effect of an update through other attributes due to underlying 
dependencies between them and succinctly defines the probability distribution after updates. 
\paratitle{Notations}
Let $D$ be a 
standard multi-relational database; 
we use $D$ for both schema and instance (as a set of tuples) where it is clear from the context. For each relation $R$ in $D$, $\attr(R)$ denotes the set of attributes of $R$ and ${\mb A} = \cup_{R \in D}\attr(R)$ denotes the 
set of attributes in $D$. For attributes $A$ appearing in multiple relations, we use $R.A$ for disambiguation.  For an attribute $A \in {\mb A}$, $\Dom(A)$
denotes the domain of $A$; $A_i[t] \in \Dom(A_i)$ denotes the value of the attribute $A_i$ of the tuple $t$. 
We assume that each relation $R$ has a (primary) key, that can be a single or a combination of multiple attributes.  
For easy reference, we annotate each tuple with a unique identifier as demonstrated by the identifiers $p_i, r_j$ in Figure \ref{fig:database}. 
We assume each relation can be modeled as a set of tuples (set semantics) and, for a relation $R$, we use the notation $t\in R$ to denote a tuple in $R$.
\par
For the purpose of hypothetical updates, a subset of attributes that can change values directly or indirectly in tuples is referred to as {\bf mutable attributes}, the other attributes are {\bf immutable attributes}. The attribute that is updated in hypothetical updates is called the {\bf update attribute}, and the final effect is measured on an {\bf output attribute} as specified by the user. The update and output attributes are always mutable, and the key attributes are always immutable. 


\begin{example}\label{ex:model}
In Figure \ref{tbl:amazondatabase}, the database has two relations {\tt Product} and {\tt Review} with keys $\{{\tt PID}\}$ and $\{\tt{PID, ReviewID}\}$ respectively. 
For example, suppose $\Dom(Price) = [0,500K]$.
In tuple $p_1$, $Category[p_1] = Laptop$ and $Price[p_1] = 999$ etc. 
The mutable attributes are {\tt Price, Quality, Color, Rating, and Sentiment}, whereas {\tt Brand} and {\tt Category} are immutable. 
The update attribute is {\tt Price} in relation {\tt Product}, and the output attribute is {\tt Rating} in relation {\tt Review}.
\end{example}

We assume the update and output attributes do not appear in multiple relations, but as Example~\ref{ex:model} illustrates, they can appear in two different \revb{tuples}.

\subsection{Probabilistic Hypothetical Updates}\label{sec:model-hyp-update}
\sys\ interprets hypothetical updates in terms of real world interventions that potentially influence the value of other attributes in the data due to probabilistic dependencies between the attributes and tuples. To capture such probabilistic influence, we use the notion of \emph{possible worlds} from the literature of probabilistic databases \cite{DalviS07} 
as the set of all possible instances on the same schema  with the same number of tuples in each relation that may contain different values in their mutable  attributes 
from the appropriate domains. 

\begin{definition}[Possible worlds]\label{def:possible-worlds}
Let $R$ in $D$ be a relation where in $\attr(R)$,  $A_1, \cdots, A_m$ are immutable attributes (including keys) and $B_1, \cdots, B_\ell$ are mutable attributes.
For a tuple $t \in R$, a {\bf possible world  of tuple $t$} is the set  (assuming values are associated with corresponding attribute names for disambiguation)
\begin{small}
$$PWD(t) = \{A_1[t], \cdots, A_m[t], v_1, \cdots, v_{\ell}~:~ v_i \in \Dom(B_i), i = 1 \text{ to } \ell \}.$$ 
\end{small}
\noindent
{\bf The set of possible worlds of relation $R$} is $PWD(R) = \times_{t \in R} PWD(t)$. {\bf The set of possible worlds of a database $D$} is $PWD(D) = \times_{R \in D} PWD(R)$.
\end{definition}


Next we define the notion of hypothetical updates. 


\begin{definition}[Hypothetical updates]\label{def:update}
A {\bf hypothetical update} $U = u_{R, B, f, S}$ on a database $D$ is a 4-tuple that includes a relation $R$ in $D$ containing the mutable update attribute $B \in \attr(R)$, 
a subset of tuples $S\subseteq R$ where the update will be applied, 
and a function $f: \Dom(B) \to \Dom(B)$ specifying the update for attribute $B[t]$ for tuples $t \in S$ to $f(B[t])$.

\cut{
Given a relation $R$ in $D$, an update attribute $B \in \attr(R)$, 
a subset of tuples $S\subseteq R$, 
and a function $f: \Dom(B) \to \Dom(B)$, 
a {\bf hypothetical update} \red{$u_{R, B, f, S}$ maps $B[t]$ to $f(B[t])$ for all tuples $t$ in $S$.}
}
\cut{
\amir{``and leaves all other attributes of $t\in S$ and all tuples $t\not\in S$ unchanged''?}

\sr{no it does not.. precisely that was one confusion}

\sr{not sure yet $u$ maps what to what.}  
\amir{I think it's clearer if $u$ maps tuples to tuples, $u: D \to {\DDom}^m$, i.e., $u$ takes a tuple and maps it to another tuple in the domain of possible tuples (as was the previous definition)} \sr{the updates changes the entire distribution of attributes, not just B.. it may be fine to leave it like this}
}
\end{definition}
In other words, the hypothetical update $u_{R, B, f, S}$ forces all tuples in set $S$ in relation $R$ to take the value $f(B[t])$ instead of $B[t]$. In the what-if query in Example~\ref{ex:intro1}, intuitively, $R = {\tt Product}$, $S$ defines the set of Asus laptops, $B$ is {\tt Price}, and $f$ increases the price by 10\% (see Section~\ref{sec:what-if-syntax} for details). This update, in turn, may change values of other mutable attributes in $R$ or even mutable attributes in other relations $R'$ in $D$ through causal dependencies as discussed next in Section~\ref{sec:model-PCM}, eventually (possibly) changing the output attribute. These changes are likely not deterministic (e.g., changing price of a laptop does not change its reviews or their sentiments in a fixed way), therefore, we model the state of the database after a hypothetical update as a probability distribution called the \emph{post-update distribution}.

\begin{definition}[Post-update distribution]\label{def:rel-poss-worlds}
Given a database $D$ and an update $U = u_{R, B, f, S}$ (Definition~\ref{def:update}), the {\bf post-update distribution} is a  probability distribution over possible worlds, i.e., $\pr_{D,U}: PWD(D) \to [0,1]$ such that $\sum_{I\in PWD(D)} \pr_{D,U}(I) = 1$. 
\end{definition}
While the previous definition defines the post-update distribution in a generic form, there will be restrictions imposed by the hypothetical update as well as by its effect on the distribution of other attributes (e.g., for all possible worlds  with non-zero probability, the value of attribute $B$ for tuples $t \in S$ must be $f(B[t])$). We define this post-update distribution with the help of a probabilistic relational causal model in Section~\ref{sec:model-PCM}.

\input{causality}

%% file: causality.tex
\subsection{Causal Model for Probabilistic Updates}\label{sec:model-PCM}
In this paper, we use causal modeling to capture probabilistic causal dependencies between attributes in relational domains, and to account for the collateral effect of hypothetical updates on other attributes. Specifically, \sys\ rests on relational causal models, recently introduced in~\cite{SalimiPKGRS20}, which are briefly reviewed next.  

\paratitle{Probabilistic Relational Causal Models (PRCM)} 
A probabilistic relational causal model (PRCM) associated with a relational instance $D$ is a tuple $(\mb \epsilon, \exo, Pr_{\mb \epsilon}, \mb \phi)$, where $\mb \epsilon$ is a set of unobserved {\em exogenous (noise)} variables distributed according to $Pr_{\mb \epsilon}$, $\exo$ is a set of {\em endogenous ground}\footnote{The endogenous variables are called {\em ground} variables since in a PRCM the attribute $A[t]$ associated with each tuple $t$ form the variables, generating multiple variables corresponding to the same attribute,  in contrast to the standard probabilistic causal model \cite{pearl2009causality} where each attribute or feature $A$ forms a unique variable.} 
variables associated with observed attribute values of each tuple $A[t]$, for all $A \in \attr(R)$, $t \in R$ and $R \in D$, and $\phi$ is a set of {\em structural equations}. The structural equations capture the {\em causal dependencies} among the attributes and are of the form $\phi_{A_i[t]} : \Dom(Pa_{\exo}(A_i[t])) \times \Dom(Pa_{\mb \epsilon}(A_i[t])) \to \Dom(A_i[t])$, where $Pa_{\mb{\epsilon}}(A_i[t]) \subseteq \mb{\epsilon}$ and $Pa_{\exo}(A_i[t]) \subseteq \exo - \{A_i[t]\}$ respectively denote the exogenous and endogenous parents of $A_i[t]$. A  PRCM is associated with a {\em ground causal graph} $G$, whose nodes are the endogenous variables $\exo$ and whose edges are all pairs $(X,Y)$ (directed edges) such that $X \in \exo$  and $Y \in Pa_{\exo}(A_i[t])$. In this paper we assume the underling causal model is acyclic. 
Due to uncertainty over the unobserved noise variables, the structural equations can be seen a set of probabilistic dependencies\footnote{\revc{Note that it is not necessary to have relational connections through database constraints like foreign key dependencies or functional dependencies for causal dependencies and vice versa.\label{footnote:constraints} }} of the form $\pr(A[t] \mid Pa_{\mb \exo}(A[t]))$ between the attributes.  From now on, we will use $A[t]$ interchangeability to refer to both an attribute value and the ground variable associated with it.   

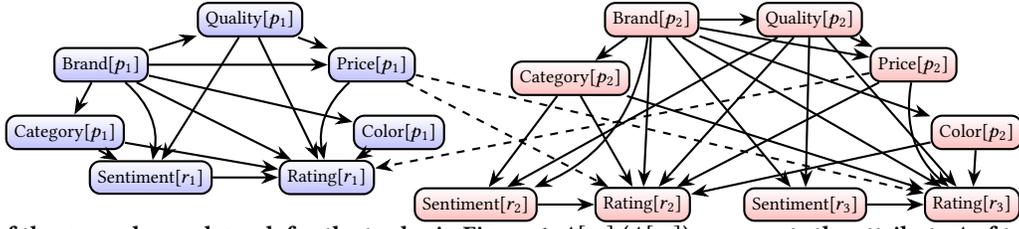
\begin{figure*}[t]
    \centering
\begin{tikzpicture}[scale=0.6]
\begin{scope}[every node/.style={shape=rectangle, rounded corners,thick,draw, top color=white, bottom color=blue!25}]
    \node (Q) at (-9.7,6.5) {\footnotesize Quality[$p_1$]};
    \node (A) at (-13.8,4) {\footnotesize Category[$p_1$]};
    \node (B) at (-13,5.5) {\footnotesize Brand[$p_1$]};
    \node (K) at (-6.4,4) {\footnotesize Color[$p_1$]};
    \node (C) at (-7,5.5) {\footnotesize Price[$p_1$]};
    \node (F) at (-8,3) {\footnotesize Rating[$r_1$]};
    \node (G) at (-11.9,3) {\footnotesize Sentiment[$r_1$]};
\end{scope}

\begin{scope}[>={Stealth[black]},
              every edge/.style={draw=black, thick}]
    \path [->] (A) edge node {} (F);
    \path [->] (B) edge node {} (F);
    \path [->] (K) edge node {} (F);
    \path [->,bend right] (C) edge node {} (F);
    \path [->] (A) edge node {} (G);
    \path [<-] (F) edge node {} (G);
    \path [->] (B) edge node {} (A);
    
    \path [->] (Q) edge node {} (C);
    \path [->] (Q) edge node {} (F);
    \path [->] (B) edge node {} (Q);
    \path [->] (Q) edge node {} (G);
    \path [->] (B) edge node {} (C);
    \path [->] (B) edge node {} (K);
    \path [->,bend left] (B) edge node {} (G);
\end{scope}
\begin{scope}[every node/.style={shape=rectangle, rounded corners,thick,draw, top color=white, bottom color=red!25}]
    \node (Q2) at (2.7,6.5) {\footnotesize Quality[$p_2$]};
    \node (A) at (-2.6,5.2) {\footnotesize Category[$p_2$]};
    \node (B) at (-.8,6.5) {\footnotesize Brand[$p_2$]};
    \node (K) at (6.4,4) {\footnotesize Color[$p_2$]};
    \node (P) at (5,5.5) {\footnotesize Price[$p_2$]};
    \node (R) at (-1,2.4) {\footnotesize Rating[$r_2$]};
    \node (G) at (-4.7,2.4) {\footnotesize Sentiment[$r_2$]};
    \node (R3) at (6.3,2.4) {\footnotesize Rating[$r_3$]};
    \node (S3) at (2.6,2.4) {\footnotesize Sentiment[$r_3$]};
\end{scope}

\begin{scope}[>={Stealth[black]},
            every edge/.style={draw=black, thick}]
    \path [->] (A) edge node {} (R);
    \path [->] (B) edge node {} (R);
    \path [->] (K) edge node {} (R);
    \path [->] (P) edge node {} (R);
    \path [->] (A) edge node {} (G);
    \path [->] (B) edge node {} (A);
    
    \path [->] (Q2) edge node {} (P);
    \path [->] (Q2) edge node {} (R);
    \path [->] (B) edge node {} (Q2);
    \path [->] (Q2) edge node {} (G);

    \path [->] (Q2) edge node {} (R3);
    \path [->] (Q2) edge node {} (S3);
    
    \path [->] (A) edge node {} (R3);
    \path [->] (B) edge node {} (R3);
    \path [->] (K) edge node {} (R3);
    \path [->,bend right] (P) edge node {} (R3);
    
    \path [->,dashed] (C) edge node {} (R);
    \path [->,dashed] (C) edge node {} (R3);
    \path [->,dashed] (P) edge node {} (F);
    
    \path [<-] (R) edge node {} (G);
    \path [<-] (R3) edge node {} (S3);
    
    \path [->] (B) edge node {} (P);
    \path [->] (B) edge node {} (K);
    \path [->,bend left] (B) edge node {} (G);
    \path [->] (B) edge node {} (S3);
    
    
\end{scope}
\end{tikzpicture}
    \caption{Part of the ground causal graph for the tuples in Figure \ref{fig:database}. 
    $A[p_i]$ ($A[r_j]$) represents the attribute $A$ of tuple $p_i$ ($r_j$). Blue nodes are related to $p_1$, red nodes are related to $p_2$, and dashed edges represent cross-tuple dependencies. Cross-tuple edges between Quality and Rating are dropped.}
    \label{fig:grounded-causal-graph}
\end{figure*}

\begin{example}
Reconsider the database in Figure \ref{fig:database} and the causal diagram in Figure \ref{fig:causal-graph}. Part of its ground version w.r.t. the database is depicted in Figure \ref{fig:grounded-causal-graph}, where the blue nodes are related to the tuple $p_1$ and the red nodes are related to the tuple $p_2$. Cross-attribute dependencies within the same tuple are illustrated as solid edges and cross-tuple dependencies between the tuples are shown as dashed edges. 
\end{example}

To be able to estimate the conditional probability distributions $\pr(A[t] \mid Pa_{\mb \exo}(A[t]))$, for $t \in R$, from the relational instance $D$, we make the following assumptions that are common in causal inference from relational data~\cite{SalimiPKGRS20,vanderweele2013social}. First, since $Pa_{\mb \exo}(A[t])$, the set of parents of $A[t]$ may have variable cardinality for each $t \in R$, we assume there exists a distribution preserving summary function $\psi$ that projects $Pa_{\mb \exo}(A[t])$ into a fixed size vector such that $\pr(A[t] \mid Pa_{\mb \exo}(A[t]))= \pr(A[t] \mid \psi(Pa_{\mb \exo}(A[t])))$, for each $t \in R$ .   Second, we assume the conditional probability distributions $\pr(A[t] \mid \psi\big(Pa_{\mb \exo}(A[t]))\big)$ are the same for all $t \in D$, i.e., the conditional probability distributions $\pr(A_i[t] \mid \psi\big(Pa_{\exo}(A_i [t]))\big)$ are independent of a particular $t \in R$ and can be readily estimated from $D$, hence we denote them by unified notation $\pr_D(A_i \mid \psi(Pa(A_i)))$. For more discussion on these assumptions, please see~\cite{SalimiPKGRS20}.

\begin{example}\label{ex:summary-func}
Continuing Example \ref{ex:intro1}, suppose we want to update attribute Price and examine its effect on Rating. Since each product has one price but several review ratings 
in Figure \ref{fig:database}, we will summarize the Rating attribute into the Product table by, e.g., averaging the Rating for each product and price. Thus, for $p_2$, we will have $Price = 529$ and $Rating = Average(4,2) = 3$ (the average over tuples $r_2$ and $r_3$).
\end{example}


\paratitle{Post-update distribution by PRCM} We describe how the post-update distribution (Definition~\ref{def:rel-poss-worlds}) is defined using a PRCM in \sys. Given a relation $R$ in $D$, an update attribute $B \in \attr(R)$, 
a hypothetical update $U = u_{R, B, f, S}$ (Definition~\ref{def:update}) can be interpreted as an {\em intervention} that modifies the underlying PRCM and replaces the structural equation associated with the variables $B[t]$ for all $t \in S$ with the constant $f(B[t])$. Updating $B[t]$ propagates through all relations, tuples and attributes according to the underlying PRCM. The {\em post-update} state of a tuple $t' \in R'$ in a relation $R'$ in $D$ is the {\em solutions} to each ground variable $A[t']$, for $A \in \attr(R')$, in the modified set of structural equations.  Now, the uncertainty over unobserved noise variables $\epsilon$ 
induces uncertainty over post-update states of all tuples $t'$ captured by their post-update distribution on the possible worlds (Definition~\ref{def:possible-worlds}): $\pr_{D,U}(\tau)$ for $\tau \in PWD(t')$, and 
in turn, 
the post-update distribution of the entire database $\pr_{D, U}(I)$ for $I \in PWD(D)$. 
As we will show in Section~\ref{sec:whatif-algo}, to answer what-if and how-to queries in \sys, it suffices to estimate 
the post-update conditional distributions of the form $\pr_{D,U}(Y=y \mid B=b, \mb C=c)$, where $Y, B, \mb C \in \attr(R)$, that measures the probabilistic influence of the update $U$ on subset of tuples for which $B=b$ and $\mb C=\mb c$. It is known that if $\mb C$ satisfies a graphical criterion called {\em backdoor-criterion} (see Section~\ref{sec:whatif-algo}) w.r.t. $B$ and $Y$ in the causal model $G$, then the following holds: 
\begin{small}
\begin{align} \scriptsize
    \pr_{D, U}(Y=y \mid B=b,\mb C=\mb c )= \pr_{D}(Y=y \mid B=f(b), \mb C=\mb c )  \label{eq:back}
\end{align}
\end{small}
Where, the RHS of \eqref{eq:back} can be estimated from $D$ using standard techniques in causal inference and Machine Learning. 
\amir{added:}
Equation \eqref{eq:back} also extends to multi-relation databases (see Section \ref{sec:appendix}).

\par
 \reva{\textbf{Background knowledge on causal DAG.~} While in this paper we assume the underlying causal model is available, \sys{} is designed to work with any level of background knowledge. 
 If the causal DAG is not available, \sys{} assumes a canonical causal model in which all attributes affect both the output and the updated attribute. In other words, 
\sys\ assumes \eqref{eq:back} holds for $\mb C= \attr(R)$,
i.e., all attributes are considered in the backdoor set in Equation \ref{eq:back}, ensuring that the ground truth backdoor set is a subset of $\attr(R)$.
We also examine this case experimentally in Section \ref{sec:experiments}. } 



%% file: whatif.tex

\section{Probabilistic What-If queries}\label{sec:queries}
In this section we describe the syntax of probabilistic what-if queries supported by \sys\ (Section~\ref{sec:what-if-syntax}), describe their semantics as expected value from the post-update distribution on possible worlds (Section~\ref{sec:whatif-semantics}), and present efficient algorithms and optimizations to compute the answers to what-if queries (Section~\ref{sec:whatif-algo}). 

\input{whatif-syntax}

\input{whatif-semantics}
\input{whatif-algo}

%% file: whatif-syntax.tex
\begin{figure}[t]
\vspace{-5mm}
    \small
	\centering
\begin{align*}
    &\use\  RelevantView\ \as\ \\
     & \quad \quad  (\select\ T1.PID, T1.Category, T1.Price, T1.Brand, \\
    & \quad \quad \quad \avg(Sentiment) \as\ Senti, \avg(T2.Rating) \as\ Rtng\\
    & \quad  \quad  \from\ Product~ \as\ T1, Review~ \as\ T2 \\
    & \quad \quad  \where\ T1.PID= T2.PID \\
    & \quad \quad  \groupby\ T1.PID, T1.Category, T1.Price, T1.Brand)\\
    &\when\ Brand = 'Asus' \\
    &\update(Price)=1.1 \times \pre(Price) \\
    &\outputw\ \avg(POST(Rtng)) \\
    &\forw\ \pre(Category) = `Laptop'~ \sqland\ \pre(Brand) = `Asus'\\ &\sqland\ \post(Senti) > 0.5
\end{align*}
    \caption{What-if query asking ``{\em If} the prices of all Asus products is increased by by 10\%, {\em what} would the effect on average ratings of Asus laptops having average sentiments in the reviews > 0.5 after the update?'' 
    }\label{fig:whatif-query}
    \vspace{-5mm}
\end{figure}

\subsection{Syntax of Probabilistic What-If Queries}
\label{sec:what-if-syntax}
A what-if query has two parts (see Figure~\ref{fig:whatif-query}): 
\begin{itemize}[leftmargin=0.15in]
\item  The required \use\ operator in the first part defines a single table as the {\bf \augmentedtable} 
with relevant attributes including the update and the output attribute to be used in the second part. The \use\ operator can simply mention the table name if no transformation is needed, and both update and output attributes belong to this table (e.g., `\use\ Review'). Otherwise, a standard SQL query within the \use\ operator can define this \augmentedtable\ as discussed below. 
 \item The second part includes the new operators for hypothetical what-if queries supported by \sys: the required \update\ and \outputw\ clauses for specifying the update and outcome attribute from the \augmentedtable, and optional \when\ and \forw\ clauses.
\end{itemize}
The second part takes as input the \augmentedtable, denoted $\cview$ 
(named as \emph{RelevantView} in Figure~\ref{fig:whatif-query}), as defined by the required {\bf \use\ operator} in the first part containing all relevant attributes, and therefore does not mention any table name for disambiguation in its operators. 
Recall that a hypothetical update
in \sys\ is of the form $U = u_{R, B, f, S}$, where the updated attribute $B\in \attr(R)$ in $D$, and is changed for all tuples $t \in S$ in $R$ according to the function $f$ (Definition~\ref{def:update}). In the what-if query, the \augmentedtable\ $\cview$ defined by the first part combines the update and outcome attributes (${\tt Price}$ and {\tt Rating} in Figure~\ref{fig:whatif-query}) along with other attributes used in the second part.
\cut{
in one of   
these two forms: 
(1) if $R$ contains both the update and outcome attributes as well as all other attributes $A_1, \ldots, A_k$ that are referred to in the second part of the query, then $\cview$ as well as the SQL query inside \use\ operator is $\select\ {\bf key}, A_1, \ldots, A_k~ \from\ R$, where ${\bf key}$ denotes the key of $R$.
\sr{check if `USE R' is simpler}
or (2) 
}
In particular, the SQL query defining $\cview$ includes the update attribute $B$ in the \select\ clause along with the key of $R$ (here ${\tt PID}$), and other attributes from $R$ and (in aggregated form) from other relations in $D$ that are used in the second part of the query. A group-by is performed on the attributes coming from relation $R$ 
Note that the first part always outputs a view having the same number of tuples as in $R$, which is ensured as the \select\ and \groupby\ clauses include the key of $R$. 
\par
The required {\bf \update\ operator} mentions the update attribute $B$ along with the function $f$. \sys\ allows hypothetical update functions $f$ of the form $Update(B) = <const>$, $Update(B) = <const>~ \times~ \pre(B)$, and $Update(B) = <const>~ +~ \pre(B)$, where $<const>$ is a constant specified by the user (here 1.1 models a 10\% price increase). 
{\bf $\pre(A)$ and $\post(A)$} respectively denote the value of an attribute $A$ before the hypothetical update (i.e., as given in the database instance $D$) and after the update according to the PRCM (see Sections~\ref{sec:model-PCM} and \ref{sec:whatif-semantics});
except in the operator as `\update($B$)' which defines updating the value of $B$, $\pre$ is assumed by 
default if $\pre$ or $\post$ is not explicitly mentioned in the query. 
\revc{\update\ is always performed w.r.t. the $\pre$ value of an attribute, rather than the $\post$ value which is the result of the update}. 
The optional SQL query in the \use\ 
operator defining the \augmentedtable\ can only have $\pre$ values of attributes, so $\pre$ is omitted in the 
query. Note that for immutable attributes $A$, $\pre(A) = \post(A)$.
\par
The optional {\bf \when\ operator} specifies the set $S$ in Definition \ref{def:update}; any valid SQL predicate can be used here that is defined for each tuple in the  \augmentedtable\ $\cview$, and allows selection of a subset of tuples from $\cview$, e.g., $A = <const>$, $A \in\ (\select\ \cdots \as\ A \cdots)$ etc. If the \when\ operator is not specified we assume $S = R$ and the hypothetical update is applied to all tuples in $R$. Since the update is applied to the original attribute values, it can only use $\pre(A)$ value for an attribute $A$, and therefore $\pre$ is omitted.
\par
The required {\bf \outputw\ operator} mentions the output attribute $Y$ (here ${\tt Rtng}$) on which we want to measure the effect of the hypothetical update. If $Y$ belongs to another table $R' \neq R$, the SQL query in the \use\ operator describes how $R$ and $R'$ are combined in the join condition, and a SQL aggregate operator $aggr_1$ ($SUM, AVG, COUNT$) 
is used to aggregate $Y$ (here $\avg(T2.Rating)$) to have a unique value for each tuple in $R$ identified by its key in the \augmentedtable. Note that the effect of an update is outputted as a single value, so another SQL aggregate operator $aggr$ is used in the \outputw\ clause (here again $\avg$). If the user wants to measure effects on different subsets of tuples, it can be achieved by the use of the optional \forw\ operator described below. The \outputw\ operator can only use $\post(A)$ values of attributes after the update. 
\par
\cut{
The $\use$ operator allows users to define the view with the relevant attributes for their what-if query. If attributes with many-to-many relations may have to be joined, some attributes may need to be aggregated in the view.  
Since the update can be performed on a subset of tuples $S\subseteq D$ (see Definition \ref{def:update}), the $\when$ clause describes the condition that the tuples that will be updated must satisfy, where $g$ is a predicate that selects certain tuples to be in $S$. 
$\update\ A_{c}\ BY\ f(A_{c})$ 
performs a update on the database. The operator changes the all values of $A_c$  to $f(A_{c})$ in the tuples in the set $S$, determined by the $\when$ clause, where $f_1: \DDom(A_{c1}) \to \DDom(A_{c1})$.
This hypothetical update has a collateral effect on other attributes due to the probabilistic dependency between the attributes. 
$\outputw Agg(A_t)$ defines the aggregated output of the query.
}
The output specified in the \outputw\ operator is 
computed only considering the tuples in the \augmentedtable\ $\cview$ that satisfy the conditions in the optional {\bf $\forw$ operator} (details in Section~\ref{sec:whatif-semantics}).
If no \forw\ operator is provided, all tuples in $\cview$ are used to compute the output. $\forw$ can contain both $\pre(A)$ and $\post(A)$ values of attributes, and $\pre$ can be optionally provided for clarity. Further, like \when, any valid SQL predicate can be used that is defined on individual tuples in \augmentedtable\ $\cview$.
\par
\cut{
$\use$ clarifies the view from which all attributes are obtained 
$\pre(A_j)=a_j$ filters the tuples based on their values before the update. 
In the presence of this condition, only tuples $t_i \in D$ that have $t_i.A_j = a_j$ will be selected. 
$\post(A_i) = a_i$ is similar to $\pre(A_j)=a_j$, but is performed after the update occurs (and possibly changes the values of the tuples). 
The $\select, \where, \groupby$ operators define a standard query that creates a view based on the values before and after the update. 
}

\begin{example}\label{eg:full-what-if}
Consider the \revc{what-if query statement} shown in Figure \ref{fig:whatif-query}. It checks the effect of hypothetically updating the price by 10\% (\update) on {\tt Brand = 'Asus'} (\when). The effect is measured on their average of average ratings (\outputw) -- the first average on ratings of the same type of Asus products, and the second 
average is on different types of Asus products, but only for {\tt Category = `Laptop'} (i.e., does not include phones for instance), and where the post-update average sentiment is still above $0.5$. Since {\tt Rating} and {\tt Sentiment} come from the {\tt Review}  table whereas the update attribute {\tt Price} belongs to the {\tt Product} table, they are aggregated in the SQL query in the \use\ operator for each {\tt Product} tuple. 
\end{example}
\revb{\sys\ supports multiple updates in a what-if query with attributes $B_1, B_2, \cdots$, e.g.,
$\update(Price) = 500\ \sqland\ \update(Color) = Red$, provided there are no paths from any $B_i[t]$ to any $B_j[t']$ for any two tuples $t, t'$ - a fact that we will use in Section~\ref{sec:howto} for how-to queries; we discuss other extensions in Section~\ref{sec:conc}. Here, we discuss single-attribute updates for simplicity.}

%% file: whatif-semantics.tex
\subsection{Semantics of Probabilistic What-If Queries}\label{sec:whatif-semantics}

Here we define the semantics of what-if queries described in Section~\ref{sec:what-if-syntax} as the \emph{expected}  value of the output attribute over possible worlds consistent with a what-if queries. 
\par
The operators in the what-if queries are evaluated in this order: \use\ $\rightarrow$ \when $\rightarrow$ \update\ $\rightarrow$ \forw\ $\rightarrow$ \outputw.
\par
(1) The \use\ operator outputs the \augmentedtable\ $\cview$ that contains all relevant attributes for the what-if query 
by a standard group-by SQL query. 
\par
(2) The \when\ operator takes $\cview$ as input, and defines the set $S$ in the update $U = u_{R, B, f, S}$. Suppose this operator uses an SQL predicate $\pred_{\when}$ defined on a subset of attributes of $\cview$. Then the output of the \when\ operator is the view $\cview_w = \{t \in \cview~:~ \pred_{\when}(t) = true\}$. Note that in both \use\ and \when\ operators, the pre-update values ({\pre} values are assumed by default) from the given database $D$ are used.
\par
(3) Then the `\update\ $B = f(\pre(B))$'  operation is applied to the tuples $t \in \cview_w$ on attribute $B$. 
As described in Section~\ref{sec:model-PCM}, this update is equivalent to modifying the structural equation $\phi_{B[t]}$ in the PRCM by replacing them with a constant value $f(\pre(B))$. Due to uncertainty induced by the noise variables, at this point, we get a set of possible worlds $PWD(D)$ (Definition~\ref{def:possible-worlds}) along with a post-update distribution $\pr_{D, U}$ on $PWD(D)$ induced by the update $U$. Clearly, some possible worlds $I$ have $\pr_{D, U}(I) = 0$, e.g., if for a tuple $t$ in relation $R$ of $I$ such that $t$ corresponds to a tuple in $\cview_w$ with the same key, $B[t] \neq f(\pre(B[t]))$. 
\par
(4 and 5) 
For the remaining \forw\ and \outputw\ operators, let us first fix a possible world $I \in PWD(D)$ obtained from the previous step. Let $\cviewpwd$ be the output of the SQL query in the \use\ operator on $I$. Suppose the predicate in the \forw\ operator is $\pred_{\forw}$, which may include  $\pre(A)$ and $\post(A')$ values for different attributes $A, A'$. For every tuple $t$ (in any relation in $D$) and attribute $A$, consider two values of $A[t]$: $\pre(A[t])$ of $t$ in $D$ and $\post(A[t])$ of $t$ in $I$ (some values remain the same in $\pre$ and $\post$, e.g., if $A$ is immutable or if there is no effect of updating $B$ for $S$ tuples on $A$). 
Using these values, we evaluate the predicate $\pred_{\forw}$, and using tuples from $R$ that satisfy this predicate, we compute the aggregate $aggr_Q$ ($\avg(Rating)$ in Figure~\ref{fig:whatif-query}) mentioned in the \outputw\ operator using their values in $I$ (i.e., $\post$ values). 




This aggregate $aggr_Q$ is computed on attribute values $Y[t]$ for $t \in \cviewpwd$, where $Y$ itself can be an aggregated attribute $Y = aggr_{\use}(Y')$ if it is coming from a different relation than the one containing the update attribute as defined by the SQL query in the \use\ operator (in Figure~\ref{fig:whatif-query}, $Y = Rtng$, $Rtng = \avg(Review.Rating)$, and both $aggr_Q$ and $aggr_{\use}$ are $\avg$). Hence, when a possible world $I \in PWD(D)$ is fixed, the what-if query answer is computed as follows:
\begin{definition}[What-if query result on a possible world]\label{def:whatif-result-posworld}
Given a what-if query $Q$ and a database $D$, 
the answer to $Q$ on a given possible world $I \in PWD(D)$ is the aggregate $aggr_Q$ over $Y_{I}[t]$ values using the notations above: 
\begin{small}
\begin{equation}\label{eq:whatif-posworld}
    \valwhatif(Q, D, I) = aggr(\{Y_{I}[t]~:~ \pred_{\forw}(t) = true, t \in \cview\})
\end{equation}
\end{small}
where $Y_I[t]$ denotes the value of attribute $Y$ for tuple $t$ in the possible world $I$. Here $t$ is tuple in the \augmentedtable\ $\cview$ and therefore corresponds to a unique tuple in relation $R$. 
\end{definition}


Then the final value of the what-if query is the expected query result on all possible worlds of $D$:
\begin{definition}[What-if query result]\label{def:whatif-result}
Given a what-if query $Q$ and a database $D$, 
the result of $Q(D)$ is the expected value of $\valwhatif(Q, D, I)$ over all possible worlds $I \in PWD(D)$, using the post-update probability distribution $\pr_{D, U}$: 
\begin{small}
\begin{align}\label{equn:whatifanswer}
\valwhatif(Q, D) & = & \mathbb{E}_{I \in PWD(D)}[\valwhatif(Q, D, I)]\nonumber\\
& = & \sum_{I \in PWD(D)} \valwhatif(Q, D, I) \cdot \pr_{D, U}(I)
\end{align}
\end{small}
\end{definition}

%% file: whatif-algo.tex
\subsection{Computation of What-If Queries}\label{sec:whatif-algo}
The semantics presented in Section~\ref{sec:whatif-semantics} does not directly lead to an efficient algorithm to compute the answer to what-if queries by Definition~\ref{def:whatif-result}, since (1) the number of possible worlds can be exponential in the size of the database $D$, and (2) computation of post-update distribution $\pr_{D, U}$ is non-trivial. 
In this section, we present our algorithm for computing what-if query answers that use two key ideas to address these challenges:
(a) Instead of computing the what-if query over the entire database, we decompose it into smaller problems and compute modified queries on  
subsets of tuples that are `independent' of each other (as fewer tuples make the computation more efficient). Then we combine the results to get the result of the original query over the entire database. 
(b) 
To compute the distribution $\pr_{D, U}$ needed for estimating the query result, we use techniques from the {\em observational causal inference} and the {\em graphical causal model} literature \cite{pearl2009causality} when the post-update distribution is determined by a PRCM. 
\noindent
\paragraph{Decomposing the computation}
The decomposition, and subsequently the composition of answers, is achieved by the use of 
\emph{block-independent databases} and \emph{decomposable aggregate functions} supported by \sys\ (SUM, COUNT, AVERAGE) described below.

\paratitle{Block-independent database decomposition}
We adapt the notion of block-independent database model that has been used in \emph{probabilistic databases} \cite{ReS07,DalviRS09} 
and \emph{hypothetical reasoning} \cite{JampaniXWPJH08}. 
First, we need the notion of independence in our context. We say that two tuples $t,t' \in D$ are {\bf independent} if there are no paths in the ground causal graph $G$ (ref. Section~\ref{sec:model-PCM}) between $A[t]$ and $A'[t']$ for any two attributes $A, A'$.







Given a database $D$ and 
a PRCM with a ground causal graph $G$, 
${\mathcal B} = \{D_1, \ldots, D_\ell\}$ is called a 
{\bf block-independent decomposition of $D$} if (i) $\{D_1, \ldots, D_\ell\}$  forms a partition of $D$, i.e., each $D_i \subseteq D$, $\cup_{i=1}^l D_i = D$, and $D_i \cap D_j = \emptyset$ for $i \neq j$, and (ii) for each $t \in D_i$ and $t' \in D_j$ where $i \neq j$, $t$ and $t'$ are independent. Note that these tuples $t$ and $t'$ can come from the same or different relations of $D$.


We compute block-independent decomposition of database $D$ given a causal graph $G$  as follows. 
\revb{The block decomposition process performs a topological ordering of the nodes in the causal graph and then performing a DFS or BFS on it, and is therefore linear in the size of the causal DAG. 
The causal DAG has at most $n \times  k$ nodes where $n$ is the number of tuples in $D$ and $k = |\attr(D)|$. In particular, the decomposition does not depend on the structure or complexity of the query.}
Block-independent decomposition provides an optimization in our algorithms; in the worst case, all tuples may be included in a single block. 

\begin{example}\label{ex:blocks}
Consider the causal graph of the PRCM (Figure~\ref{fig:grounded-causal-graph}) defined on the database presented in Figure \ref{fig:database}. 
\revb{
The procedure first performs a topological sort of the nodes. For example, in Figure \ref{fig:causal-graph}, the node $Brand[p_1]$ is first, and then the node $Quality[p_1]$ etc.
Then, the algorithm performs a BFS to detect the connected components of the graph which are all tuples belonging to the same category, along with their reviews. 
}
The block-independent decomposition of the database $D$ in Figure~\ref{fig:database} is then ${\mathcal B} = \{D_1, D_2, D_3\}$ where $D_1 = \{p_1, p_2, p_3, r_1, r_2, r_3, r_4, r_5\}$, $D_2  = \{p_4, r_6\}$, and $D_3 = \{p_5\}$ corresponding to laptops, camera, and books along with their reviews.

\end{example}
\paratitle{Decomposable functions}
The aggregate functions supported by \sys\ are \emph{decomposable} as defined below, which allows us to combine results from each block after a block-independent decomposition to compute the answer to a what-if query. Since the immutable attributes include keys that are unchanged in all possible worlds $I \in PWD(D)$ of $D$ (Definition~\ref{def:possible-worlds}), given a block-independent decomposition $\mathcal B$ of $D$, we will use the corresponding decomposition ${\mathcal B}_{I}$ of $I$ where the same tuples identified by their keys go to the same blocks in $\mathcal B$ and ${\mathcal B}_{I}$. The aggregate functions $f_{Q, D}, f'_{Q, D}$ below map a set of tuples to a real number whereas $g$ maps a set of real numbers to another real number. 

\begin{definition}[Decomposable aggregate function]\label{def:decomposable}
Given a database $D$, a block-independent decomposition $\mathcal{B} = \{D_1, \ldots, D_\ell\}$ of $D$, a what-if query $Q$, and any possible world  $I \in PWD(D)$ of $D$, an aggregate function $f_{Q, D}$ 
is {\bf decomposable} if there exist aggregate functions $f'_{Q, D}$ 
and $g$ 
such that: 
\begin{itemize}
    \item $f_{Q, D}(I) = g(\{f'_{Q,D}(D_i) \mid \forall D_i\in \mathcal{B}_I\})$ where $\mathcal{B}_I$ is the block partition of $I$ corresponding to $\mathcal{B}$, 
    \item $\alpha g(\{x_1,\ldots,x_l\}) = g(\{\alpha x_1,\ldots,\alpha x_l\})$, $\forall \alpha \geq 0$, and
    \item $ g(\{x_1,\ldots,x_l\}) + g(\{y_1,\ldots,y_l\}) = g(\{x_1+y_1,\ldots, x_l+y_l\})$
\end{itemize}\label{assume:homogenous}


\end{definition}

When the aggregate function $aggr$ given in Equation (\ref{eq:whatif-posworld}):\\ $\valwhatif(Q, D, I)$ = $aggr(\{Y_{I}[t]~:~ \pred_{\forw}(t) = true, t \in \cview\})$ is decomposable, 
we show that the computation can be performed on the blocks $\mathcal{B}_I$ and then aggregated 
to compute $\valwhatif(Q, D, I)$. 
We note that every supported aggregate function in this paper  
($\sumsql$, $\avg$, $\ct$) is decomposable. We  demonstrate this for $\avg$ below. 
\begin{example}\label{ex:decomposable}
\begin{sloppypar}
Reconsider the what-if query in Figure \ref{fig:whatif-query}. Suppose the database can be partitioned into blocks by Category as demonstrated in Example \ref{ex:blocks}. 
In this case, $aggr = \avg$ and $Y = Rtng = \avg(T2.Rating)$, and for any $I \in PWD(D)$,
$\valwhatif(Q, D, I) = \avg(\{Rtng_I[t] \mid t\in \cview, Category[t] = Laptop, Brand[t] = Asus, \post(Senti[t]) > 0.5\})$
We use the standard formula for decomposing average: $\avg(D)$ = $\frac{1}{|D|}\sum_{i=1}^\ell  \sumsql(D_i)$. 
For each block $D_i \in {\mathcal B}_I$, 
$f'_{Q, D}(D_i)$ = $\frac{1}{|D|}$ $\sumsql(\{Rtng_I[t] \mid t\in \cviewpwd \cap D_i$, $Category[t] = Laptop$, $Brand[t] = Asus$, $\post(Senti[t]) > 0.5\})$
Here, $g = \sumsql$, and $\sumsql$ satisfies the properties in Definition \ref{def:decomposable}. 
\end{sloppypar}

\end{example}



In the proof of the following proposition, we leverage the ability to marginalize the distribution $\pr_{D,U}$ over the possible worlds of the database $D$ (Definition \ref{def:rel-poss-worlds}) given a what-if query $Q$  to get a distribution and a set of possible worlds for any block $D_i\in {\mathcal B}$, which we denote by $PWD(D_i) \subseteq PWD(D)$. $PWD(D_i)$ are all instances where all tuples $t' \notin D_i$ remain unchanged and all mutable attributes of $t\in B_i$ get all possible values from their respective domains. 
We further 
denote $\overline{PWD}(D_i)$ as the set of possible worlds of $D_i$ that only includes the 
tuples in $D_i$; i.e., $\overline{PWD}(D_i)$ is the projection of $PWD(D_i)$ on $D_i$. All proofs are deferred to the appendix (Section \ref{sec:appendix}) due to space constraints. 

\begin{proposition}[Decomposed computation]\label{prop:blocks}
Given a database $D$, its block-independent decomposition $\mathcal{B} = \{D_1, \ldots, D_\ell\}$, and a what-if query $Q$ whose result on a possible world $I \in PWD(D)$ is $\valwhatif(Q, D, I) = aggr(\{Y_{I}[t]~:~ \pred_{\forw}(t) = true, t \in \cview\})$ (Definition~\ref{def:whatif-result-posworld}), if 
$aggr$ is a decomposable function, i.e., if there exist  functions $g$ and $f'_{Q, D}$ according to Definition~\ref{def:decomposable}, then 
\begin{small}
\begin{equation}\label{eq:whatif-block-total}
\valwhatif(Q, D) = g(\{\valwhatif(Q', D_i) \mid \forall D_i\in \mathcal{B}\})
\end{equation}
\end{small}
where $Q'$ is the same query as $Q$ with $f'_{Q, D}$ replacing $aggr$ and
\begin{small}
\begin{equation}\label{eq:whatif-block}
    \valwhatif(Q', D_i) = \expectation_{I_j \in \overline{PWD}(D_i)}[\valwhatif(Q', D_i, I_j)]
    \end{equation}
\end{small}
\label{lem:general}
\end{proposition}
\paragraph*{Computing results with causal inference}
We show the connection between the what-if query results and techniques in observational causal inference. This connection will allow us to compute the results for each block as given in Equation (\ref{eq:whatif-block}). 
\amir{Changed from here:}
Specifically, we show how the computation in each block is done by the post-update probabilities, which we further reduce to pre-update probabilities.

\begin{proposition}[Connection to causal inference for $\ct$]\label{prop:causal-results}
Given a database $D$ with its block independent decomposition $ \mathcal{B}_D$, a block $D_i\in \mathcal{B}_D$, a ground causal graph $G$, a what-if query $Q'$ where $Agg=\ct$, and the $\forw$ operator is denoted by $\pred_{\forw}$, the following holds.
\begin{footnotesize}
\begin{equation*}
\valwhatif(Q', D_i) = \sum_{t\in D_i} \left(\sum_k\left( \pr_{D_i,U}( \mu_{\forw,\post}^k(t)=\true | \mu_{\forw,\pre}^k(t)=\true)\right)\right)
\end{equation*}
\end{footnotesize}
In this equation, $\pr_{D_i,U}( \mu_{\forw,\post}^k(t)=\true  | \pred_{\forw,\pre}^k(t)=\true)$ denotes the sum of probabilities of all possible worlds of $D_i$ such that the tuple $t$ that satisfied $\pred_{\forw,\pre}^k(t)=\true$ before the update $U$ also satisfies $\pred_{\forw,\post}^k(t)$ after the update.
\end{proposition}

The proof of the proposition relies on the fact that the sum of probabilities of all possible worlds is $1$ and the fact that a $\forw$ clause can be represented as a CNF of $\pre$ and $\post$ conditions. 
Proposition~\ref{prop:causal-results} assumes $Agg=\ct$, however, a similar result for $Agg=\sumsql/\avg$ can be found in the appendix (Section \ref{sec:appendix}). 

\input{whatif-optimization}

%% file: whatif-optimization.tex
\paratitle{Estimating the probability values}
The expression in Proposition~\ref{prop:causal-results}  relies on the post-update distribution to evaluate conditional probability of certain attribute values. For example, we need a way to estimate $\pr_{D,U}(A_i=a_i \mid A_j=a_j, \pred_{\when})$ when $aggr=\ct$. 
Our goal is to find a way to estimate these probability values from the input database $D$, assuming we have a PRCM. 

To do so, we leverage the notion of {\bf backdoor criterion} from causal inference \cite{pearl2009causality}. 
A set of attributes $\mb C$ satisfies the backdoor criterion w.r.t. $A_i$ and $B$ if no attribute $C\in \mb C$ is a descendant of $A_i$ or $B$ and all paths from $B$ to $A_i$ which contain an incoming edge into $A_i$ are blocked by $\mb C$. For example,  in Figure~\ref{fig:grounded-causal-graph}, Brand[$p_1$], Quality[$p_1$], and Category[$p_1$] satisfy the backdoor criterion with respect to Sentiment[$p_1$] and Rating[$p_1$].
Using this criterion, we show (in the full version) that the element $\pr_{D,U}(A_i=a_i \mid B=b, C=c, A_j=a_j, \pred_{\when})$ in the query result expression in Proposition \ref{prop:causal-results} can be estimated from $\pr_D$ using the following calculations.
\begin{small}
\begin{align*}
    &\pr_{D,U}(A_i=a_i \mid A_j=a_j, \pred_{\when})= \\
    &\sum_{c\in \Dom(C)}\pr_{D,U}(A_i=a_i \mid C=c, A_j=a_j, \pred_{\when})
    \pr_D(C= c|A_j=a_j, \pred_{\when})
\end{align*}
\end{small}
The first probability term can be simplified as follows.
\begin{small}
\begin{align*}
    &\pr_{D,U}(A_i=a_i \mid C=c, A_j=a_j, \pred_{\when})= \\
    &\sum_{b\in \Dom(B)}\pr_{D,U}(A_i=a_i \mid B=b, C=c, A_j=a_j, \pred_{\when})\cdot\\
    & \quad\quad\quad\quad\pr_D(B=b|C=c, A_j=a_j, \pred_{\when})
\end{align*}
\end{small}
This shows that the query output relies on $\pr_{D,U}(A_i=a_i \mid B=b, C=c, A_j=a_j, \pred_{\when})$, which can be estimated from $\pr_D$ using equation~ (\ref{eq:back}).
Using these probability calculations, we estimate the query output from the input data distribution $\pr_D$. 
The equations require that we iterate over the values in the domain of $\mb B$ and $\mb C$, which can be inefficient as the domain set size increases exponentially with the number of attributes in the set. However, the majority of the values in $\Dom(\mb C)$ would have zero-support in the database $D$, implying $\pr_D(C= c|A_j=a_j, \pred_{\when})=0$ for $\mb C=c$. Therefore, we build an index of values in $\Dom(\mb C)$ to efficiently identify the set of values that would generate a positive probability-value. This optimization ensures that the runtime is linear in the database size.

%% file: howto.tex
\section{Probabilistic How-To queries}\label{sec:howto}
How-to queries support \emph{reverse data management} (e.g., \cite{MeliouGS11Reverse}), and suggest how a given mutable attribute can be updated to optimize the output attributes subject to various constraints. 
In this section we describe the syntax of probabilistic how-to queries supported by \sys\  (Section~\ref{sec:howto-syntax}), describe their semantics  (Section~\ref{sec:howto-semantics}), and present algorithms to compute their answers (Section~\ref{sec:howto-algo}). How-to queries are computed by solving an optimization problem over several relevant what-if queries. 

\begin{figure}[t]
    \centering \small
        \begin{align*}
    &\use\ (\ldots) \quad \texttt{/* same as Figure~\ref{fig:whatif-query} */}\\
    &\when\ Brand = `Asus'~ \sqland\ Category = `Laptop'\\
    &\howto\ Price, Color\\
    &\suchthat\ 500 \leq Post(Price) \leq 800~ \sqland\ \\
    &\quad \quad L1(\pre(Price), \post(Price)) \leq 400\\
    & \maxi\ \avg(\post(Rtng))\\
    &\forw\  (\pre(Category) = 'Laptop'~ \sqlor\\ & \quad \quad \pre(Category) = 'DSLR\ Camera')~ \sqland\ Brand = 'Asus'
    \end{align*}
    \caption{How-to query asking
    ``how to maximize the average rating of Asus laptops and cameras over the determined view by changing the price and/or color of Asus laptops so that it will not drop below 500 and increase above 800, and will be at most 400 away from it original value?'' 
    }\label{fig:howto-query}
    \vspace{-5mm}
\end{figure}

\input{howto-syntax}

\input{howto-semantics}

\input{howto-algo}

%% file: howto-syntax.tex
\subsection{Syntax of Probabilistic How-To Queries}\label{sec:howto-syntax}


The syntax of how-to queries in \sys\ is similar to that of what-if queries (see Figures~\ref{fig:whatif-query} and \ref{fig:howto-query}, and Section~\ref{sec:what-if-syntax}). 
How-to queries have two parts. 
The first part uses the required $\use$ operator and is identical to the {\bf \use\ operator} in the what-if queries in its functionality -- it defines the \augmentedtable\ $\cview$ that contains the key of the relation containing the update attribute, and includes all attributes used in the second part of the query; attributes coming from other relations are aggregated.
\par
In the second part, the optional {\bf $\when$} and {\bf $\forw$ operators} have the same functions as the what-if queries. Then $\when$ operator specifies the set $S$ on which an update $U = u_{R,  B, f, S}$ can be applied, whereas the $\forw$ operator defines the subset on which the effect is estimated. Like what-if queries, \when\ only includes pre-update values $\pre(A)$, whereas \forw\ can include both pre- and post-update values $\pre(A), \post(A)$.
\par
The required {\bf $\howto$ operator} corresponds to the $\update$ operator of what-if queries, and uses $\pre(A)$,  but instead of specifying an attribute (or a set of attributes) to update, it specifies the set of mutable attributes that can be updated. 
In Figure~\ref{fig:howto-query}, {`$\howto$ Price, Color'} states that any combination of these three attributes can be updated, and some attributes can be left unchanged as well. To ensure that the updates on these attributes are valid, our algorithms assume that, for any pair of the attributes mentioned in this clause $A_1, A_2$, there are no paths in the ground causal graph of the PRCM between $A_1[t]$ and $A_2[t']$ for any $t, t'\in D$. 
\par
Possible {\bf outputs of the how-to queries} are of these forms for each attribute $A$ specified in the $\howto$ operator: (i) $\update(B) = <const>$, (ii) $\update(B) = <const>~ \times~ \pre(B)$, (iii) $\update(B) = <const>~ +~ \pre(B)$, and $\update(B)$ = {\tt no change}, where $<const>$ is a constant found by our algorithms from the search space. One example output of this \howto\ query is
\begin{small}
$$\{\texttt{Price: 1.1x,~
Color: no change}\}$$
\end{small}
stating the price should be increased by 10\%, the color should be changed to red, and the category should not be changed. 

\cut{
Price: 1.1x
Color: Red
Quality:no change
}



The optional {\bf Limit operator} states the constraints for optimization, i.e., it defines the conditions that restrict the post-update values of update attributes specified in the $\howto \update$ operator for tuples in $\cview$ that satisfy the \when\ operator. 
In particular, if an attribute $A$ is numeric, its updates can be bounded by numeric limits, e.g., $l \leq \post(A) \leq h$, $l \leq \post(A)$, $\post(A) \leq \pre(A) + <const>$, $\post(A) \leq \pre(A) \times <const>$, etc., and if $A$ is categorical or numeric, the user can specify the permissible values as a set, e.g., $\post(A)\ \inht\ (v_1, v_2, v_3)$. Furthermore, this operator allows users to specify the maximal or minimal $L1$ distance between the original attribute values ($\pre(A)$) and the updated ones ($\post(A)$) for attributes $A$ in the \howto\ operator for the tuples satisfying the condition in the \when\ operator: $L1(\post(A), \pre(A))$ takes a vector of values $V_{u}$ and $V_{u}[i]$ is an update value of the $i$'th attribute mentioned in the $\suchthat$ operator, and returns the normalized $L1$ distance between the original value vector the vector of update values $|V_{u} - V_{orig}|$.
The $L1$ operator helps model the {\bf cost of an update} (with suitable weights) as some updates can be more expensive than the others.  



Finally, the how-to query needs to include a required {\bf $\maxi$ or $\mini$ operator}, which specifies an aggregated value of an attribute from the \augmentedtable\ $\cview$ that is to be maximized or minimized using the updates on the attributes specified in the $\howto$ operator. Only post-update values $\post(A)$ of attributes are allowed in $\maxi$ and $\mini$.

\begin{example}\label{ex:howto}
Consider the query in Figure \ref{fig:howto-query}. 
It asks for the maximum value of the average value of Rtng (\howto) by updating the tuples with {\tt Brand = `Asus'}, {\tt Category = `Laptop'} (\when). The attributes allowed to be updated are~ {\tt Price, Color} (\howto). The update to the {\tt Price} attribute is restricted to $[500, 800]$, where distance between the original values and the updated values in this attribute has to be $\leq 400$. 
The average of Rtng is computed over the view defined by the $\forw$ operator.
\end{example}

%% file: howto-semantics.tex
\subsection{Semantics of Probabilistic How-To Queries}\label{sec:howto-semantics}

We next define the results of how-to queries in terms of what-if queries. 
Intuitively, every how-to query optimizes over a set of what-if queries, where each what-if query contains a possible update allowed in the how-to query. Assuming, without losing generality, that the how-to query contains a $\maxi$ operator, the result of the how-to query is then the what-if query that yields the maximum result of the output attribute in the $\maxi$ operator of the how-to query, subject to the constraints on post-update values of attributes specified in the \suchthat\ operator. 

\cut{
\paratitle{What-if queries with multiple updates}
\amir{Not sure where this should go:}
Since we use what-if queries to define the results of how-to queries, 
we first extend what-if queries to contain a set of updates. 
The syntax of what-if queries that we define in Section \ref{sec:what-if-syntax} and the post-update distribution in Definition \ref{def:rel-poss-worlds} can be easily extended to accommodate updates to multiple attributes. 
In practice, what-if queries will be able to contain $\update(B_1) = f_1(B_1)~ \sqland\ \ldots, \sqland\ \update(B_c) = f_c(B_c)$, and the distribution will change accordingly. 
The restriction of this scenario being that for any two attributes $B_1, B_2$ in the $\update$ operator, there cannot be a path in the causal graph between $B_1[t]$ and $B_2[t']$ for any $t, t' \in D$. 
Within the context of our running example, a what-if query can contain the clause $\update(Price) = 500\ \sqland\ \update(Color) = Red$.
}

\begin{definition}[Candidate what-if query]\label{def:candidate}
Given a how-to query $Q_{HT}$ that includes (i) a $\maxi$ operator of $Agg(Post(Y))$, (ii) a $\howto$ operator with update attributes $B_1, \ldots, B_c$, and (iii) a $\suchthat$ operator that without loss of generality specifies permissible ranges ${\mathcal R_i}$ and $L1(\pre(B_i), \post(B_i)) < \theta_i$ for all $i \in [1, c]$ (if there are no constraints on the range in $Q_{HT}$ for $B_i$,  $\mathcal{R}_i = \Dom(B_i)$ and if no $L1$ constraint is specified, $\theta_i = \infty$), 
a {\bf candidate what-if query} is a what-if query $Q_{WI}$ such that:
\begin{itemize}
    \item The $\use$, $\when$, and $\forw$ operators in $Q_{WI}$ are identical to the ones in $Q_{HT}$,
    \item $Q_{WI}$ contains $\update\ B_{j_1} = b_1, \ldots, B_{j_k} = b_k$, where $\{j_1, \ldots, j_k\}$ $\subseteq$ $\{1, \ldots, c\}$, $b_i \in {\mathcal R_{j_i}}$, 
    and $L1(\pre(B_{j_i}), \post(B_{j_i})) < \theta_{j_i}$. 
    \item The $\outputw$ operator in $Q_{WI}$ specifies the attribute $Agg(Post(Y))$ from the $\maxi$ operator in $Q_{HT}$.
\end{itemize}
This query is denoted as $Q_{WI}((B_{i_1}, b_1), \ldots, (B_{i_c}, b_c))$. 
The {\bf set of all candidate what-if queries} for a how-to query $Q_{HT}$ is denoted by $\candwhatif(Q_{HT})$.
\end{definition}

\begin{example}\label{ex:candidate-whatif}
A candidate what-if query $Q_{WT}((Price, 500))$ for the how-to query depicted in Figure \ref{fig:howto-query} is given below (\use\ operator is the same as that in Figure~\ref{fig:whatif-query}):
\begin{small}
\begin{align*}
&\use\ (\ldots)\\
&\when\ Brand = `Asus'~ \sqland\ Category = `Laptop'\\
&\update\ Price = 500 \\
& \outputw\ \avg(\post(Rating))\\
&\forw\ (\pre(Category) = 'Laptop'~ \sqlor\\ & \quad \quad \pre(Category) = 'DSLR\ Camera')~ \sqland\ Brand = 'Asus'
\end{align*}
\end{small}
In particular, the update on the Price attribute is in $[500,800]$ and satisfies the L1 distance since the original price of the Asus laptop is $529$, and the rest of the query is identical to the query in Figure \ref{fig:howto-query}. 
\end{example}


We now define the result if a how-to query that optimizes over the result of all candidate what-if queries. 

\begin{definition}[How-to query result]\label{def:howto-result}
Given a database $D$ and a how-to query $Q_{HT}$ with a $\maxi$ operator, the result of $Q_{HT}$ is defined as follows:
\begin{equation}
    argmax_{Q_{WI} \in \candwhatif(Q_{HT})} \valwhatif(Q_{WI}, D)
\end{equation}
where $\valwhatif(Q_{WI}, D)$ denotes the result of the what-if query $Q_{WI}$ on D as defined in Definition~\ref{def:whatif-result}; \mini\ is defined similarly. 
\cut{\begin{equation*}
\begin{split}
argmax(\{Q_{WI}(b_1, \ldots, b_c) \mid l_1 \leq b_1 \leq h_1, \ldots, l_c \leq b_c \leq h_c,\\ 
Cost(\pre(B_i), \post(B_i)) < \theta_i\})
\end{split}
\end{equation*}
i.e., the candidate what-if query that yields the maximum value of the attribute in the $\maxi$ operator of $Q_{HT}$. 
}
\end{definition}

We take the argmax of $\candwhatif(Q_{HT})$ since a how-to query asks about {\em the manner in which the database needs to be updated} and not about the result. This corresponds to the output we defined and demonstrated in Section \ref{sec:howto-syntax}.
Definition \ref{def:howto-result} requires taking the maximum over a large set of 
candidate what-if queries, which can even be infinite if the domain is continuous. In the next section, we provide optimizations to make their computation feasible. 



%% file: howto-algo.tex
\subsection{Computation of How-to queries}\label{sec:howto-algo}

The naive approach to computing the result of a how-to query by Definition~\ref{def:howto-result} is inefficient as it evaluates a large number of 
candidate what-if queries. 
Instead, we model the problem of computing the result of how-to queries as an Integer Program (IP). Denote by $\mb U = \{B_1, \cdots, B_{c}\}$ the set of  update attributes in the $\howto$ operator. For each  attribute  $B_i \in \mb U$, we enumerate all permissible updates (denoted by $S_{B_i}$) and define an indicator variable $\delta_{b_i}$ for every $b_i$ which denotes the potential updated value of attribute $B_i$. 
For example, the set $S_{Price}$ can consist of the following updates: 
\begin{small}
\begin{align*}
   S_A\equiv& \{\texttt{1.1x\pre (Price)},~\texttt{1.2x\pre (Price)},\ldots, ~\texttt{2.5x\pre (Price)}\\
   & ~\texttt{100+\pre (Price)}, ~\texttt{200+\pre (Price)},\ldots,~\texttt{500+\pre (Price)},\\
   & ~ \texttt{250 },\texttt{300},\ldots,\texttt{600}\}
\end{align*}
\end{small}
The elements of set $S_A$ are defined such that all these updates satisfy the constraints mentioned in \suchthat\  operator.
If the set of potential updates is continuous, we bucketize them so that we can treat their values as discrete. 
Given a set $S_{B_i}$ and variables $\delta_{b_i}$ for all ${b_i}\in S_{B_i}$, we add a constraint for each attribute that $\sum_{b_i \in S_{B_i}}\delta_{b_i} \leq 1$ to ensure that at most one of the updates is performed. If $\delta_{b_i}$ is zero for all values in $S_{B_i}$, then $B_i$ is not updated. 
Given this formulation, the corresponding what-if query is estimated as a linear expression by using Proposition~\ref{prop:causal-results} and  training a regression function over the dataset $D$. Let this linear function be $\phi: \Dom(\mb U)\rightarrow O$, where $O$ is the range of the output of candidate what-if queries. 
The following IP models the solution to the how-to query using the variables $\delta_{b_i}$.
{\small  
\begin{align}
    \argmax \quad \phi(D, \sum_{b_1 \in S_{B_1}}\delta_{b_1} b_1, \ldots, \sum_{b_c \in S_{B_c}} \delta_{b_c} b_c)&
    \label{eq:opt1_obj} \\
    \text{subject to } \quad \quad \quad \quad \quad 
    \sum_{b_i \in S_{B_i}} \delta_{b_i} &\leq 1, \quad \forall i = 1~ to~c \label{eq:opt1_c2} \\
    \quad \delta_{b_i} \in \{0, 1\}, \quad \forall b_i \in S_{B_i}, \forall i = 1~ to~c \label{eq:opt1_c3}
\end{align}
}
\cut{\small  
\begin{align}
    \argmax \quad \phi(D, \sum_{a \in S_{A}}\delta_a a, \ldots, \sum_{a \in S_{X}} \delta_{x} x)&
    \label{eq:opt1_obj} \\
    \text{subject to } \quad \quad \quad \quad \quad 
    \sum_{a \in S_A} \delta_{a} &\leq 1, \quad \forall A \in \mb A \label{eq:opt1_c2} \\
    \quad \delta_{a} \in \{0, 1\}, \quad \forall a \in S_A, A \in \mb A \label{eq:opt1_c3}
\end{align}
}
In addition to these constraints, additional constraints are added to the IP based on the constraints in the \suchthat\ operator. 
\amir{A proposition ``If the domains are discrete, the solution to the IP is the solution to the solution to the how-to query''?}
Since all constraints and the objective function are linear equations, we leverage standard IP solvers to calculate the output of the \howto\ query\footnote{As an alternate formulation, our framework allows to optimize the cost (L1 distance between the original attribute and the updated value) while adding a constraint on the aggregated attribute. We discuss more details in Section \ref{sec:appendix}.}. Note that the number of constraints in the IP grows linearly with the number of attributes in $\mb U$ and the number of variables grows linearly in the number of possible updates for each attribute.

\paratitle{\revb{Extension to preferential multi-objective optimization}} \revb{\sys{} can be adapted to the settings where an user aims to optimize multiple objectives that are lexicographically ordered based on preference. Consider a ordered set of preferences $p_1,\ldots,p_t$ where each preference $p_i$ is less important than $p_j$ for $j<i$. In this case, we propose to solve IP iteratively as follows. First, we can solve the single objective optimization problem for the first preference $p_1$ as described above, ignoring other preferences. In the subsequent iteration, the identified objective value of the first considered objective is added as a constraint to maximize the second preference $p_2$. In this way, all previously solved objectives are added as constraints while optimizing for a preference $p_i$. The solution to the last integer program that optimizes for $p_t$ where all other preferences are added as constraints is returned as the final solution to the preferential multi-objective optimization.
}
\revb{
\begin{example}\label{ex:multiple_obj}
Consider the database in Figure \ref{fig:database} and a how-to query that aims to maximize the average ratings as a first priority and the average sentiment as a second priority. 
In the first IP, we will solve for the clause $\maxi\ \avg(\post(Rtng))$, where $Rtng$ are the ratings. Suppose the maximum average rating  we get is $c$. We then solve the IP for the clause $\maxi\ \avg(\post(Sentiment))$ and add the constraint that 
($\avg(\post(Rtng))$) will equal $c$.
\end{example}
}

%% file: experiments.tex
\section{experiments}\label{sec:experiments}
We evaluate the effectiveness of \sys\ {and its variants} on various real-world and synthetic datasets and answer the following questions: 
\begin{compactenum}
    \item Do the results provided by \sys\ make sense in real-world scenarios?
    \item How does \sys\ compare to other baselines for hypothetical reasoning when the ground truth is available? 
    \item How does the runtime of \sys\ depend on query complexity and dataset properties like number of tuples, the causal graph structure, \common{discretization of continuous attributes, and the number of attributes in different operators of the query}?
    \item \common{How does combining a sampling approach with \sys\ influences runtime performance and the quality of the results?}
\end{compactenum}
Our experimental study includes 5 datasets and 3 baselines that are either inspired by previous approaches or simulate the absence of a causal model. 
We provide a qualitative and quantitative evaluation of \sys, showing that it gives logical results in real-world scenarios and achieves interactive performance in most cases. 

\paratitle{\revb{Implementation} and setup} We implemented the algorithms in Python. \sys\ was run on a MacOS laptop with 16GB RAM and 2.3 GHz Dual-Core Intel Core i5 processor. We used random forest regressor~\cite{randomforest} to estimate conditional probabilities.

\subsection{Datasets and Baselines}
We give a short description of the datasets and baselines used in this section. 

\paratitle{Datasets} The following datasets and causal models were used.
\begin{itemize}
    \item The {\bf Adult} income dataset \cite{Adult} comprises demographic information of individuals along with their education, occupation, hours of work, annual income, etc. It is composed of a single table. We used the causal graph from prior studies~\cite{chiappa2019path}. 
    \item {\bf German} dataset \cite{Dua2019} contains details of bank account holders including demographic and financial information along with their credit risk. It composed of a single table and the causal graph was used from~\cite{chiappa2019path}.
    \item {\bf Amazon} dataset \cite{HeM16} is a relational database consisting of two types of tables, as described in Figure~\ref{fig:database}, and the causal graph is presented in Figure \ref{fig:causal-graph}. We identified product brand from their description, used Spacy~\cite{spacy} for sentiment analysis of reviews and estimated quality score from expert blogs~\cite{pcmag}. 
    \item {\bf German-Syn} is a synthetically generated dataset using the same causal graph as \texttt{German} dataset~\cite{Dua2019}. It consists of a single table. \reva{We consider two different versions for our analysis, one with {20K records} and the other with {1 million} records.}
    \item {\bf Student-Syn} dataset contains two different tables (a) Student information consisting of their age, gender, country of origin and their attendance. (b) Student participation attributes like discussion points, assignment scores, announcements read and overall grade. Each student was considered to enroll in $5$ different courses and their overall grade is an average over respective courses. This data was generated keeping in mind the effect of attendance on class discussions, announcements and grade. The causal model has student age, gender and country of origin as the root nodes, which affect their attendance and other performance related attributes. 
\end{itemize}
\paratitle{Variations}
In the experiments, \sys\ is run assuming that background knowledge about the causal graph is known a priori. 
We consider one variation where the causal model is not available (denoted by \sys-NB), \common{and another where we perform sampling for training the regressor (denoted \sys{}-sampled).}
\begin{itemize}
    \item {\bf \sys-NB:} when no causal model is available, all attributes are assumed to affect the updated attribute and the output.
    \item \reva{{\bf \sys{}-sampled:} is an optimized version of \sys{}  that considers a randomly chosen subset of $100$k records for the calculation of conditional probabilities of Proposition~\ref{prop:causal-results}. The choice of sample size is discussed in Section~\ref{sec:samplesize}}
\end{itemize}
\paratitle{Baselines.}We consider two different baselines of \sys\ to evaluate hypothetical queries: 
\begin{itemize}
    \item {\bf Indep:} baseline inspired by previous work on provenance updates \cite{DeutchIMT13}: this approach ignores the causal graph and assumes that there is no dependency between different attributes and tuples.
    \item {\bf Opt-HowTo:} baseline for how-to analysis where we compute the optimal solution by enumerating all possible updates, evaluating what-if query output for each update and choosing the one that returns the optimal result.
   
\end{itemize}


\subsection{\sys{} and its sampling variant\label{sec:samplesize}}
\begin{figure}
    \centering
    \subcaptionbox{Solution quality 
    \label{fig:exp:variance}}{
    \includegraphics[width=0.5\columnwidth]{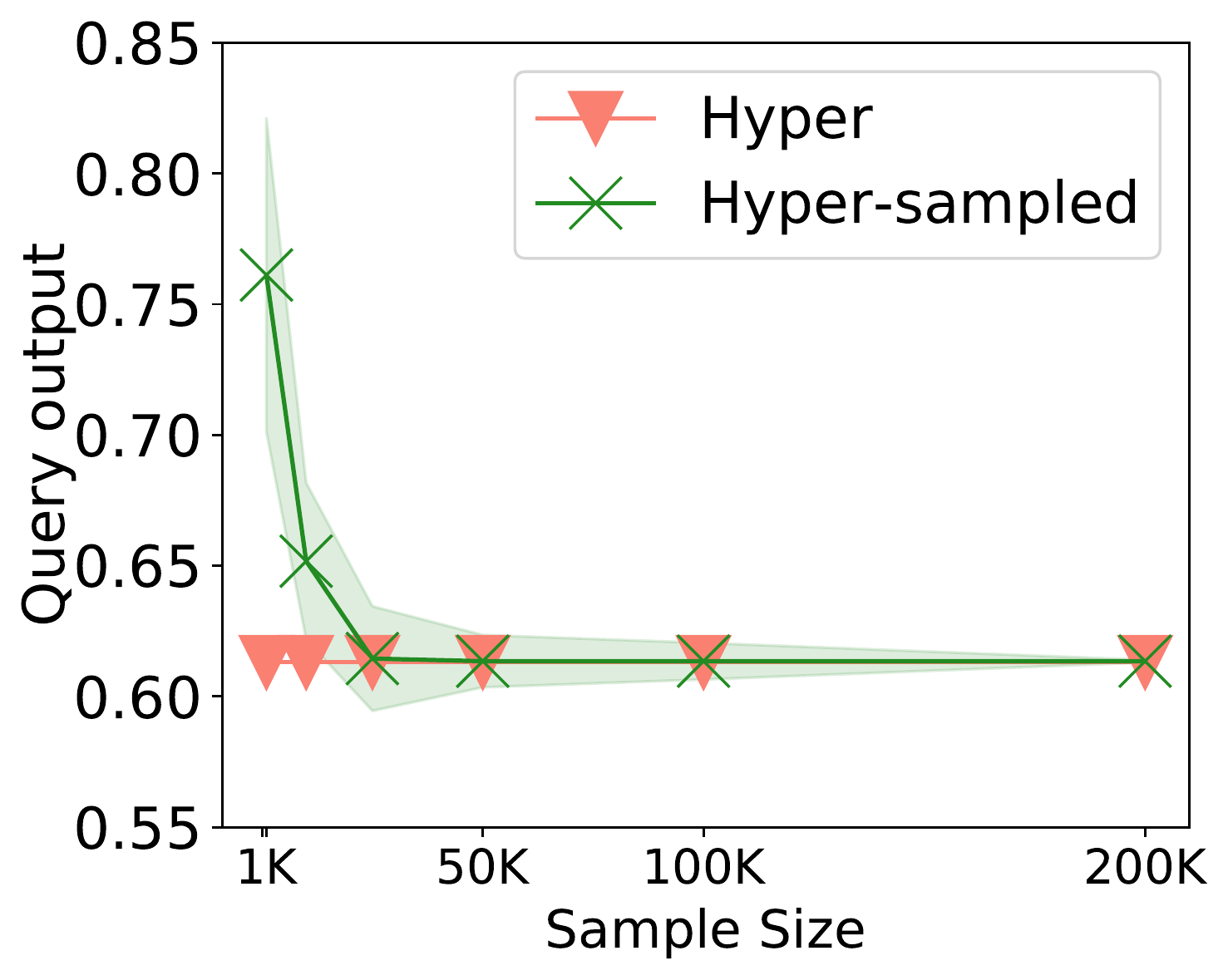}}
     \subcaptionbox{Running time 
     \label{fig:exp:sampletime}}{
    \includegraphics[width=0.41\columnwidth]{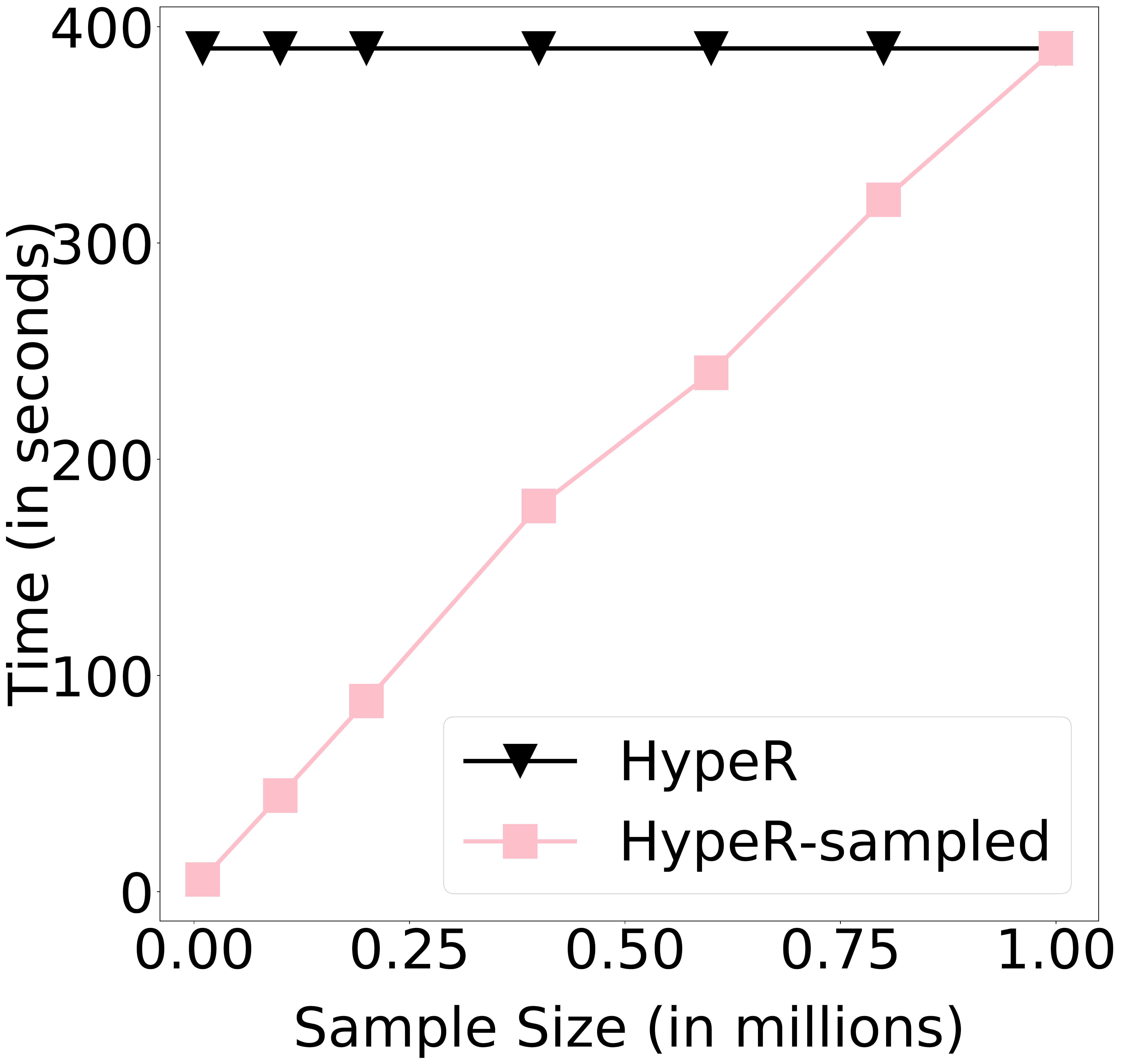}}
    \caption{\reva{Effect of varying sample size on \sys{}-sampled output and running time for \texttt{German-Syn} (1M) dataset}}
    \label{fig:samplesize}
    \vspace{-5mm}
\end{figure}

First, we evaluate the effectiveness of \sys{} with its variant \sys{}-sampled to understand the tradeoff between quality and running time.
\reva{Figure~\ref{fig:samplesize} compares the effect of changing the sample size on the quality of output generated (Figure~\ref{fig:exp:variance}) and running time (Figure~
\ref{fig:exp:sampletime}) by \sys{}-sampled. Figure~\ref{fig:exp:variance} shows that the standard deviation in query output of \sys-sampled reduces with an increase in sample size and is within $1\%$ of the mean whenever more than $100k$ samples are considered. In terms of running time, we observe a linear increase in time taken to calculate query output. Due to low variance of \sys{}-sampled for $100k$ samples and reasonable running time, we consider 100k as the sample-size for subsequent analysis. }

\subsection{What-If Real World Use Cases}

\begin{figure}[t]
    \centering \small
\begin{subfigure}{\linewidth}
	\centering
\begin{align*}
        \use\ D~~
        \update(B)= b~~
        \outputw\ \ct(Credit=Good)~~
        \forw\ \pre(A) = a
    \end{align*}
 \cut{
    \begin{align*}
        & \use\ D\\
        &\update(B)= b\\
        &\outputw\ \ct(Credit=Good)\\
        &\forw\ \pre(X_2) = x_2
    \end{align*}
    }
    \vspace{-4mm}
    \caption{What-if query (German dataset): What fraction of individuals will have good credit if $B$ is updated to $b$?}\label{fig:whatif:endtoend}
\end{subfigure}
\begin{subfigure}{\linewidth}
	\centering \small
	\vspace{-2mm}
    \begin{align*}
       \use\ D \quad \quad
        \update(B)= b \quad \quad
        \outputw\ \ct(*) \\
        \forw\   \post(Income)>50k~  \sqland\ \pre(A) = a
    \cut{
        & \use\ D\\
        &\update(B)= b\\
        &\outputw\ \ct(*)\\
        &\forw\ \post(Income)>50k~ \sqland\ \pre(a)A= a
        }
    \end{align*}
    \vspace{-5mm}
    \caption{What-if query (Adult dataset): How many individuals with attribute $A=a$ will have income $\geq 50K$ if $B$ is updated to $b$?}\label{fig:whatif:adult}
\end{subfigure}
\caption{What-if queries for real world use cases}\label{fig:real-world-queries}
\end{figure}
{
\begin{table}[t] \centering \small
\vspace{-2mm}
\caption{\textmd{Average Runtime in seconds for $\ct$ query to evaluate the effect of a hypothetical update on target for what-if queries. \reva{The time in (..) in the last row is by \sys(-NB)-sampled, which takes the same time as \sys(-NB)\ on all other datasets with $<100k$ tuples.}
}
}
	\label{tbl:data}
	{\footnotesize
		\begin{tabular}{@{}lrrrrrrrrr@{}}\toprule
			{Dataset} & {Att. [$\#$]} & {Rows[$\#$]} &  \sys & \sys-NB &  Indep   \\ \midrule
			\textbf{Adult}~\cite{Adult} & 15 & 32k & 45s  &105s & 3s&  \\ 
			\textbf{German}~\cite{Dua2019}& 21  & 1k & 1.2s & 12.5s&0.4s & \\ 
		    {\textbf{Amazon}}~\cite{HeM16}& {5,3} &{3k, 55k}  & 1.7s & 10.5s&0.8s & \\ 
		   \textbf{Student-syn}& 3,6  & 10k,50k & 4.5s  &12.3s & 1.2s & \\
		   
		   \reva{\textbf{German-Syn} (20k)}& 6  & 20k & 7.2s & 22.45s & 1.4s & \\ 
			
		\reva{\textbf{German-Syn} (1M)}& \reva{6}  & \reva{1M} & \reva{390s (44.5s)} & \reva{1173s (132s)} & \reva{73s} & 
		\end{tabular}
		}
		\vspace{-4mm}
	
\end{table}
}

\begin{figure}
    \centering
    \subcaptionbox{\texttt{German} \label{fig:exp:germanreal}}{
    \includegraphics[width=0.47\columnwidth]{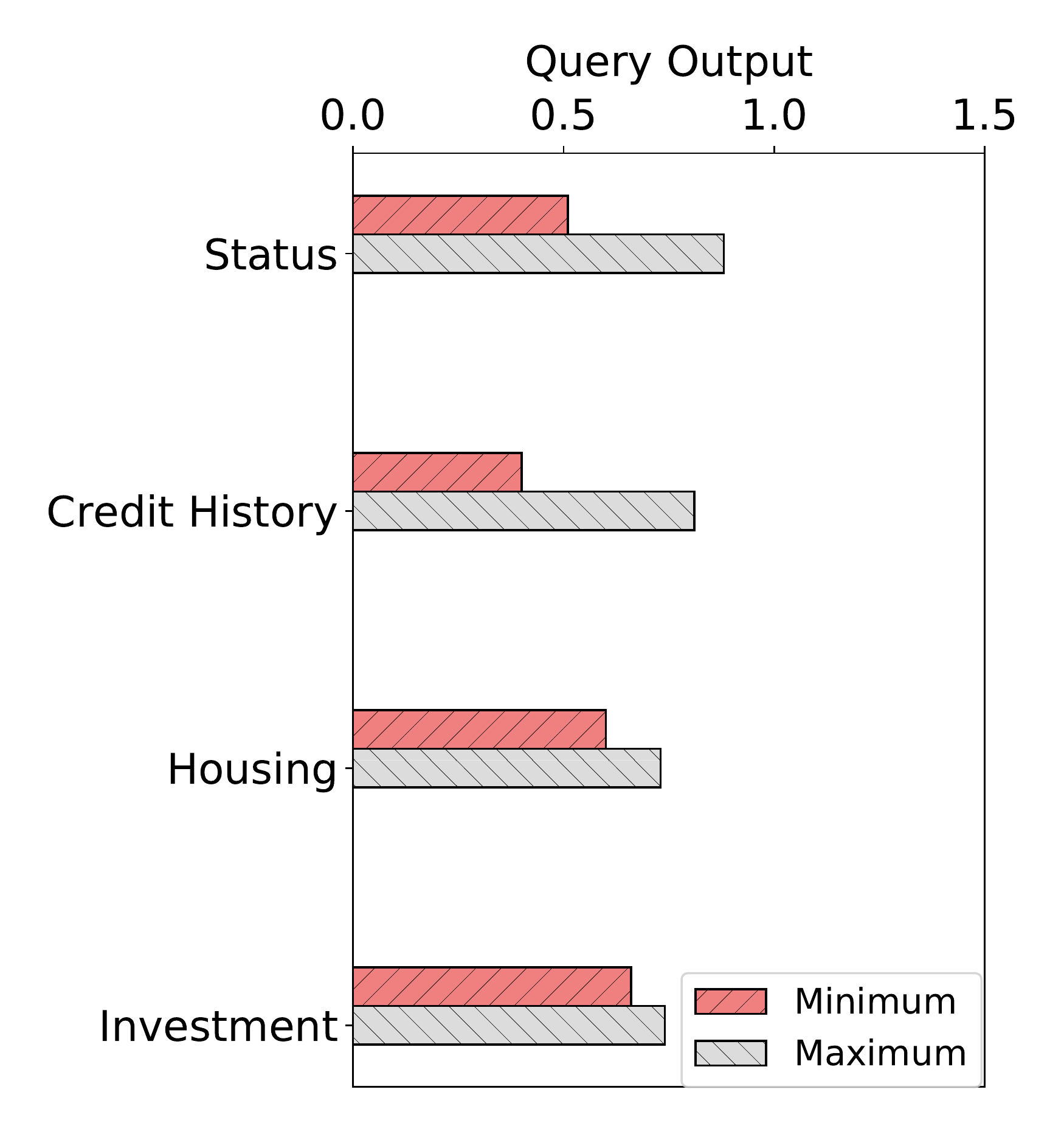}}
     \subcaptionbox{\texttt{Adult} \label{fig:exp:adultreal}}{
    \includegraphics[width=0.47\columnwidth]{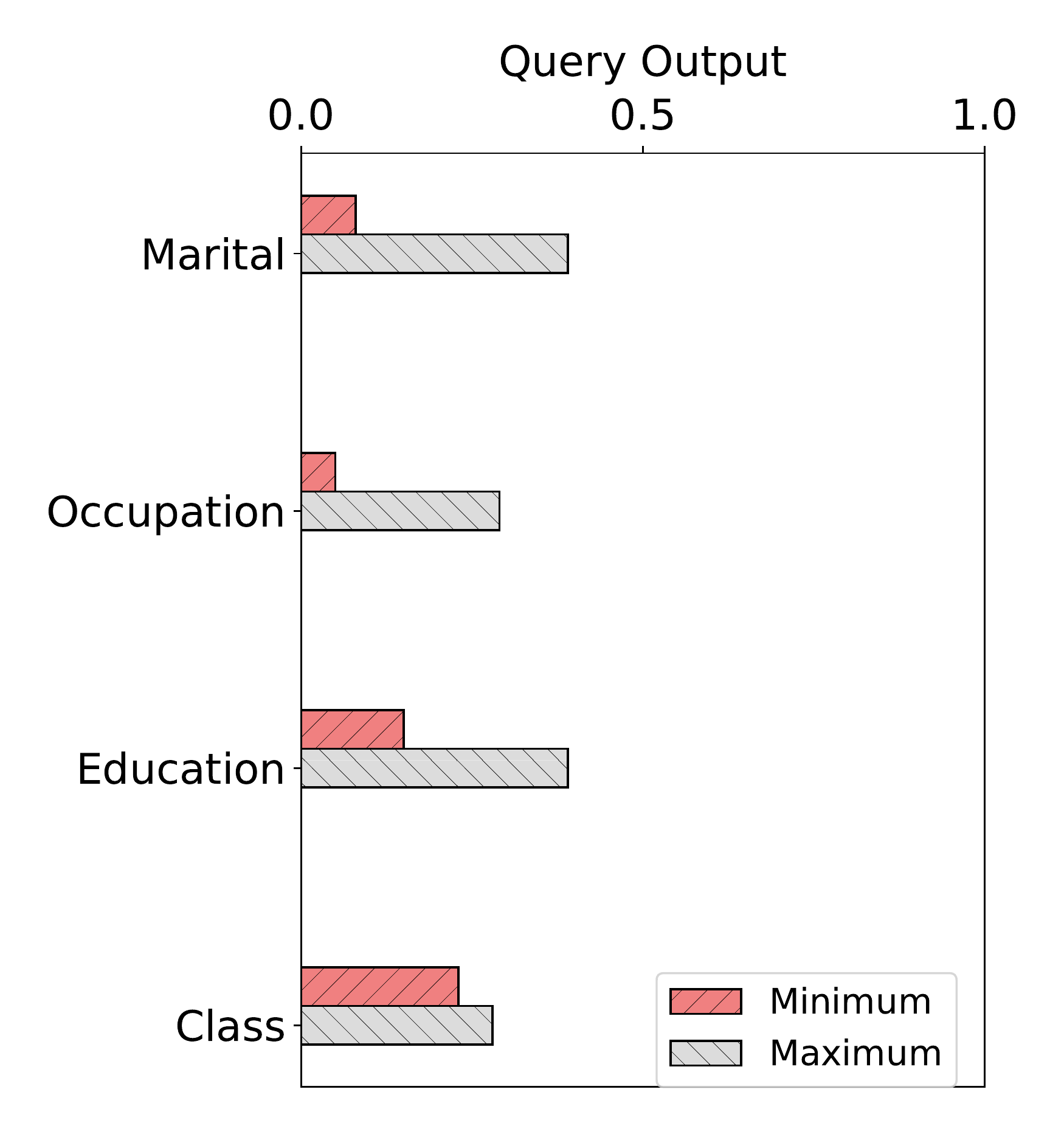}}
    \caption{What-if query output for \texttt{German} and \texttt{Adult} datasets on updating each attribute to their min and max values; a larger gap 
    denotes higher attribute importance. 
    }
    \label{fig:real}
    \vspace{-5mm}
\end{figure}
In this experiment, we evaluate the output of \sys\  on a diverse of hypothetical queries on various real-world datasets. Due to the absence of ground-truth, we discuss the coherence of our observations with intuitions from existing literature.

\paratitle{German} We considered a hypothetical update of fixing attributes `Status', `Credit history', and `housing' to their respective minimum and maximum values to evaluate the effect of these attributes on individual credit. Figure~\ref{fig:whatif:endtoend} demonstrates the query template where $X,x, X_2,x_2$ are varied to evaluate the effect of different updates. Whenever status or credit history are updated to the maximum value, more than $81\%$ of the individuals have good credit. Similarly, updating these attributes to the minimum value reduces the credit rating of more than $30\%$ individuals. 
On the other hand, updating other attributes like `housing' and `investment' affects the credit score of less than $20\%$ individuals. Figure~\ref{fig:exp:germanreal} presents the effect of updating these attributes to their minimum and maximum value. Larger gap in the query output for Status and credit history shows that these attributes have a higher impact on credit score.
We also tested the effect of updating pairs of attributes and observed that {\bf updating `credit history' and `status' at the same time can affect the credit score of more than $70\%$ individuals.}
These observations are consistent with our intuitions that credit history and account status have the maximum impact of individual credit. 

\paratitle{Adult} This dataset has been widely studied in the fairness literature to understand the impact of individual's gender on their income. It has a peculiar inconsistency where married individuals report total household income demonstrating a strong causal impact of marital status on their income \cite{DBLP:conf/sigmod/SalimiGS18,TAGH+17,10.1109/ICDM.2011.72}. We ran a hypothetical what-if query to analyze the fraction of high-income individuals when everyone is married (Figure~\ref{fig:whatif:adult}). We observed that $38\%$ of the individuals have more than $50$K salary. Similarly, {\bf if all individuals were unmarried or divorced, less than $9\%$ individuals have salary more than $50$K}. This wide gap in the fraction of high-income individuals for two different updates of marital status demonstrate its importance to predict household income. Figure~\ref{fig:exp:adultreal} shows the effect of updating the attributes with the minimum or the maximum value in their domain.
Additionally, updating class of all individuals has a smaller impact on the fraction with higher income. These observations match the observations of existing literature~\cite{GalhotraPS21}, where marital status, occupation and education have the highest influence on income.

\paratitle{Amazon} We evaluated the effect of changing price of products of different brands on their rating.  When all products have price more than the $80^{th}$ percentile, around $32\%$ of the products have average rating of more than $4$. {\bf On further reducing the laptop prices to $60^{th}$ and $40^{th}$ percentiles, more than $60\%$ of the products get an average rating of more than $4$}. This shows that reducing laptop price increases average product ratings. Among different brands, we observed that Apple laptops have the maximum increase in rating on reducing laptop prices, followed by Dell, Toshiba, Acer and Asus. These observations are consistent with previous studies on laptop brands~\cite{amazonstudy}, which mention Apple as the top-quality brand in terms of quality, customer support, design, and innovation.

\vspace{-2mm}
\subsection{Solution Quality Comparison}\label{sec:sol_quality}
In this experiment, we analyzed the quality of the solution generated by \sys\ with respect to the ground truth and baselines over synthetic datasets. The ground truth values are calculated using the structural equations of the 
causal DAG for the synthetic data. 

\begin{figure}
    \centering
    \subcaptionbox{\reva{\texttt{German-Syn (1M)}} \label{fig:exp:germansyn}}{
    \includegraphics[width=0.45\columnwidth]{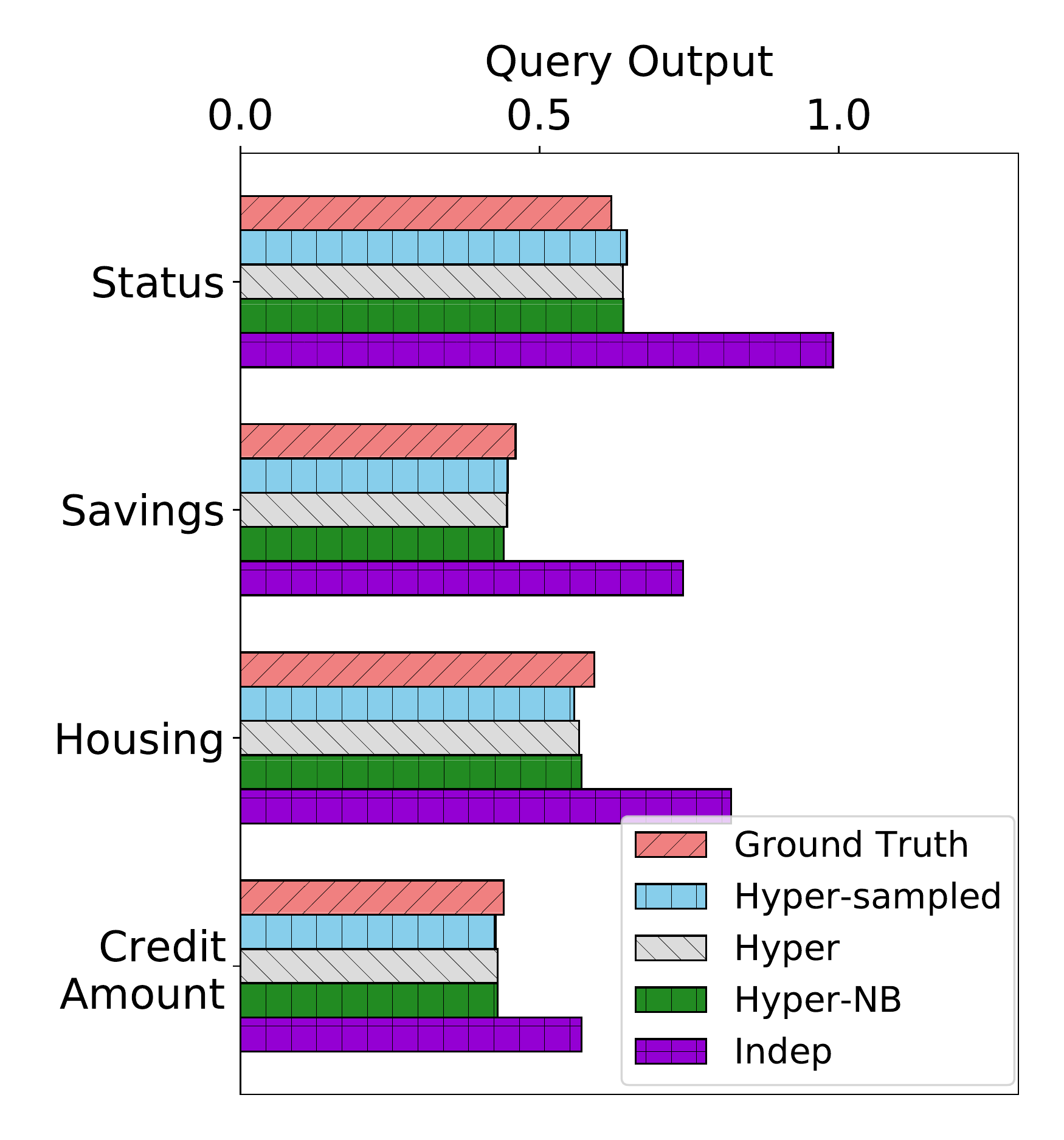}}
     \subcaptionbox{\texttt{Students-Syn} \label{fig:exp:studentsyn}}{
    \includegraphics[width=0.45\columnwidth]{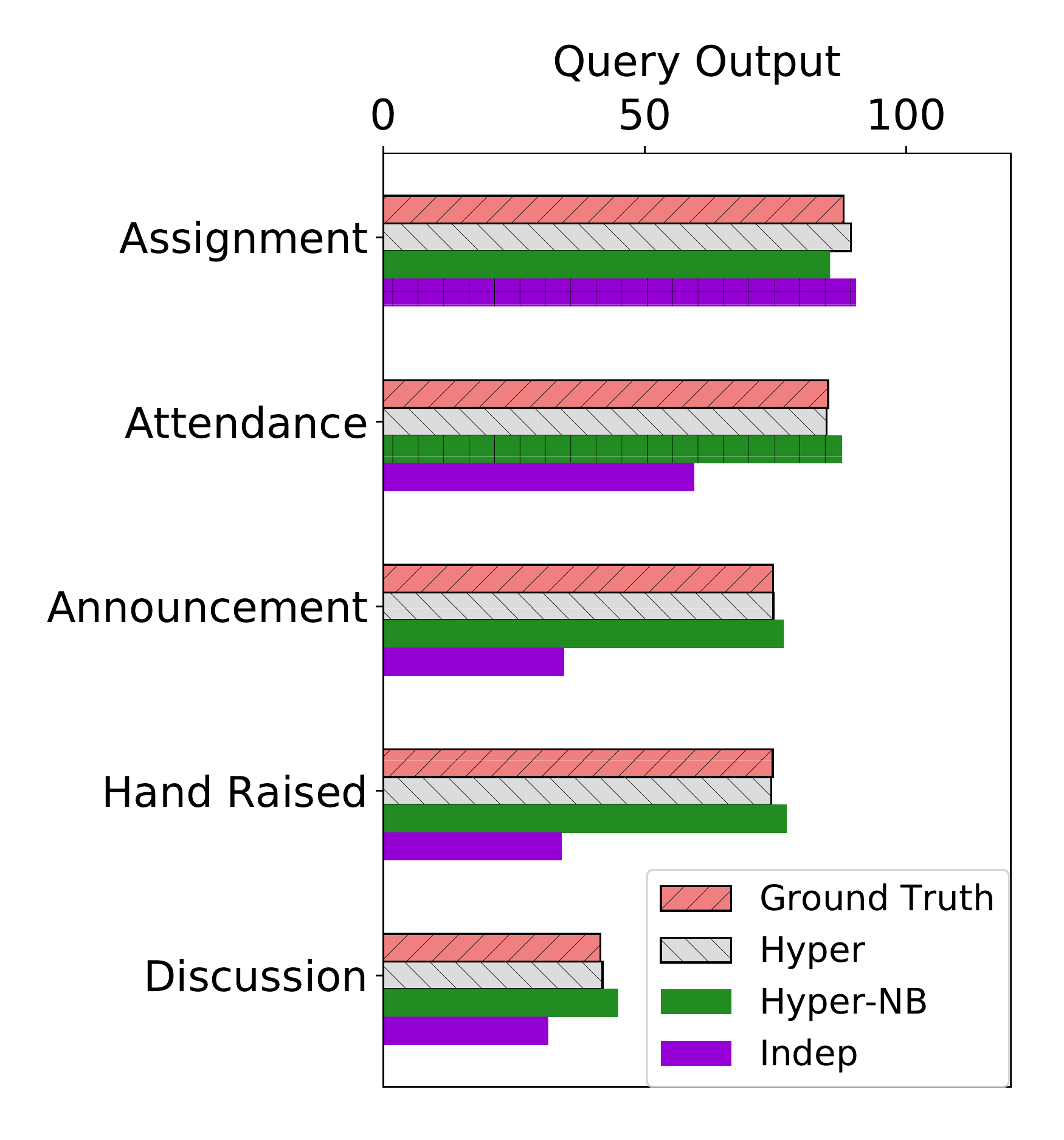}}
    \caption{\reva{What-If Query output.
    }}
    \label{fig:corr}
    \vspace{-5.5mm}
\end{figure}

\paratitle{What-if}
For the \reva{German-Syn (1M)} dataset, Figure~\ref{fig:exp:germansyn} presents the output of a query that updates different attributes related to individual income and evaluates the probability of achieving good credit. For all attributes, \reva{\sys, \sys-sampled, and \sys-NB estimate the query output accurately with an error margin of less than $5\%$.}  In contrast, Indep baseline ignores the causal structure and relies on correlation between attributes to evaluate the output. Since, the individuals with high status are highly correlated with good credit, Indep incorrectly outputs that updating Status would automatically improve credit for most of the individuals.

For the Student-Syn dataset, Figure~\ref{fig:exp:studentsyn} presents the average grade of individuals on updating different attributes that are an indicator of their academic performance. In all cases, \sys\ and \sys-NB output is accurate while Indep is confused by correlation between attributes and outputs noisy results.
In addition to these hypothetical updates, we considered complex what-if queries that analyzed the effect of assignment and discussion attributes on individuals that read announcements and have high attendance. In these individuals, we observed that improving assignment score has the maximum effect on overall grade of individuals.

\paratitle{How-to}
For the \reva{German-Syn (20k)} dataset, we considered a how-to query that aims to maximize the fraction of individuals receiving good credit. We provided Status, Savings, Housing and Credit amount as the set of attributes in the \howto\ operator. \revb{\sys\ returned that updating two attributes i)  account status, and ii) housing attributes is sufficient to achieve good credit. This showed that updating a single attribute would not maximize the fraction of individuals with good credit. We evaluated the ground truth (Opt-HowTo) by enumerating all possible update queries  and  used the structural equations of the causal graph to evaluate the post-update value of the objective function for each update. We identified that \sys's output matches the ground truth update. }

For the Student-Syn dataset, we evaluated a how-to query to maximize average grades of individuals with a budget of updating atmost one attribute. \sys\ returned that improving individual attendance provide the maximum benefit in average grades.
This output is consistent with ground truth calculated by evaluating the effect of all possible updates (Opt-HowTo). 

\begin{figure}
    \centering
    \subcaptionbox{Solution quality 
    \label{fig:exp:bucketquality}}{
    \includegraphics[width=0.45\columnwidth]{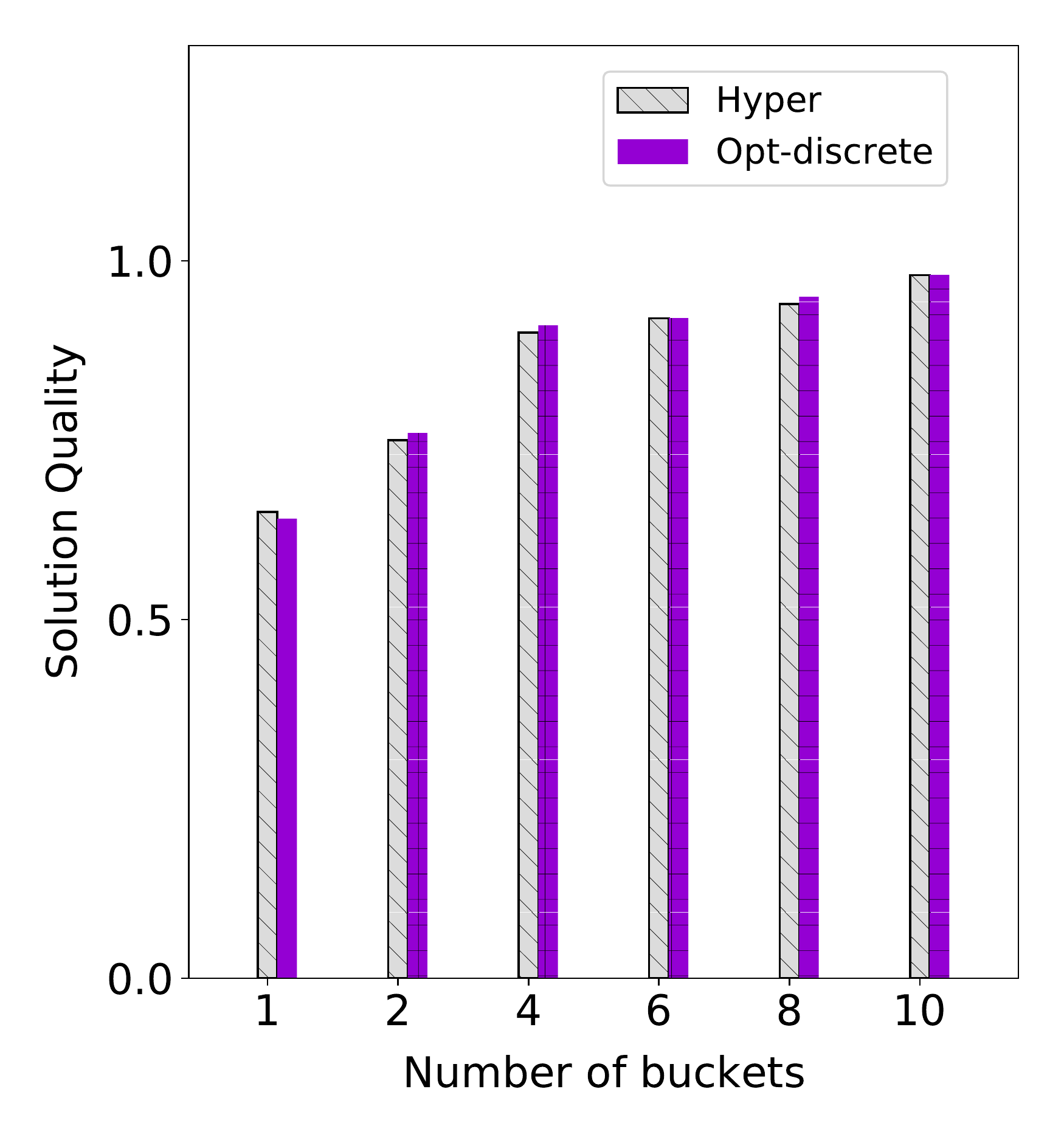}}
     \subcaptionbox{Running time
     \label{fig:exp:buckettime}}{
    \includegraphics[width=0.45\columnwidth]{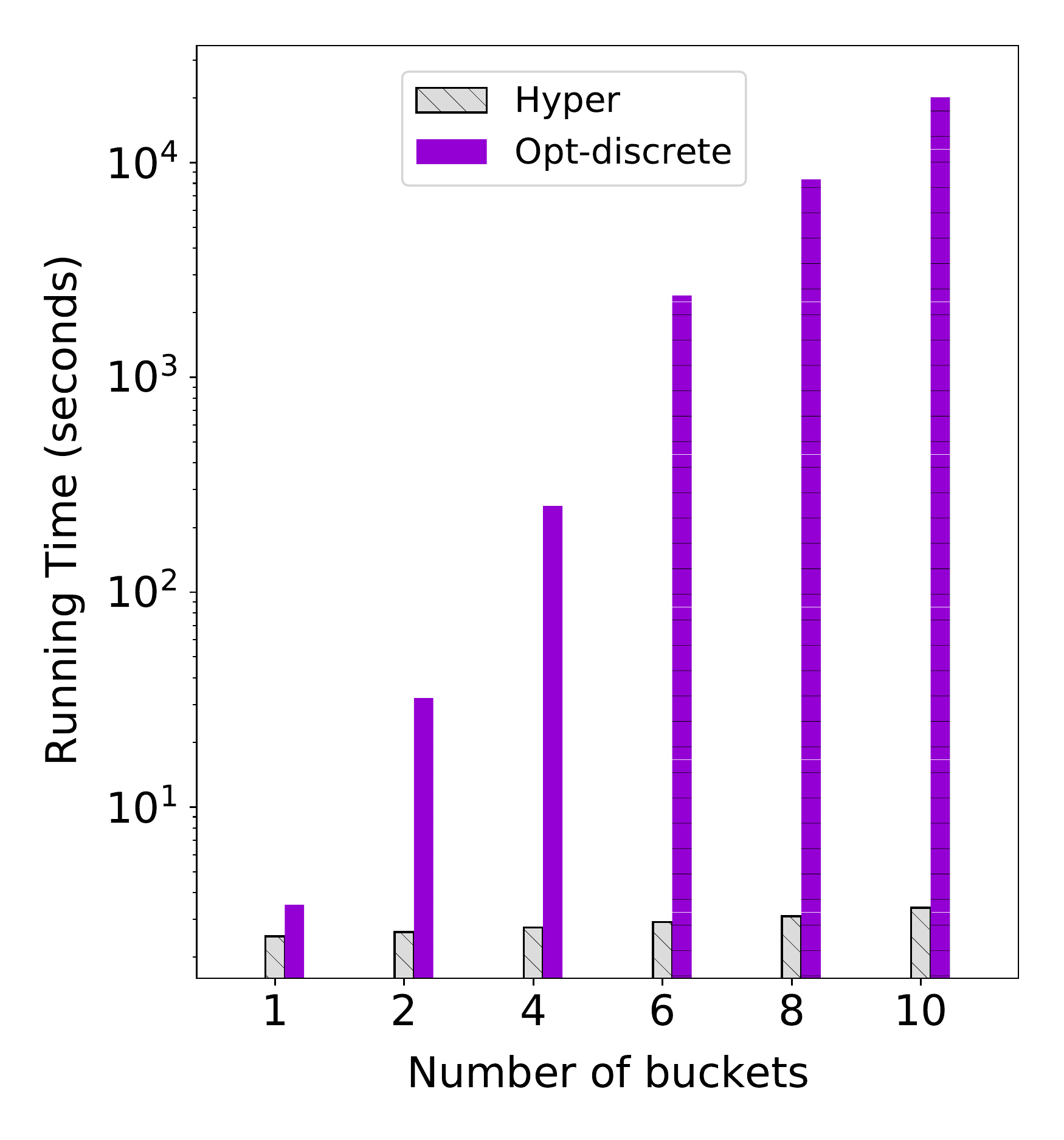}}
    \vspace{-0.5mm}
    \caption{\revc{How-to Query output for \texttt{German-Syn (20k)} with varying number of buckets.}}
    \label{fig:buckettradeof}
    \vspace{-5mm}
\end{figure}

\paratitle{\revc{Effect of discretization}}
\revc{\sys{} bucketizes all continuous attributes before solving the integer program. In this experiment, we evaluate the effect of number of buckets on the solution quality and running time on a modified version of \texttt{German-Syn} (20k) dataset that contains continuous attributes.  We partitioned the dataset into equi-width buckets and compared the solution returned by 
\sys{} and the optimal solution calculated after discretization (Opt-discrete) with the ground truth solution (OptHowTo). Figure~\ref{fig:exp:bucketquality} compares the quality of \sys{} and Opt-discrete as a ratio of the optimal value. We observe that the solution quality improves with the increase in the number of buckets and the returned solution is within $10\%$ of the optimal value whenever we consider more than $4$ buckets. The solution returned by Opt-discrete is similar to that of \sys{}. The time taken by Opt-discrete increases exponentially with the number of buckets. In contrast, time taken by \sys{} does not increase considerably as the number of variables in the integer program depends linearly on the number of buckets. This shows that running \sys{} over a bucketized version of the dataset leads to competitive quality in reasonable amount of time.
}

\subsection{Runtime Analysis and Comparison}\label{subsec:runtime}
In this section, we evaluate the effect of different facets of the input on the runtime of \sys. \reva{Note that our approach comprises two steps: (a) creating the aggregate view on which the query should be computed (done using a join-aggregate query), and (b) training regression functions to calculate conditional probability in the calculation of query output (the mathematical expression is in Proposition~\ref{prop:causal-results}). This training is performed over a subset of the attributes of the view computed in the previous step. 
Training a regression function is more time-consuming than computing the aggregate view in step (1). Therefore, \sys{} is as scalable as prior techniques for regression (we use a random forest regressor from the sklearn package).  Hence the parameters we consider } include (1) database size, (2) backdoor set size (see Section \ref{sec:whatif-algo}), and (3) query complexity.
Since the effect of (1), (2) on the runtime of what-if query evaluation is directly translated to an effect on the runtime of how-to query evaluation, for how-to queries, we focus on the effect of the number of attributes in the $\howto$ operator which will change the optimization function $\phi$ (see Section \ref{sec:howto-algo}). We use the synthetic datasets \texttt{German-Syn} and \texttt{Student-Syn}. 

\begin{figure}
    \centering
    \subcaptionbox{ What-if query  \label{fig:exp:scalabilitywhatif}}{
    \includegraphics[width=0.43\columnwidth]{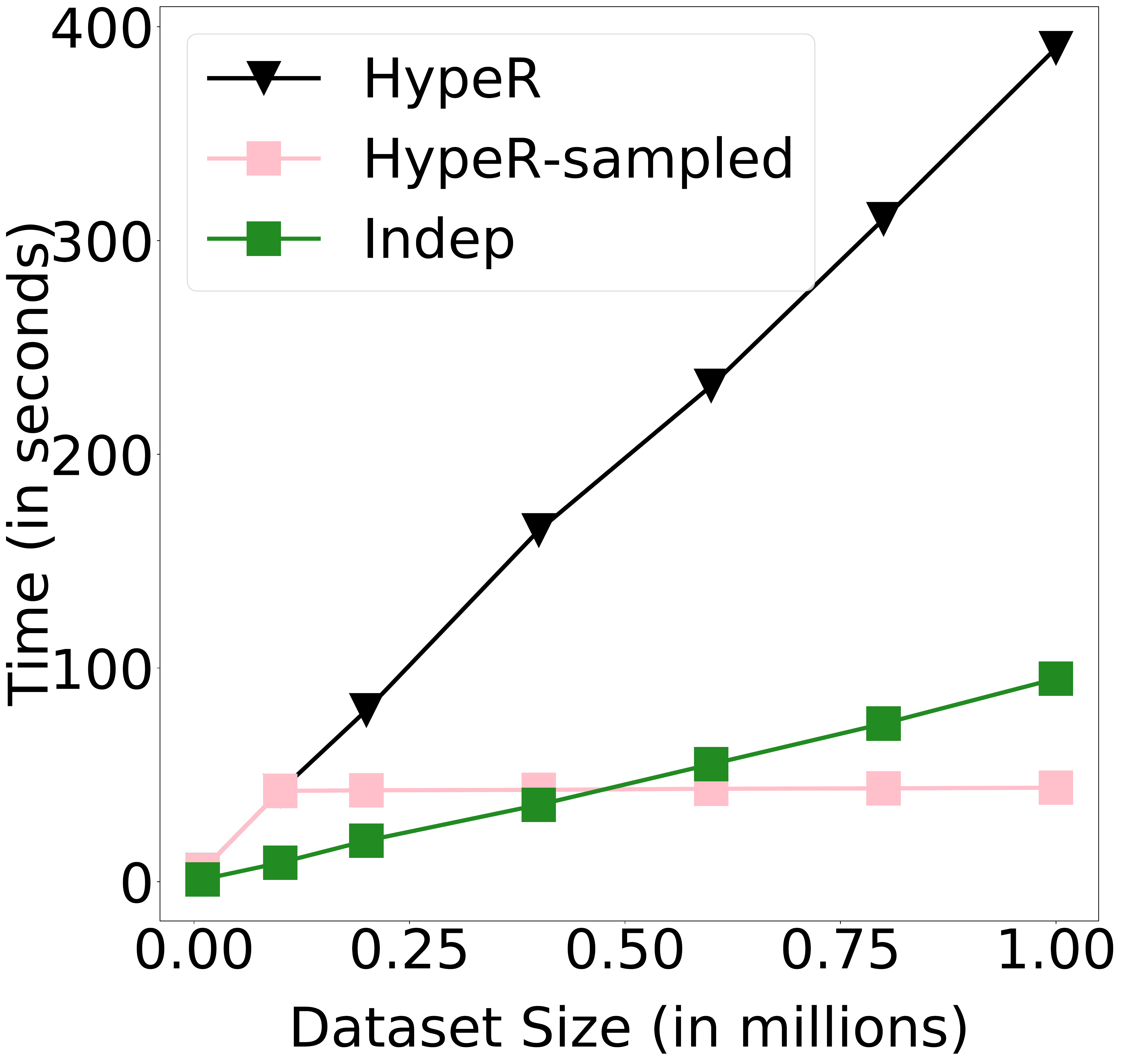}}
     \subcaptionbox{ How-to query \label{fig:exp:scalabilityhowto}}{
    \includegraphics[width=0.45\columnwidth]{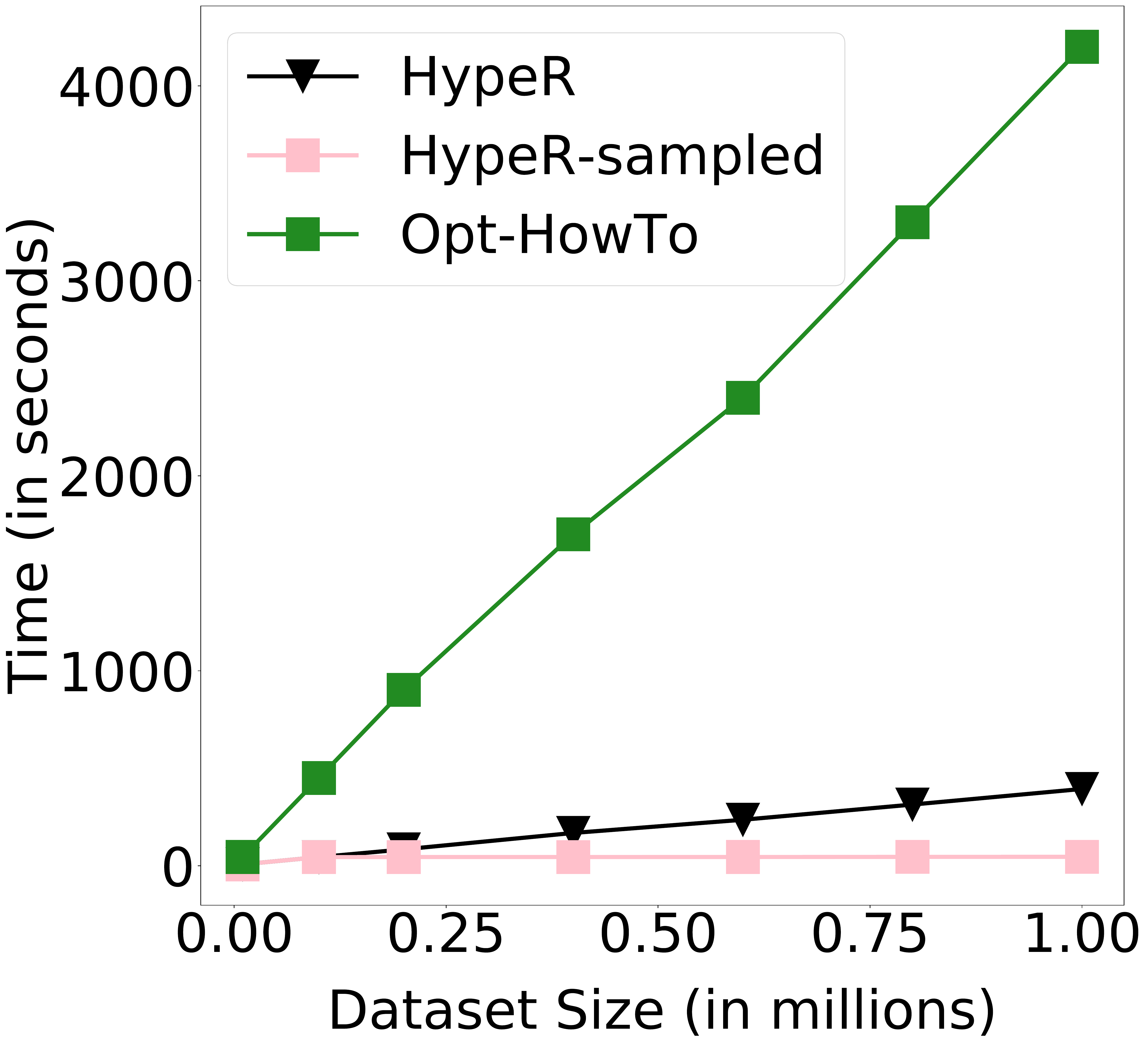}}
    \caption{\reva{Running Time comparison on varying dataset size for \texttt{German-Syn} dataset averaged over five different queries.}}
    \label{fig:runtime-sample}
    \vspace{-5mm}
    \label{fig:scalability}
\end{figure}

\paratitle{What-if: database size} Table~\ref{tbl:data} presents the average running time to evaluate the response to a what-if query in seconds. \reva{To further evaluate the effect of database size on running time, we considered \texttt{German-Syn} dataset and varied the number of tuples from $10K$ to $1M$. In this experiment we consider a new variation of \sys{}, denoted by \sys-sampled, which considers a randomly chosen subset of $100K$ records for the calculation of conditional probabilities of Proposition~\ref{prop:causal-results}. Figure~\ref{fig:scalability} compares the average time taken by \sys{}, \sys{}-sampled with Indep for five different What-If queries and Opt-HowTo for How-to queries.
We observed a linear increase in running time with respect to the dataset size for all techniques except \sys{}-sampled.  The increase in running time is due to the time taken to train a regressor which is used to estimate conditional probabilities for query output calculation
To answer a what-if (or how-to) queries,  aggregate view calculation requires less than $1\%$ of the total time. The majority of the time is spent on calculating the query output using the result in Proposition~\ref{prop:causal-results}.
Therefore, the time taken by \sys-sampled does not increase considerably when the dataset size is increased beyond 100K. 

}

\paratitle{What-if: backdoor set size} This experiment changed the background knowledge to increase the backdoor set from $2$ attributes to $6$ attributes. The running time to calculate expected fraction of high credit individuals on updating account status increased from $7.2$ seconds when backdoor set contains age and sex to $22.45$ seconds when the backdoor set contains all attributes. 

\paratitle{What-if: query complexity} 
In this experiment, we synthetically add multiple attributes in the \texttt{Student-syn} dataset and the different operators of the query to estimate their on running time.

On adding multiple attributes in the \use\ operator, the time taken to compute the \augmentedtable\ increases minutely. For \texttt{Student-Syn}, \use\ operator was evaluated in less than $0.5$ seconds when $5$ different attributes are added from other datasets. The increase in these attributes do not affect the running time of subsequent steps unless the attributes in \forw\ operator increase.

\begin{figure}
    \centering
    \subcaptionbox{What-if (\forw\ operator)  \label{fig:exp:timewhatif}}{
    \includegraphics[width=0.45\columnwidth]{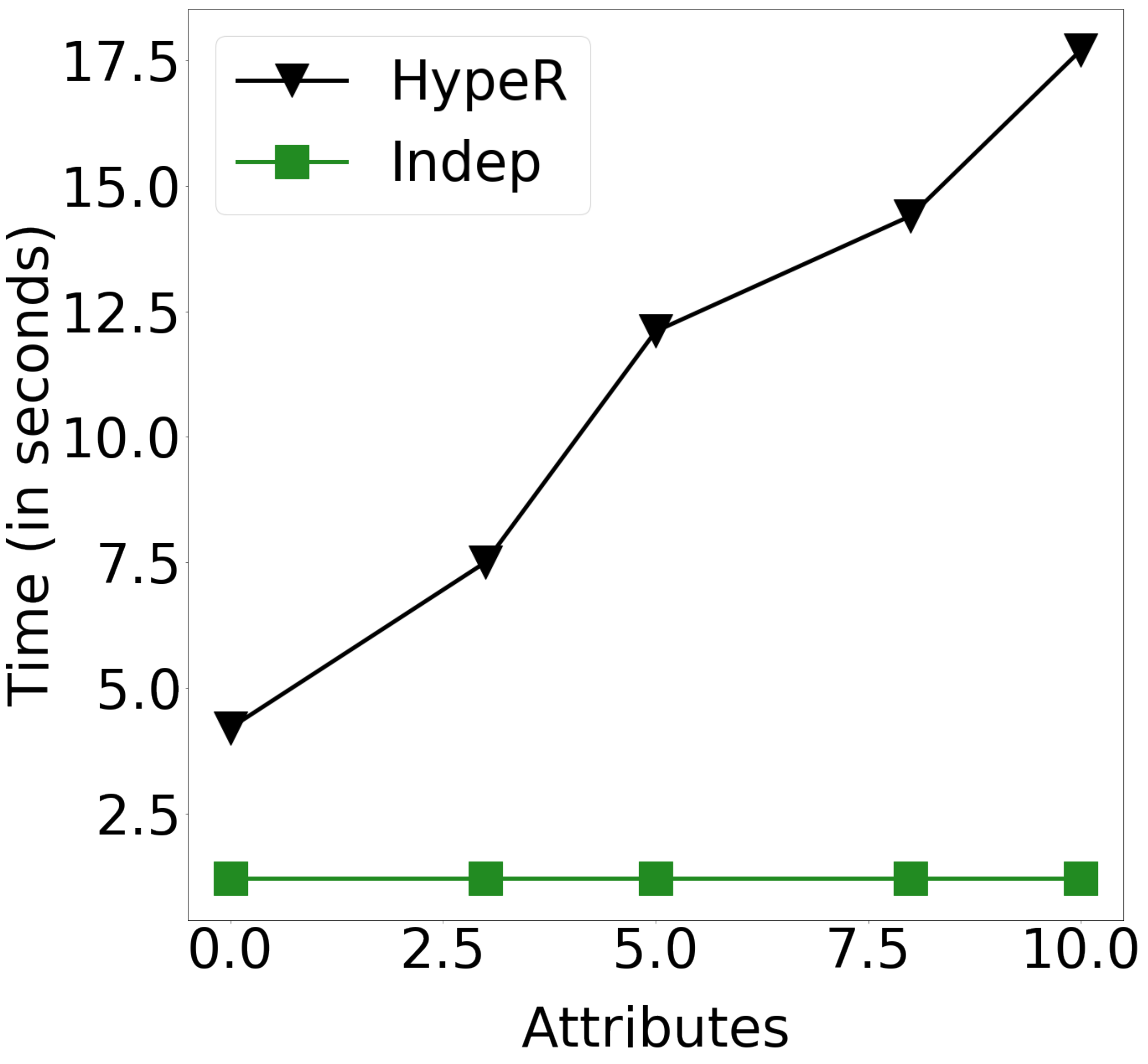}}
     \subcaptionbox{How-to (\howto) \label{fig:exp:timehowto}}{
    \includegraphics[width=0.45\columnwidth]{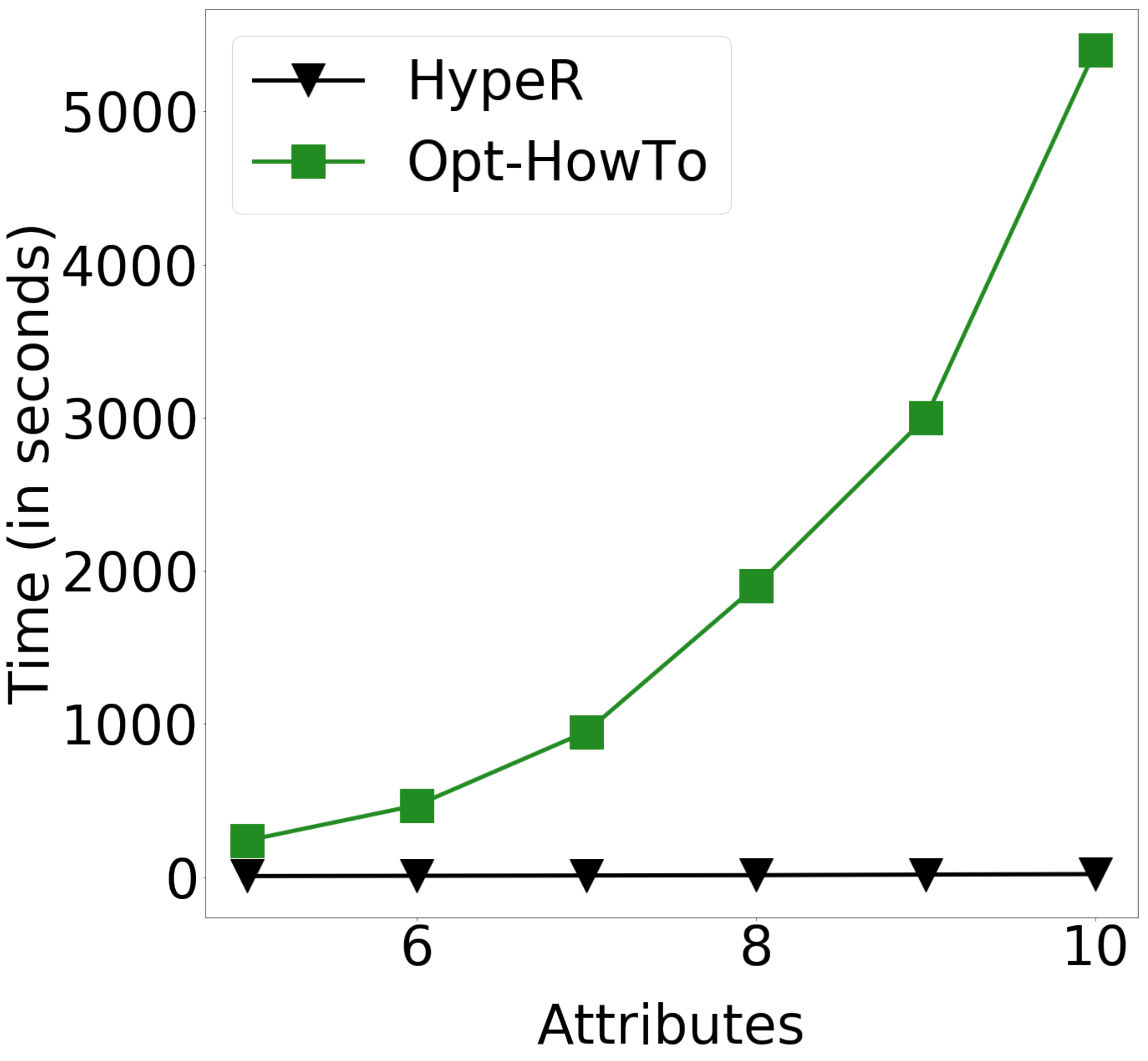}}
    \caption{Running Time comparison on varying number of attributes in different operators for \texttt{Student-Syn} dataset.}
    \label{fig:corr:time}
\end{figure}

We now compare the effect of adding multiple attributes in the \forw\  operator of a \ct{} query.  Adding conditions involving \pre\ value of randomly chosen attributes increases the number of attributes used to train the regressor, which increases the running time (Figure~\ref{fig:exp:timewhatif}).  Running time increased from $4.2$ seconds when \forw\  operator is empty to $12.1$ seconds and $17.7$ seconds when it contains $5$ and $10$ attributes, respectively. In contrast, Indep is more efficient as it does not use additional attributes to compute query output.
However, if the added attribute is in the backdoor set, then the output is evaluated faster. To understand the effect of adding such attributes, we considered a query where the backdoor set contained $10$ binary attributes. To evaluate the output, probability calculation iterated over the domain of backdoor attributes and required $49.7$ seconds. The running time reduced to $7.4$ seconds when $5$ conditions on these attributes are added to the \forw\ operator.


\paratitle{How-to: query complexity} Figure~\ref{fig:exp:timehowto} presents the effect of the number of attributes in \howto\ operator on the time taken to process the query. Increasing attributes leads to a linear increase in the number of variables in the integer program.  Therefore, the time taken by \sys\ increases from $7$ seconds for $5$ attributes in \howto\ operator to $20$ seconds for $10$ attributes. In contrast, Opt-HowTo considers all possible combinations of attribute values in the domain of attributes in the \howto operator. It takes around $4$minutes for $5$ attributes and more than $90$ minutes for $10$ attributes. This shows that the Integer Program based optimization provides orders of magnitude improvement in running time.

%% file: related.tex
\section{related work}\label{sec:related}
Here we review relevant literature in 
hypothetical reasoning in databases, probabilistic databases, and causality. 
{\em The main distinction of this paper from previous work is a framework that allows for hypothetical reasoning over relational databases using a post-update distribution 
over possible worlds that is able to capture both direct and indirect probabilistic dependencies between attributes and tuples using a probabilistic relational causal model.}

Previous work has focused on What-if and How-to analysis mainly in terms of provenance and view updates. 
Due to its practicality, and real applications like evaluating business strategies, there have been several works that developed support for \emph{hypothetical what-if reasoning} in SQL, OLAP, and map-reduce 
environments \cite{BalminPP00,LakshmananRS08,ZhouC09,HerodotouB11,NievaSS20}. 
What-if reasoning through provenance updates 
have been studied in \cite{DeutchIMT13,DeutchMT15,ArabG17,DeutchMR19} 
to efficiently measure the direct effect of updating values in the database on a view created by the query. 
Nguyen et. al. \cite{Nguyen0WZYTS18} study the problem of efficiently performing what-if analysis with conflicting goals 
using data grids. 
Other works have considered models for hypothetical reasoning in temporal databases \cite{ArenasB02,HartmannFMRT19}, where 
Arenas et. al. \cite{ArenasB02} focused on a logical model in which each transaction updates the database and the goal is to answer a query about the generated sequence of states, without performing the update on the whole database, and GreyCat \cite{HartmannFMRT19} focused on  time-evolving graphs. 
Christiansen et. al. \cite{ChristiansenA98} 
propose an approach that considers a single possible world and then modifies the query evaluation procedure within a logic-based framework. 
Another part of hypothetical reasoning is \emph{how-to queries} which have been explored mostly in terms of provenance updates \cite{MeliouGS11Reverse,MeliouGS11,MeliouS12} that compute their results with hypothetical updates modeled as a Mixed Integer Program. 
MCDB \cite{JampaniXWPJH08} allows users to create an uncertain database that has randomly generated values in the attributes or tuples (that may be correlated with other attributes or tuples). These are generated using variable generation functions that can be arbitrarily complex. It then evaluates queries over this database using Monte Carlo simulations. 
Eisenreich et. al. \cite{EisenreichR10} propose a data analysis system allowing users to input attribute-level uncertainty and correlations using histograms and then perform operations on the data such as aggregating or filtering uncertain values. 
%
We note that uncertainty in databases has been studied in previous work on {\em probabilistic databases} \cite{AgrawalBSHNSW06,DalviS07,AntovaKO07a,DalviRS09,Suciu20} where each tuple or value has a probability or confidence level attached to it, and in stochastic package queries \cite{BrucatoYAHM20} that allow for optimization queries on stochastic attributes. We adapt and use the concept of block-independent database model from probabilistic databases \cite{ReS07,DalviRS09} in this paper. 
The framework suggested in this paper uses a {\em probabilistic relational causal model} \cite{SalimiPKGRS20} to model updates as interventions and generate the post-update distribution that describes the dependencies between the attributes and tuples. There is a vast literature on 
{\em observational causal inference} on stored data
in AI and Statistics (e.g., \cite{greenland1999relation, robins1989probability, greenland1999epidemiology, tian2000probabilities, robertson1996common, cox1984probability, pearl2009causality,angrist1996identification, rubin2005causal}), and we use standard techniques from this literature to compute query output.


%% file: conclusions.tex
 \vspace{-2mm}
\section{conclusions}\label{sec:conc}
We have defined a probabilistic model for hypothetical reasoning in relational databases. 
While the post-update distribution can stem from any probabilistic model, we focus here on causal models. 
We develop \sys: a novel framework that supports what-if and how-to queries and performs hypothetical updates on the database, measures their effect, and computes the query results. 
Our framework includes new SQL-like operators to support these queries for testing a wide variety of hypothetical scenarios. 
We prove that the results of our queries can be computed using causal inference and we further devise an optimizations by  block-independent decompositions. 
We show that our approach provides query results that are rational and account for implicit dependencies in the database. 
\common{In future work, we plan to add support for multi-attribute updates consisting of dependent attributes and also account for database constraints and other semantic constraints.
Extensions to cyclic dependencies of attributes in causal graphs is an intriguing future work. One idea that can be explored is `unfolding' cyclic dependencies  between attributes A and B by using a time component on attributes, and adding edges from $A[t]$ to $B[t']$ and $B[t]$ to $A[t']$  where time $t' > t$ (called \emph{`chain graphs'}, e.g., \cite{DBLP:conf/uai/ShermanS19, ogburn2020causal}).
We also plan to 
develop an interactive UI where users can pose and explore hypothetical queries.} 

%% file: proofs/decomposition-proof.tex
\onecolumn
\appendix
\section{Appendix: Computation of What-if queries and Proofs}\label{sec:appendix}

The computation of what-if queries in the most general form uses a number of techniques including decomposable aggregates, block-independent decompositions (when available), and causal graphs (when available) and backdoor condition from the causal inference literature. For readability, we decompose the computations and their correctness proofs in the following steps:
\begin{enumerate}
    \item {\bf (Section~\ref{sec:to-blocks})} Computation of what-if queries for a single-relation database with a block-independent decomposition can be reduced to computation of (modified) what-if queries on individual blocks using properties of decomposble aggregate functions (see Proposition~\ref{prop:single-relation-decompose}). This step is omitted if there are no block-independent decomposition, i.e., if the entire database forms a single block.
    \item {\bf (Section~\ref{sec:single-block})} Computation of what-if queries for a single block within a single-relation database. This calculation leverages the causal graph $G$ and the set of attributes that satisfy the backdoor criterion to estimate the query output for a block.
    \item {\bf (Section~\ref{sec:muti-relation})} Extends the analysis of single-relation database to multi-relation database.
    \item {\bf (Section~\ref{sec:implementation})} Presents the key ideas used to estimate the conditional probability distribution from the original database $D$ in our algorithms.
\end{enumerate}



\subsection{Reduction from Block-Independent Decomposition to Individual Blocks}\label{sec:to-blocks}
First we give a proof that the computation of a what-if query can be computed as the aggregate of the results of what-if queries over each block where the database $D$ has a single relation $R$, such that both the update attribute $B$ and the outcome attribute $Y$ belong to $\attr(R)$ for any given what-if query $Q$. In particular,  on such a database, we can assume without loss of generality that the \augmentedtable\ $\cview = R = D$, although some of the attributes of $R$ may not be used in the second part of query $Q$. Further, both the update attribute $B$ and the outcome attribute $Y$ belong to $\attr(R)$. {In Section~\ref{sec:single-block} we show how what-if queries are answered on each block that cannot be decomposed further.} 
\begin{proposition}\label{prop:single-relation-decompose}
Given a single-relation database $D = R = \cview$ containing both the update attribute $B$ and outcome attribute $Y$, its block-independent decomposition $\mathcal{B} = \{D_1, \ldots, D_\ell\}$, and a what-if query $Q$ whose result on a possible world $I \in PWD(D)$ is $\valwhatif(Q, D, I) = aggr(\{Y_{I}[t]~:~ \pred_{\forw}(t) = true, t \in \cview\})$ (Definition~\ref{def:whatif-result-posworld}), if 
$aggr$ is a decomposable function, i.e., if there exist  functions $g$ and $f'_{Q, D}$ according to Definition~\ref{def:decomposable}, then 
\begin{small}
\begin{equation}\label{eq:whatif-block-total}
\valwhatif(Q, D) = g(\{\valwhatif(Q', D_i) ~:~ \forall D_i\in \mathcal{B}\})
\end{equation}
\end{small}
where $Q'$ is the same query as $Q$ with $f'_{Q, D}$ replacing $aggr$, $\overline{PWD}(D_i)$ denotes the possible worlds for tuples in $D_i$, and
\begin{small}
\begin{equation} 
    \valwhatif(Q', D_i) = \expectation_{I_i \in \overline{PWD}(D_i)}[\valwhatif(Q', D_i, I_i)]
    \end{equation}
\end{small}
\end{proposition}

\begin{proof}





Recall the query result in Definition \ref{def:whatif-result}: 
\begin{align}\label{equn:whatifanswer}
\valwhatif(Q, D) = \mathbb{E}_{I \in PWD(D)}[\valwhatif(Q, D, I)]\nonumber\\
 = \sum_{I \in PWD(D)} \pr_{D, U}(I) \times \valwhatif(Q, D, I) 
\end{align}

Using the assumption that $\valwhatif(Q,D,I) = g(\{ \valwhatif(Q', D_i, I_i)~:~I_i\in \mathcal{B}_I\})$, we get the following.
\begin{align}
\scriptsize
  \valwhatif(Q,D) = \sum_{I\in PWD(D)}\left(\pr_{D,U}(I) \cdot g(\{  \valwhatif(Q', D_i, I_i)~:~I_i\in \mathcal{B}_I\} )\right)\label{eq:13}
 \end{align}

Assuming block-level independence, we substitute $\pr_{D,U}(I)$ for $\prod_{I_i\in \mathcal{B}_I} \pr_{D_i,U}(I_i)$, where $\pr_{D_i,U}$ denotes the post-update probability distribution of $D_i$. 
Therefore, 
 
 \begin{align}
  (\ref{eq:13})  &=\sum_{I\in PWD(D)}\left(\left(\prod_{I_j\in \mathcal{B}_I} \pr_{D_j,U}(I_j)\right)  \left( g(\{ \valwhatif(Q', D_i, I_i)~:~I_i\in \mathcal{B}_I\})\right)\right) 
 &=\sum_{I\in PWD(D)} g\left(\left\{\left(\prod_{I_j\in \mathcal{B}_I} \pr_{D_j,U}(I_j)\right)  \valwhatif(Q', D_i, I_i)~:~I_i\in \mathcal{B}_I\right\}\right)
 \end{align}
 
 Note that for the second transition, we used the property of function $g$ in Definition \ref{def:decomposable}:
 $\alpha g(\{x_1,\ldots,x_l\}) = g(\{\alpha x_1,\ldots,\alpha x_l\})$, $\forall \alpha \geq 0$.

Now, suppose that for the block $I_i\in \mathcal{B}_I$ of $I \in PWD(D)$, the corresponding block in $D$ is $D_i\in \mathcal{B}_D$, {with tuples having the same key. }\\
Separating out $\pr_{D_i,U}(I_i)$ from $\prod_{I_j\in \mathcal{B}_I} \pr_{D_j,U}(I_j)$, we get the following.  

 \begin{align}
 &=\sum_{I\in PWD(D)} g \left\{ \left(\left(\prod_{I_j\in \mathcal{B}_I\setminus\{I_i\}} \pr_{D_j,U}(I_j)\right) \times \pr_{D_i,U}(I_i) \times  \valwhatif(Q', D_i, I_i)\right) ~:~ ~I_i\in \mathcal{B}_I\right\} \\
  &=\sum_{I\in PWD(D)}g\left\{ \left(\left(\prod_{I_j\in \mathcal{B}_I\setminus\{I_i\}}\pr_{D_j,U}(I_j)\right) \times \pr_{D_i,U}(I_i) \times \valwhatif(Q', D_i, I_i)\right) ~:~ \forall D_i\in\mathcal{B}_D\right\} \\
&=g\left\{\sum_{I\in PWD(D)} \left(\left(\prod_{I_j\in \mathcal{B}_I\setminus\{I_i\}}\pr_{D_j,U}(I_j)\right) \times \pr_{D_i,U}(I_i) \times \valwhatif(Q', D_i, I_i)\right) ~:~ \forall D_i\in \mathcal{B}_D\right\}\label{eq:19}
\end{align}
In the last step, we used the property of function $g$ from Definition~\ref{def:decomposable}: $g(\{x_1,\ldots,x_l\}) + g(\{y_1,\ldots,y_l\}) = g(\{x_1+y_1,\ldots, x_l+y_l\})$.\\
Substituting $PWD(D)$ as the Cartesian product of $\overline{PWD}(D_k)$ over blocks, $PWD(D)= \bigtimes_{D_k\in \mathcal{B}_D} \overline{PWD}(D_k)$, hence,
\begin{align}
(\ref{eq:19})&=g\left\{\sum_{\substack{I\in \bigtimes_{D_k\in \mathcal{B}_D}  \overline{PWD}(D_k)}} \left(\left(\prod_{I_j\in \mathcal{B}_I\setminus\{I_i\}}\pr_{D_j,U}(I_j)\right)  \times \pr_{D_i,U}(I_i)  \times \valwhatif(Q', D_i, I_i)\right) ~:~ \forall D_i\in \mathcal{B}_D\right\}
\end{align}

Substituting ${\bigtimes_{D_k\in \mathcal{B}_D}  \overline{PWD}(D_k)} =  \overline{PWD}(D_i)\times \left(\bigtimes_{D_k\in \mathcal{B}_D\setminus \{D_i\}}  \overline{PWD}(D_k)\right)$

\begin{align}
&=g\left\{\sum_{\substack{I\in\overline{PWD}(D_i)\times \\\left(\bigtimes_{D_k\in \mathcal{B}_D\setminus \{D_i\}}  \overline{PWD}(D_k)\right) }} \left(\left(\prod_{I_j\in \mathcal{B}_I\setminus\{I_i\}}\pr_{D_j,U}(I_j)\right)  \times \pr_{D_i,U}(I_i)  \times \valwhatif(Q', D_i, I_i)\right) ~:~ \forall {D_i\in \mathcal{B}_D}\right\}
\end{align}
Let $I=I_i\cup I_i'$ where  $I_i'\in \left(\bigtimes_{D_k\in \mathcal{B}_D\setminus \{D_i\}}  \overline{PWD}(D_k)\right)$.

\begin{align}
&=g\left\{\sum_{\substack{I_i\in \\\overline{PWD}(D_i) }} \quad \sum_{\substack{I_i'\in \\\left(\bigtimes_{D_k\in \mathcal{B}_D\setminus \{D_i\}}  \overline{PWD}(D_k)\right) }} \left(\left(\prod_{I_j\in \mathcal{B}_I\setminus\{I_i\}}\pr_{D_j,U}(I_j)\right) \times  \pr_{D_i,U}(I_i) \times  \valwhatif(Q', D_i, I_i)\right)  ~:~  \forall {D_i\in \mathcal{B}_D}\right\}\label{eq:22}
\end{align}


Separating out the terms that depend on $D_i$ and $I_i$ from the rest.
\begin{align}
&=g\left\{\sum_{\substack{I_i\in \\ \overline{PWD}(D_i) }}\left(\left(\pr_{D_i,U}(I_i) \times \valwhatif(Q', D_i, I_i)\right) \quad \times \sum_{\substack{I_i'\in \\ \left(\bigtimes_{D_k\in \mathcal{B}_D\setminus \{D_i\}}  \overline{PWD}(D_k)\right) }} \left(\prod_{I_j\in \mathcal{B}_I\setminus\{I_i\}}\pr_{D_j,U}(I_j)\right) \right) ~:~ \forall {D_i\in \mathcal{B}_D}\right\} 
\label{eq:24}
\end{align}

Blocks $I_j\in \mathcal{B}_I\setminus\{I_i\}$ are independent. Therefore, $\left(\prod_{I_j\in \mathcal{B}_I\setminus\{I_i\}}\pr_{D_j,U}(I_j)\right)=\pr_{D\setminus D_i,U}(I_i')$, where $I_i' = \cup_{I_j \in \mathcal{B}_I\setminus\{I_i\}} I_j$,
which denotes the post-update probability of all blocks except $I_i$. Hence, 
\begin{align}
(\ref{eq:24})&=g\left\{\sum_{\substack{I_i\in \\ \overline{PWD}(D_i) }}\left(\left(\pr_{D_i,U}(I_i) \times \valwhatif(Q', D_i, I_i)\right) \quad \times \sum_{\substack{I_i'\in \\ \left(\bigtimes_{D_k\in \mathcal{B}_D\setminus \{D_i\}}  \overline{PWD}(D_k)\right) }} \left(\pr_{D\setminus D_i,U}(I_i')\right) \right) ~:~  \forall {D_i\in \mathcal{B}_D}\right\}\label{eq:22}
\end{align}
Since $\sum_{\substack{I_i'\in\left(\bigtimes_{D_k\in \mathcal{B}_D\setminus \{D_i\}}  \overline{PWD}(D_k)\right) }} \left(\pr_{D\setminus D_i,U}(I_i')\right)$ is $1$ (the sum of probabilities of possible worlds of all blocks except $D_i$),
\begin{align}
(\ref{eq:22})&=g(\{\sum_{\substack{I_i\in\overline{PWD}(D_i) }}\left(\pr_{D_i,U}(I_i) \times  \valwhatif(Q', D_i, I_i)\right) ~:~ \forall D_i\in \mathcal{B}_D \})
\end{align}
Notice that the term $\sum_{\substack{I_i\in\overline{PWD}(D_i) }}\left(\pr_{D_i,U}(I_i) \times \valwhatif(Q', D_i, I_i)\right)$ denotes the expected value of $f'_{Q, D_i}$ over the post-update distribution, denoted by $\expectation_{I_i \in \overline{PWD}(D_i)}[\valwhatif(Q', D_i, I_i)]$, and thus $\valwhatif(Q',D_i) = \expectation_{I_i \in \overline{PWD}(D_i)}[\valwhatif(Q', D_i, I_i)]$,
and $\valwhatif(Q,D) = g(\{\valwhatif(Q',D_i) ~:~ \forall D_i \in \mathcal{B}\})$ as stated in the proposition. 
\end{proof}

%% file: proofs/pre-post-split-lemma.tex
\subsection{Computation for a single-block \label{sec:single-block}}
In this section we show how to compute $\valwhatif(Q',D_i)$ using the causal graph of the block $D_i$ given a (possibly modified) what-if query $Q'$. 
First, in Section~\ref{sec:singleblock-simple}, we consider the case where the predicate in the $\forw$ operator ($\mu_{\forw}$) is a disjunction of different $\forw$ sub-operators explained below. 
We then show that a $\forw$ clause that does not satisfy a disjoint property can be modified using the principle of inclusion-exclusion.
Lastly, we show that any $\forw$ clause can be represented as a disjunction that satisfies these properties in Section~\ref{sec:generalfor}. 

\subsubsection{$\forw$ operator has Disjunction of Conjunctions of $\pre$ and $\post$ operators, and $Agg = \ct$ in the what-if query}\label{sec:singleblock-simple}
Here we assume that the aggregate operator $Agg = \ct$ in the what-if query. Further, we assume that the $\forw$ operator ($\mu_{\forw}$) is a disjunction of different $\forw$ sub-operators denoted by $\lor_k \pred_{\forw}^k$ and these sub-operators  satisfy the following conditions.
\begin{enumerate}
    \item Each sub-operator $\pred_{\forw}^k$ can be decomposed into a conjunction over two $\forw$ clauses, one denoting $\forw$ condition on pre-update values of the tuples, and the other referring to the post-update values of the tuples. This condition is required to separate out the conditions applied by the $\forw$ operator on the original/pre-update value of a tuple $t\in D$ and its post-update values.
    \item Disjointness: Each pair of tuple (t,t'), where $t\in D,t'\in I$ for any $I\in PWD (D)$ satisfies at most one of the sub-operators $\pred_{\forw}^k$. 
\end{enumerate} 
For example, consider a $\forw$ clause, $$\left(\pre(A_1)=1 \right) \lor \left(\pre(A_1)\in\{2,3,4\} \wedge \post(A_2)=2 \right) \lor \left(\pre(A_1)>4 \wedge \post(A_2)=5 \right).$$ 

\cut{
$\pre(A_1)=1  \lor \pre(A_1)\in\{2,3,4\},  \post(A_2)=2  \lor \pre(A_1)>4,  \post(A_2)=5 .$ 
}

It consists of three different sub-clauses separated by disjunctions: (a) $\pre(A_1)=1$, (b) $\pre(A_1)\in\{2,3,4\} \wedge \post(A_2)=2$, and (c) $\pre(A_1)>4 \wedge \post(A_2)=5$. In this case a tuple $t\in D$ and its post-update tuple $t'\in I$ can satisfy only one of the three sub-clauses.\\

\noindent
\textbf{(A) Computation of what-if query in a block in terms of the post-update probabilities of tuples}.~ Proposition~\ref{prop:singlerel-simple-backdoor-count} shows how the computation in each block is done by the post-update probabilities, which we further reduce to pre-update probabilities in step (B) below.   
To prove Proposition~\ref{prop:singlerel-simple-backdoor-count}, we augment the notation presented in Definition \ref{def:whatif-result-posworld} for the $\forw$ operator to be more fine-grained and define $\pred_{\forw,\pre}$ and $\pred_{\forw,\post}$ as the conditions in the $\forw$ operator that are defined with the $\pre$ and $\post$ operators, respectively. 
The Boolean representation of disjoint $\forw$ clauses is denoted as $\lor_k (\pred_{\forw,\pre}^k\wedge \pred_{\forw,\post}^k)$ where any tuple $t\in D$ and the corresponding tuple $t'$ sharing the same key (denoted by $\key[t] = \key[t']$, where $\key$ refers to all attributes defining the primary key of the tuple) in any possible world $I\in PWD(D)$ satisfies at most one of the sub-clauses $(\pred_{\forw,\pre}^k\wedge \pred_{\forw,\post}^k)$. 

%% file: proofs/singleblockproof.tex
\begin{proposition}\label{prop:singlerel-simple-backdoor-count}
Given a single-relation  database $D$ with its block independent decomposition $ \mathcal{B}_D$, a block $D_i\in \mathcal{B}_D$, a ground causal graph $G$, a what-if query $Q'$ where $Agg=\ct$, and $\forw$ operator is denoted by $\pred_{\forw}$ where $\pred_{\forw}$ can be represented as a disjunction of conjunction of disjoint $\forw$ conditions,  $\lor_k(\pred_{\forw,\pre}^k\wedge\pred_{\forw,\post}^k)$, the following holds.
\begin{small}
\begin{equation} \label{eq:whatif-singlerel-simple-backdoor-count}
    \valwhatif(Q', D_i) =  \sum_{t\in D_i} \left( \sum_k\left( \pr_{D_i,U}( \pred_{\forw,\post}^k(t)=\true | \pred_{\forw,\pre}^k(t)=\true)\right)\right)
    \end{equation}
\end{small}
In this equation, $\pr_{D_i,U}( \mu_{\forw,\post}^k(t)=\true  | \pred_{\forw,\pre}^k(t)=\true)$ denotes the sum of probabilities of all possible worlds of $D_i$ such that the tuple $t$ that satisfied $\pred_{\forw,\pre}^k(t)=\true$ before the update $U$ also satisfies $\pred_{\forw,\post}^k(t)$ after the update.

\end{proposition}
\begin{proof}
Using equation (\ref{eq:whatif-block}) in Proposition~\ref{prop:blocks}, we expand $\valwhatif(Q',D_i)$ as follows. Here $\mathbbm{1}$ denotes the indicator function. 
\begin{align}
\valwhatif(Q',D_i) &= \expectation_{I_i \in \overline{PWD}(D_i)}[\valwhatif(Q', D_i, I_i)]\label{eq:25}\\
& = 
\sum_{\substack{I_j\in\overline{PWD}(D_i) }}\left(\pr_{D_i,U}(I_j) \times \valwhatif(Q', D_i, I_j)\right) \label{eq:26} \\
&= \sum_{\substack{I_j\in\overline{PWD}(D_i) }}\left(\pr_{D_i,U}(I_j) \times \sum_{t\in D_i,t'\in I_j ~:~ \key[t]=\key[t']}\left(\mathbbm{1}\{\lor_k\left(\pred_{\forw,\pre}^k(t)=\true \wedge \pred_{\forw,\post}^k (t')=\true\right)\}\right)\right)\label{eq:27}
\end{align}
Since $\lor_k(\pred_{\forw,\pre}^k\wedge\pred_{\forw,\post}^k)$ consists of disjoint $\forw$ conjunctive predicates, a pair of tuples $(t,t')$ having the same $\key$ can satisfy atmost one of the sub-predicates. Therefore, $\left(\mathbbm{1}\{\lor_k\left(\pred_{\forw,\pre}^k(t)=\true \wedge \pred_{\forw,\post}^k (t')=\true\right)\}\right)$ can be written as a sum of different indicator random variables.

\begin{align}
&= \sum_{\substack{I_j\in\overline{PWD}(D_i) }}\left(\pr_{D_i,U}(I_j) \times \sum_{t\in D_i,t'\in I_j ~:~\key[t]=\key[t']}\left(\sum_k\mathbbm{1}\left\{\pred_{\forw,\pre}^k(t)=\true \wedge \pred_{\forw,\post}^k (t')=\true\right\}\right)\right)
\end{align}
By splitting the inner indicator into a product of the indicators of the two conjunctions and extracting the sum over $k$:
\begin{align}
&= \sum_{\substack{I_j\in\overline{PWD}(D_i) }}\sum_{k}\left(\pr_{D_i,U}(I_j) \times  \sum_{t\in D_i} \left(\mathbbm{1}\{\pred_{\forw,\pre}^k(t)=\true\} \times \mathbbm{1}\{ \pred_{\forw,\post}^k(t')=\true, \text{where } \key[t]=\key[t'], t'\in I'\} \right)\right) \\
&=\sum_{t\in D_i}\sum_k\left(\mathbbm{1}\{ \pred_{\forw,\pre}^k(t)=\true \} \times \sum_{\substack{I_j\in\overline{PWD}(D_i) }}\left(\pr_{D_i,U}(I_j) \times  \sum_{t'\in I_j ~:~ \key[t]=\key[t']}\mathbbm{1}\{ \pred_{\forw,\post}^k(t')=\true \}\right)\right)\\
&=\sum_{t\in D_i}\left(\sum_k\mathbbm{1}\{ \pred_{\forw,\pre}^k(t)=\true\} \times \left( \sum_{\substack{I_j\in\overline{PWD}(D_i)\\ t'\in I_j, \key[t]=\key[t']} }\left(\pr_{D_i,U}(I_j) \times  \mathbbm{1}\{   \pred_{\forw,\post}^k(t')=\true\}\right)\right)\right) \\
&=\sum_{t\in D_i}\left(\sum_k\left( \sum_{\substack{I_j\in\overline{PWD}(D_i)\\ t'\in I_j, \key[t]=\key[t']} }\left(\pr_{D_i,U}(I_j) \times  \mathbbm{1}\{  \pred_{\forw,\pre}^k(t)=\true \wedge \pred_{\forw,\post}^k(t')=\true\}\right)\right)\right) \\
&=\sum_{t\in D_i} \left(\sum_k\left( \pr_{D_i,U}( \mu_{\forw,\post}^k(t)=\true | \mu_{\forw,\pre}^k(t)=\true)\right)\right)\label{eq:postupdate-sum-t}
\end{align}

Note that if a tuple $t$ is not affected by the update, $\pr_{D_i,U}( \mu_{\forw,\post}^k(t)=\true | \mu_{\forw,\pre}^k(t)=\true) = \mathbbm{1} \{\mu_{\forw,\pre}^k(t)=\true \wedge \mu_{\forw,\post}^k(t)=\true\}$.\\

\end{proof}

\noindent
\textbf{(B) Reduction of post-update probability in equation (\ref{eq:whatif-singlerel-simple-backdoor-count}) of Proposition~\ref{prop:singlerel-simple-backdoor-count} in terms of the causal graph of given database $D$.~} 
The expression in equation (\ref{eq:whatif-singlerel-simple-backdoor-count}) in Proposition~\ref{prop:singlerel-simple-backdoor-count} relies on the post-update probability distribution of the block $D_i$, denoted by $\pr_{D_i,U}$. 
We now use the {\bf backdoor criterion} from causal inference literature \cite{pearl2009causality} to simplify these expressions and estimate the probability from the input database $D$, which we review briefly. 
A set of attributes $\mb C$ satisfies the backdoor criterion w.r.t. $B$ and $Y$ if no attribute $C\in \mb C$ is a descendant of $Y$ or $B$ and all paths from $B$ to $Y$ which contain an incoming edge into $Y$ are \emph{blocked} by $\mb C$. A path is considered to be blocked by $\mb C$ if there is a non-collider attribute\footnote{A collider is a vertex in the causal graph with two incoming edges. For example, $A\rightarrow B\leftarrow C$ has $B$ as a collider.} on the path that is present in $\mb C$ or if a collider attribute is not in $\mb C$ then none of the descendant of the collider is in $\mb C$.   
With the help of the backdoor criterion, we leverage the following property for our simplification~\cite{pearl2009causality}, which reduces post-update probability $\pr_{D, U}$ to the pre-update distribution $\pr_D$. 
    \begin{small}
    \begin{align} \scriptsize
        \pr_{D, U}(Y=y \mid B=b,\mb C=\mb c )= \pr_{D}(Y=y \mid B=f(b), \mb C=\mb c )  \label{eq:back2}
    \end{align}
    \end{small}
{where $f(b)$ denotes the post-update value of $B = b$. } 

\emph{Computation of blocking set $C$:~} Let $\mb C$ denote a set of attributes that satisfy the backdoor criterion with respect to the update attribute $B$ and the attributes in $\pred_{\forw,\post}^k$. We use the ground causal graph $G$ to identify the minimal subset of all ancestors of $B$ and attributes in $\pred_{\forw,\post}^k$ that block all backdoor paths~\cite{pearl2009causality} by a greedy procedure: we start with all non-descendants of $B, Y$ excluding $B, Y$ as  $\mb C$, and the remove one node at a time until we reach a minimal set for blocking that cannot be reduced further.   In case $G$ is not known, we consider all attributes of all tuples in the block $D_i$ to satisfy the backdoor criterion\footnote{This design choice guarantees that the set  $\mb C$ is always a superset of the optimal set of backdoor attributes and is commonly used as a proxy in causal inference~\cite{GalhotraPS21}}.

\emph{Computation of post-update probability for $Agg = \ct$}
We will use $\mb C_k$ to denote the backdoor set for sub-predicate $\pred^k$, and $\mb{c}_k \in \Dom(\mb{C}_k)$ to denote a combination of values from the domain of these nodes.  
Then
\begin{small}
\begin{align}
\allowdisplaybreaks
&\pr_{D_i,U}(\mu_{\forw,\post}^k(t)=\true | \mu_{\forw,\pre}^k(t)=\true)\label{eq:35}\\
 &=  \sum_{\mb c_k\in~\Dom(\mb C_k)}\left( \pr_{D_i,U}( \pred_{\forw,\post}^k(t)=\true ~|~ \pred_{\forw,\pre}^k(t)=\true,\mb C_k[t]=\mb c_k) \times \pr_{D_i, U}(\mb C_k[t]=\mb c_k~|~\pred_{\forw,\pre}^k(t)=\true)\right)\label{eq:36}
 \end{align}
 Since the second component only involves non-descendants of the update attribute $B$ in the set $\mb C_k$, therefore for these $\mb C_k[t]$, post-update probability $\pr_{D_i, U}$ is the same as the pre-update probability $\pr_{D_i}$. Hence,
 \begin{align}
 (\ref{eq:36}) &=  \sum_{\mb c_k\in~\Dom(\mb C_k)}\left( \pr_{D_i,U}( \pred_{\forw,\post}^k(t)=\true ~|~ \pred_{\forw,\pre}^k(t)=\true,\mb C_k[t]= \mb c_k) \times \pr_{D_i}(\mb C_k[t]=\mb c_k~|~\pred_{\forw,\pre}^k(t)=\true)\right) \label{eq:37}
  \end{align}
 We now use the same simplification to split the first term into two terms, 
 using conditional probabilities with respect to the value $b$ of $B$ before the update.
  \begin{align}
  (\ref{eq:37}) &= \sum_{\mb c_k\in~\Dom(\mb C_k)} ( \sum_{b\in~ \Dom(B)}\left( \pr_{D_i,U}( \pred_{\forw,\post}^k(t)=\true ~|~ \pred_{\forw,\pre}^k(t)=\true,B[t]=b, \mb C_k[t]=\mb c_k) \times \pr_{D_i,U}(B[t]=b~|~\pred_{\forw,\pre}^k(t)=\true,\mb C_k=\mb c_k) \right)\label{eq:38} 
 \nonumber\\ & \quad \quad \quad \quad\quad \quad
 \times  \pr_{D_i}\left(\mb C_k[t]=\mb c_k~|~\pred_{\forw,\pre}^k(t)=\true\right) ) 
  \end{align}
 Since $B[t] = b$ refers to the pre-update value of attribute $B$, the second term $\pr_{D_i,U}(B[t]=b~|~\pred_{\forw,\pre}^k(t)=\true,\mb C_k=\mb c_k) $ is the same as $\pr_{D_i}(B[t]=b~|~\pred_{\forw,\pre}^k(t)=\true,\mb C_k=\mb c_k) $. Hence,
   \begin{align}
(\ref{eq:38}) &= \sum_{\mb c_k\in~\Dom(\mb C_k)}( \sum_{b\in~\Dom(B)}\left( \pr_{D_i,U}( \pred_{\forw,\post}^k(t)=\true ~|~ \pred_{\forw,\pre}^k(t)=\true,B[t]=b, \mb C_k[t]=\mb c_k) \times \pr_{D_i}(B[t]=b~|~\pred_{\forw,\pre}^k(t)=\true,\mb C_k=\mb c_k) \right)  \nonumber\\
 & \quad \quad \quad \quad\quad \quad \times \pr_{D_i} \left(\mb C_k[t]=\mb c_k|\pred_{\forw,\pre}^k(t)=\true\right))    \label{eq:eq39}
\end{align}
Using, equation (\ref{eq:back2}), we replace the post-update probability $\pr_{D_i,U}$ in the first term with $\pr_{D_i}$ and $B[t]=b$ with $B[t]=f(b)$ as specified in the update $U$:

  \begin{align}
(\ref{eq:eq39}) &= \sum_{\mb c_k\in~\Dom(\mb C_k)}( \sum_{b\in~\Dom(B)}\left( \pr_{D_i}( \pred_{\forw,\post}^k(t)=\true ~|~ \pred_{\forw,\pre}^k(t)=\true,B[t]=f(b), \mb C_k[t]=\mb c_k) \times \pr_{D_i}(B[t]=b~|~\pred_{\forw,\pre}^k(t)=\true,\mb C_k=\mb c_k) \right)  \nonumber\\
 & \quad \quad \quad \quad\quad \quad \times \pr_{D_i} \left(\mb C_k[t]=\mb c_k|\pred_{\forw,\pre}^k(t)=\true\right)) 
 \label{eq:eq40}
\end{align}
\end{small}

Replacing (\ref{eq:eq40}) in equation (\ref{eq:postupdate-sum-t}) and summing over all tuples $t$ in $D_i$ and all disjoint sub-predicates $\pred_{\forw,\pre}^k \wedge \pred_{\forw,\post}^k$, we get the final expression for computing the post-update probability for $Agg = \ct$.\\

\noindent
\textbf{Complexity} The computation of (\ref{eq:eq40}) iterates over all values in the domain of attributes ${\mb C_k}\cup \{B\}$ and computes three different probability values for each value of these attributes. Each  probability calculation expression is estimated from the input database $D$ using regression analysis and runs in time linear in the number of records under the homogeneity assumption (please see Section~\ref{sec:implementation} for more details). Additionally, $\pr_{D_i} \left(B[t]=b|\mb C_k[t]=c_k,\pred_{\forw,\pre}^k(t)=\true\right)$ is $0$ if the original database contains no tuple with the value $c_k$ for ${\mb C_k}$ and $b$ for $B[t]$. Therefore, the expression contains non-zero terms only when the support of attribute values $c_k\in \Dom(\mb C_k)$ and $b$ is non-zero. Using this property, our implementation 
first identifies all values in ${\mb C_k}\cup\{B\}$ that have non-zero support and ignores the rest.
Therefore, the overall complexity is $O(n\times \gamma(B\cup \mb C_k))$ where the $\gamma$ function identifies values with non-zero support. 
This shows that $\gamma(B\cup \mb C_k)< n$ (because each value has non-zero support) and $\gamma(B\cup \mb C_k)< |\Dom(B)| \times_{A\in \mb C_k}|\Dom(A)| $ (because $\gamma$ denotes a subset of all possible values in the domain of the attributes), simplifying the overall complexity to $O(n\times \min \{n,|\Dom(B)|\times_{A\in \mb C_k}|\Dom(A)| \})$. Hence, the computation can be done in time polynomial in data complexity \cite{Vardi82} (when the size of the schema and the query is fixed), but can be exponential in query complexity depending on the size of the backdoor set $\mb C_k$.\\


\noindent
\textbf{Probability distribution $\pr_{D_i}$ } denotes the probability distribution of constructing $D_i$ which is dependent on the causal graph $G$. Even though the initial database $D$ is fixed, we assume that all tuples are generated homogeneously according to the causal graph.  

\subsubsection{Computation for $Agg = SUM$ and $AVG$}\label{sec:agg-sum} Proposition~\ref{prop:singlerel-simple-backdoor-count} showed that a disjunction of disjoint $\forw$ sub-predicates translates to a summation of 
probability values when $Agg = \ct$. The condition for $Agg = SUM$ and its proof are similar. 
We now simplify $\valwhatif$ when $Agg=SUM$ for a single sub-predicate which consists of a conjunction of $\pre$ and $\post$ predicates $(\pred_{\forw,\pre}\wedge\pred_{\forw,\post})$. In general, the final value is obtained by summing over all sub-predicates $(\pred^k_{\forw,\pre}\wedge\pred^k_{\forw,\post})$ similar to (\ref{eq:postupdate-sum-t}).

\begin{proposition}\label{prop:sum}
Given a single-relation database $D$ and a block $D_i\in \mathcal{B}_D$ and a what-if query $Q'$ with aggregate $Agg=\sumsql$, where the predicate in the $\forw$ operator is $\pred_{\forw} =  (\pred_{\forw,\pre}\wedge\pred_{\forw,\post})$, the following holds.
\begin{small}
\begin{equation} 
    \valwhatif(Q', D_i) =  \sum_{t\in D_i} \left(\sum_{y\in \Dom(Y)} \left(y \times  \pr_{D_i,U}( Y[t]=y,\mu_{\forw,\post}(t)=\true ~|~ \mu_{\forw,\pre}(t)=\true)\right)\right)
    \end{equation}
\end{small}
\end{proposition}
\begin{proof}
Similar to (\ref{eq:25})-(\ref{eq:26}), 
\begin{align}
\valwhatif(Q',D_i) &=\sum_{\substack{I_j\in\overline{PWD}(D_i) }}\left(\pr_{D_i,U}(I_j) \times  \valwhatif(Q',D_i)\right) \\
&= \sum_{\substack{I_j\in\overline{PWD}(D_i) }}\left(\pr_{D_i,U}(I_j) \times  \sum_{t\in D_i,t'\in I_j ~:~ \key{}[t]=\key{}[t']} \left(Y[t'] \times \mathbbm{1}\{\pred_{\forw,\pre}(t)=\true \wedge \pred_{\forw,\post} (t')=\true\}\right)\right) \\
&= \sum_{\substack{I_j\in\overline{PWD}(D_i) }}\left(\pr_{D_i,U}(I_j) \times \sum_{t\in D_i} \left(Y[t'] \times \mathbbm{1}\{\pred_{\forw,\pre}(t)=\true\} \times \mathbbm{1}\{ \pred_{\forw,\post}(t')=\true, \text{where } \key{}[t]=\key{}[t'], t'\in I'\} \right)\right) \\
&=\sum_{t\in D_i}\left(\mathbbm{1}\{ \pred_{\forw,\pre}(t)=\true \} \times  \sum_{\substack{I_j\in\overline{PWD}(D_i) }}\left(\pr_{D_i,U}(I_j) \times \sum_{t'\in I_j ~:~ \key{}[t]=\key{}[t']}Y[t'] \times \mathbbm{1}\{ \pred_{\forw,\post}(t')=\true \}\right)\right)\\
&=\sum_{t\in D_i} \left(\mathbbm{1}\{ \pred_{\forw,\pre}(t)=\true\} \times  \sum_{\substack{I'\in\overline{PWD}_{D_i}\\ t'\in I',\key{}[t]=\key{}[t']} }Y[t'] \times \left(\pr_{D_i,U}(I') \times \mathbbm{1}\{   \pred_{\forw,\post}(t')=\true\}\right)\right) \\
&=\sum_{t\in D_i}\left( \sum_{y\in \Dom(Y)} \left(y \times \pr_{D_i,U}(Y[[t]=y,  \mu_{\forw,\post}(t) =\true ~|~ \mu_{\forw,\pre}(t)=\true)\right)\right)
\end{align}

\end{proof}
The post-update probability distribution $\pr_{D_i,U}$ can be estimated from the input database $D_i$ using the backdoor criterion, as shown above in equations (\ref{eq:35})-(\ref{eq:eq40}).
Proposition~\ref{prop:sum} extends to the case where $Agg=\avg$ as $\avg$ is equivalent to dividing the output of $\sumsql$ by the number of tuples, $|D_i|$, which remains constant in all possible worlds of $D_i$. Similarly, Proposition~\ref{prop:sum} extends to any aggregate function that can be expressed as $c\times \sumsql$ for some constant $c$. 


\subsubsection{Relaxing the disjointness property of the $\forw$ predicate expressed as a Boolean formula.} 
When the $\forw$ operator cannot be directly expressed as a disjunction of disjoint sub-predicates but is an arbitrary Boolean formula, it can be translated into an equivalent formulation that satisfies disjointness by using the principle of \emph{inclusion-exclusion}. For example, consider $\pred_{\forw} = \pred^1_{\forw} \lor \pred^2_{\forw}$. that does not satisfy disjoint property.  Using principle of inclusion exclusion, 
$\pred_{\forw} = (\pred^1_{\forw}\wedge \bar{\pred}^2_{\forw} ) \lor (\pred^2_{\forw}\wedge \bar{\pred}^1_{\forw} ) \lor (\pred^1_{\forw}\wedge \pred^2_{\forw}) $ where $\bar{\pred}$ denotes the negation of the $\forw$ operator. In this way, any general Boolean formula can be split into different components that satisfy disjoint property. \\
\noindent \emph{Complexity:} If the Boolean formula consists of $t$ sub-predicates separated by disjunction, the disjoint sub-predicates identified by the principle of inclusion-exclusion is $2^t$ where each sub-predicate contains the same set of attributes as the ones in the original $\forw$ predicate. Note that this translation of the Boolean formula does not affect the dependence of our algorithm on the dataset size, hence the complexity still remains polynomial in data complexity. 

\subsubsection{Extension to general $\forw$ predicates\label{sec:generalfor}} In the two previous propositions, we considered the case where $\forw$ can be represented as a Boolean formula over different sub-predicates involving single tuples $t$. 
In this section, we analyze more complex $\forw$ operators. For example, consider a for clause $\pred_{\forw} \equiv \pre(A_i)-\post(A_i)<2$, where the \pre\ and the \post\ conditions are immediately not separable and 
we cannot decompose the $\forw$ operator directly. Instead, we construct a different $\forw$ predicate which captures the same set of tuples but can be represented as a disjunction of disjoint sub-predicates over \pre\ and \post\ attribute values of tuples.

\begin{proposition}
Given a what-if query $Q$ with $\forw$ operator $\mu_{For}$, the output of the query is equivalent to that of a what-if query $Q'$, where $Q'$ and $Q$ differ only in that the $\pred_{\forw}$ predicate of $Q'$ can be written as a disjunction of different $\forw$ operators $\lor_k (\pred_{\forw,\pre}^k\wedge \pred_{\forw,\post}^k)$ such that every tuple $t\in D$ or $t'\in I$, where $\key[t] = \key[t']$,
satisfies a single $\pred_{\forw,\pre}^k\wedge \pred_{\forw,\post}^k$ sub-predicate. 
\end{proposition}
\begin{proof}
The $\forw$ operator defines a subset of $D$ containing a single relation $R$,  and the instances $I\in PWD(D)$ to evaluate the query response. Let $T_I$ denote the set of pairs of tuples in $D$ and corresponding tuples in an instance $I\in PWD(D)$ that satisfy the complex $\pred_{\forw}$ operator. Formally, $T_I=\{(t,t') ~:~\forall t\in D, \forall t'\in I, \key[t] = \key[t'] \text{ and}~\pred_{\forw}(t,t')=\true \}$. We consider $T = \bigcup_{I\in PWD(D)} T_I$ and use these tuples to construct an alternative $\pred_{\forw}$ operator that is a disjunction of disjoint sub-operators, where each sub-operator $\pred_{\forw}^i$   uniquely captures a tuple $(t,t')\in T$, i.e., 
$\pred_{\forw}^i (t,t')=\true$ and $\false$ for any other pair of tuples. This sub-operator is defined as a conjunction of attribute values of the tuples $t$ and $t'$, i.e.,
{$\wedge_{A_i \in \Dom(R)} \pre(A_i)=A_i[t] \wedge_{A_j \in \Dom(R)} \post(A_j)=A_j[t]$}
In this way, any complex $\forw$ operator can be represented as a disjunction of at most $|T|$ $\forw$ sub-operators, where each sub-operator consists of conjunction of $\pre$ and $\post$ conditions. 
\end{proof}


We demonstrate the construction of $\pred_{\forw}$ for an example non-boolean predicate, $\pre(A_i)-\post(A_i)<2 \wedge \pre(A_i)\geq \post(A_i)$, where $\Dom(A_i)=\{1,2,3,4\}$. In this case, we iterate over the values to identify values that satisfy the condition. Different sets of values that satisfy the $\forw$ predicate are $\pre(A_i)=4\wedge \post(A_i)=3$, $\pre(A_i)=4\wedge \post(A_i)=4$, $\pre(A_i)=3\wedge \post(A_i)=2$, $\pre(A_i)=3\wedge \post(A_i)=3$,  $\pre(A_i)=2\wedge \post(A_i)=1$, $\pre(A_i)=2\wedge \post(A_i)=2$, $\pre(A_i)=1\wedge \post(A_i)=1$. Therefore, we represent $\pred_{\forw}\equiv  (\pre(A_i)-\post(A_i)<2) \wedge (\pre(A_i)\geq \post(A_i)) $ as a disjunction of seven different $\forw$ sub-predicates, each constraining the $\pre$ and $\post$ values of attributes in the original $\forw$ clause.
In this way, we can represent the original $\forw$ predicate as a disjunction of multiple sub-operator where each sub-operator contains a conjunctive condition on $\pre$ and $\post$ values of different attributes. The number of  sub-operators in this decomposition is dependent on the domain of attributes involved in the original $\forw$ clause, which is exponential in the query complexity. 


%% file: proofs/multi-relation-extension.tex
\subsection{Extension to Multi-Relation Database\label{sec:muti-relation}}
Recall from Section~\ref{sec:what-if-syntax} that, when we have multiple relations in the what-if query $Q$, we have a relevant view $\cview$ containing the primary keys of the tuples from the relation $R$ (= $D$ for a single-relation database) containing the update attribute $B$, and having other relevant attributes as well as an aggregated form of the output attribute $Y$. 
Here we argue that our analysis so far extends to what-if queries with
multiple relations because of following reasons.
\begin{itemize}
    \item $\cview$ has the same blocks as the 
    relation $R$ containing the update attribute $B$ (Proposition~\ref{prop:consistent-blocks} below). This shows that the query output by aggregating the output from individual blocks in 
    $R$ is equivalent to aggregating the output from individual blocks in $\cview$.
    \item The backdoor criterion analysis presented in (\ref{eq:back}) extends to multi-relation databases where attributes from different relations are embedded according to an aggregate function. To prove this condition, we leverage the analysis from prior literature on causal inference on multi-relation database~\cite{SalimiPKGRS20}.
\end{itemize}

\subsubsection{Proof that $\cview$ Has the Same Blocks as the Multi-Relation Database}

We next prove that the block decomposition procedure that we describe in Section \ref{sec:whatif-algo} places two tuples in the same block in a multi-relation database $D$ if and only if it places their aggregated version in $\cview$ in the same block if it was performed on $\cview$. 

Recall that our procedure for dividing the database $D$ into independent blocks, which includes taking a tuple $t_1$, identifying all tuples with paths to and from $t_1$ in the causal graph and add them to the same block as $t_1$.This is repeated until all tuples are included in some block. 

\begin{proposition}\label{prop:consistent-blocks}
Given a (multi-relation) database $D$, its block-independent decomposition $\mathcal{B} = \{D_1, \ldots, D_\ell\}$, and a what-if query $Q$ creating update view $\cview$, then $t,t' \in D$ are placed in the same block by the above procedure of computing blocks in Section \ref{sec:whatif-algo} if and only if their corresponding tuples in $\cview$, i.e.,   $t_{v}, t'_v \in \cview$ would have been placed in the same block, if the block decomposition procedure was performed on $\cview$. 
where $t_v$ corresponds to $t$ if it contains a subset of its attributes or an aggregated form thereof (i.e., $\key[t] = \key[t_v]$). 
\end{proposition}

\begin{proof}
($\Leftarrow$) Assume $t, t' \in D$ are not placed in the same block $D_i$ by our procedure in Section \ref{sec:whatif-algo}.
Assume further that the block $D_i \subseteq D$ contains $t$ (and not $t'$). 
If $t,t'$ do not have primary key-foreign key relationship, then we know that $t_v \neq t_v'$ in $\cview$ (since they cannot be summarized to the same tuple) and the attributes of $t_v$ and $t_v'$ are still independent in $\cview$ or dropped from $\cview$. Therefore $t_v$ and $t_v'$ will be in different blocks if we apply our procedure on $\cview$. 
Assume $t,t'$ are independent but have a key relationship possibly through other tuples. 
According to our procedure, $D_i$ {\em contains all tuples that have a path to or from $t$ in the causal graph}. In particular, $D_i$ contains all tuples that have a primary key-foreign key relationship with $t$, as the causal graph contains edges between such tuples. 
Since $t' \not\in D_i$, in particular, it does not share a primary key-foreign key relationship with $t$. 
As mentioned in Section \ref{sec:what-if-syntax}, $\cview$ is created over the relation $R$ containing the update attribute $B$ in $Q$, and other attributes from different relations that are aggregated to $R$ with respect to the tuples in $R$. 
Suppose $t_v, t_v' \in \cview$ are the tuples generated from the (possibly aggregated) attributes of $t, t'$ and $t \in R$ w.l.o.g. 
Here $t_v \in \cview$ can only contain summarized attributes of tuples that have a primary key-foreign key relationship with $t$, and thus cannot include attributes with the key of $t'$ and vice versa. 
Furthermore, if the attributes of $t$ and $t'$ were placed in different blocks in $D$, and they were summarized to $t_v \neq t_v' \in \cview$, then the attributes of $t_v$ and $t_v'$ are will also be placed in different blocks if the procedure is performed on $\cview$. 
So in $\cview$, $t_v, t_v' \in \cview$ will also be placed in different blocks.

($\Rightarrow$) Assume $t, t' \in D$ share the same block $D_i$, then there is a tuple $t'' \in R \cap D_i$ (it may be the case that $t=t''$ or $t'=t''$) and attributes $A, A', A''$ such that there is a path to/from $A[t]$ to/from $A''[t'']$ to/from $A'[t']$. 
If in $\cview$, $t$ and $t'$ are aggregated to the same tuple with the key of $t''$ (e.g., $r_2$, $r_3$ are summarized to the same tuple using $p_2$ in the view created by the what-if query in Figure \ref{fig:whatif-query} in our running example), 
then, denote this tuple by $t_v'' \in \cview$. $t_v''$ has the same key as $t''$ so, in particular, $t_v''$ will be in the same block with itself. 
Otherwise, both $t$ and $t'$ are in $R$, and clearly they will be placed in the same block if the procedure is performed on $\cview$ since they were placed in the same block when the procedure was performed on $D$.  
\end{proof}

\subsubsection{Backdoor Criterion for a multi-relation database.}

First, we discuss the construction of an augmented causal graph $G'$ which contains new nodes denoting aggregated values of attributes collected from different relations. Then, we present the analysis that backdoor criterion presented in equation~\ref{eq:back} holds with respect to $G'$, extending the previous analysis to this setting.

\noindent \textbf{Augmented causal graph.} Given the ground causal graph $G$, we construct an augmented causal graph $G'
$ following the procedure from prior literature~\cite{SalimiPKGRS20}. The augmented graph contains all nodes from the ground causal graph along with new nodes denoting aggregated attributes from different relations. These aggregated attribute nodes are a superset of the aggregated attributes in the \use\ clause of the query. Aggregated node $A'\equiv Agg(A_1,\ldots,A_t)$ is added as a child of every $A_i$ for all $i\in \{1,\ldots,t\}$ and $A'$ is added as a parent of all children of $A_i$ in $G$. Notice that each $A_i$ has same set of children under the homogeneity assumption. In addition to these new edges, all edges between $A_i$ and its children in the ground causal graph are removed.

Using this augmented causal graph, we show the backdoor criterion mentioned in equation~\ref{eq:back} holds for multi-relation database using two different properties. For this analysis, we define a
$\Vec{b}$ to denote a vector of attribute values $B$ of all units in an augmented causal graph. Under this notation, we 
first use the counterfactual interpretation of backdoor set~\cite{pearl2016causal} to simplify $\pr_{D,U}(Y|B=\Vec{b},\mb C=\mb c)= \pr_{D,f(\Vec{b})}(Y|\mb C=\mb c)$ (Proposition~\ref{prop:counterfactual}) where $f$ maps each value $b_i\in \Vec(b)$ according to the update. Second, we use the backdoor set analysis from~\cite{SalimiPKGRS20} to reduce
$\pr_{D,f(\Vec{b})}(Y|\mb C=\mb c)$ to $\pr_{D}(Y|B=f(\Vec{b}),\mb C=\mb c)$.

\begin{proposition} [Counterfactual Interpretation of Backdoor~\cite{pearl2016causal}]
Given an augmented causal graph $G'$ with an update $B\leftarrow f(\Vec{b})$, the following holds.
$$\pr_{D,U}(Y|B=\Vec{b},\mb C=\mb c)= \pr_{D,f(\Vec{b})}(Y|\mb C=\mb c),$$
where $\mb C$ denotes a set of attributes that satisfy the backdoor criterion in the augmented causal graph $G'$.
\label{prop:counterfactual}
\end{proposition}

Now, we re-state the result from \cite{SalimiPKGRS20} which is then used to simplify  $\pr_{D,f(\Vec{b})}(Y|\mb C=\mb c)$.

\begin{theorem}[Relational Adjustment Formula~\cite{SalimiPKGRS20}]\label{thm:carl}
Given an augmented relational causal graph $G'$,  treatment
and updated attribute $T$ with the update $U\equiv (B\leftarrow f(\Vec{b}))$  where all units that are not in a set $S$ are not updated (equivalent to $f$ denoting an identity function). Note that $S$ is defined by the \use\ clause of the query.
We have the following relational adjustment formula:
$$\pr_{D,U}[Y[x']| \mb Z=z] =\pr_{D} [Y[x']| \mb Z=\mb z, B = f(\Vec{b}) ]$$
where $\mb Z$ is the set of nodes in $G'$ corresponding to the groundings of a subset of attributes such that
$$Y[x'] \indep \left(\bigcup_{x\in S} Pa(B[x])\right) |_{G'} \left(\mb Z,\bigcup_{x\in S}B[x]\right) $$
\end{theorem}

To use this theorem, we show that the set of backdoor variables $\mb C$ satisfies the condition $Y[x'] \indep \left(\cup_{x\in S} Pa(T[x])\right) |_{G'} \mb C,\cup_{x\in S}T[x] $. 
\begin{proposition}
Given an augmented relational causal graph $G'$, with an update $B\leftarrow f(\Vec{b})$, the following holds.
$$\pr_{D,f(\Vec{b})}(Y|\mb C=\mb c) = \pr_{D}(Y|f(\Vec{b}),\mb C=\mb c)$$
where $\mb C$ denotes a set of attributes that satisfy the backdoor criterion in the augmented causal graph $G'$.
\label{prop:backdoormultirel}
\end{proposition}
\begin{proof}
Let $\mb C$ denote the set of backdoor variables for the update with respect to the augmented causal graph $G'$. This means that all backdoor paths from $B$ to $Y$ are blocked by $\mb C$. This means either of the two conditions hold 
\begin{enumerate}
    \item A variable $X\in \Pa(B)$ is in the set $\mb C$, 
    \item A variable $X\in \Pa(B)$ is not in the backdoor set $X\notin \mb C$ but the path from $X$ to $Y$ is blocked by the set $\mb C$.
\end{enumerate}

Now consider all paths from $\Pa(B)\setminus \mb C$ to $Y$. Among these paths, all paths through $B$ are blocked by $B$ and other paths are blocked by $\mb C$ (because of the second point above). Therefore, $Y$ is independent of $\Pa(B)$ when conditioned on $B$ and $\mb C$. Using $\mb C$ as the set of variables $Z$ in Theorem~\ref{thm:carl}, we get the following.
\begin{align}
    \pr_{D,f(\Vec{b})}(Y|\mb C=\mb c) = \pr_{D}(Y|f(\Vec{b}),\mb C =\mb c)
\end{align}
\end{proof}

Using Propositions~\ref{prop:counterfactual} and \ref{prop:backdoormultirel}, equation~(\ref{eq:back}) extends to the multi-relation database.

%% file: proofs/implementation.tex
\subsection{Algorithm Implementation\label{sec:implementation}}
Previous analysis showed that the query output can be decomposed into conditional probability distribution over the original database $D$ (or a block $D_i$). For implementation purpose, we assume that all tuples are homogeneously generated according to a causal graph $G$ (as mentioned in Section \ref{sec:model-PCM}). For example, a probability value $\pr_D(A_i[t]=a_i|A_j[t]=a_j), \forall t\in D$ is assumed to be distributed according to a distribution $\pr_D(A_i|A_j)$. In this case, \sys\ uses the input  database $D$ to learn a single regression function (with the conditioning set as features and $A_i$ as the prediction variable) to estimate the conditional probability distribution  $\pr_D(A_i|A_j)$ 
This assumption is commonly used in causal inference to estimate conditional effects of specific attributes on the outcome \cite{SalimiPKGRS20,vanderweele2013social}.

Our algorithms crucially rely on the domain of the set of attributes that satisfy the backdoor criterion 
(say $\mb C$).  Naively, the $\Dom (\mb C)$ grows exponentially in the number of attributes $|\mb C|$. However, majority of the values in the domain would have zero-support in the database $D$. To efficiently ignore such values $c\in \Dom(\mb C)$, we construct an index to process the database $D$ to store all values that have non-zero support. In this way, our algorithm complexity remains linear in the database size and does not grow exponentially with the size of $\mb C$.